\pdfoutput=1 

\documentclass[times,twocolumn,longtitle,final]{elsarticle}

\usepackage{medima}
\usepackage{framed,multirow}

\usepackage{amssymb}
\usepackage{latexsym}
\usepackage{xspace}
\usepackage{url}
\usepackage{xcolor}
\usepackage{hyperref}
\definecolor{newcolor}{rgb}{.8,.349,.1}
\newcommand{\finaltesturl}{\url{https://doi.org/10.5281/zenodo.10514185}\xspace}
\usepackage{soul}
\usepackage{amsmath}
\usepackage{multirow}
\usepackage{booktabs}
\usepackage{graphicx}
\usepackage{pifont}
\usepackage{lscape}
\usepackage{rotating} 
\usepackage{longtable}	
\usepackage{color, colortbl}
\usepackage{caption}
\usepackage{subcaption} 
\newsavebox{\twosubbox} 
\usepackage{pdflscape}
\usepackage{afterpage}
\usepackage{cleveref}
\usepackage{chngcntr} 
\usepackage{placeins}

\usepackage{adjustbox}
\usepackage{caption}
\usepackage{subcaption}
\usepackage{tabularx}
\journal{Medical Image Analysis}
\begin{document}
\verso{E. Huijben, M. Terpstra \textit{et~al.}}

\begin{frontmatter}

\title{Generating Synthetic Computed Tomography for Radiotherapy: SynthRAD2023 Challenge Report}

\author[1]{Evi M. C. \snm{Huijben}\fnref{fn1,fn2}}
\fntext[fn1]{Equally contributing first authors}
\fntext[fn2]{Challenge Organizer}
\address[1]{Department of Biomedical Engineering, Eindhoven University of Technology, Eindhoven, The Netherlands}

\author[2,3]{Maarten L. \snm{Terpstra}\fnref{fn1,fn2}}
\address[2]{Radiotherapy Department, University Medical Center Utrecht, Utrecht, The Netherlands}
\address[3]{Computational Imaging Group for MR Diagnostics $\&$ Therapy, University Medical Center Utrecht, Utrecht, The Netherlands}

\author[4]{Arthur Jr. \snm{Galapon}\fnref{fn2}}
\address[4]{Department of Radiation Oncology, University Medical Center Groningen, University of Groningen, Groningen, The Netherlands}

\author[5]{Suraj \snm{Pai}\fnref{fn2}}
\address[5]{Department of Radiation Oncology (Maastro), GROW School for Oncology, Maastricht University Medical Centre, Maastricht, The Netherlands}

\author[4,6]{Adrian \snm{Thummerer}\fnref{fn2}}
\address[6]{Department of Radiation Oncology, LMU University Hospital, LMU Munich, Munich, Germany}

\author[7]{Peter \snm{Koopmans}\fnref{fn2}}
\address[7]{Department of Radiation Oncology, Radboud University Medical Center, Nijmegen, The Netherlands}

\author[8]{Manya \snm{Afonso}\fnref{fn2}}
\address[8]{Wageningen University $\&$ Research, Wageningen Plant Research, Wageningen, The Netherlands}

\author[1]{Maureen \snm{van Eijnatten}\fnref{fn2}}
\author[9,10]{Oliver \snm{Gurney-Champion}\fnref{fn2}}
\address[9]{Department of Radiology and Nuclear Medicine, Amsterdam UMC, location University of Amsterdam, Amsterdam, The Netherlands}
\address[10]{Cancer Center Amsterdam, Imaging and Biomarkers, Amsterdam, The Netherlands}

%


\author[12]{Zeli \snm{Chen}}
\author[12]{Yiwen \snm{Zhang}}
\author[12]{Kaiyi \snm{Zheng}}
\author[12]{Chuanpu \snm{Li}}
\address[12]{School of Biomedical Engineering, Southern Medical University, Guangzhou, China}

\author[13]{Haowen \snm{Pang}}
\author[13]{Chuyang \snm{Ye}}
\address[13]{School of Integrated Circuits and Electronics, Beijing Institute of Technology, Beijing, China}

\author[14]{Runqi \snm{Wang}}
\author[15]{Tao \snm{Song}}
\address[14]{School of Biomedical Engineering, ShanghaiTech University, Shanghai, China}
\address[15]{Fudan University, Shanghai, China}

\author[16]{Fuxin \snm{Fan}}
\author[16]{Jingna \snm{Qiu}}
\author[16]{Yixing \snm{Huang}}
\address[16]{Friedrich-Alexander-Universität Erlangen-Nürnberg, Erlangen, Germany}

\author[18]{Juhyung \snm{Ha}}
\author[18]{Jong \snm{Sung Park}}
\address[18]{Indiana University, Bloomington, USA}

\author[19]{Alexandra \snm{Alain-Beaudoin}}
\author[19]{Silvain \snm{B\'eriault}}
\address[19]{Advanced Development Engineering, Elekta Ltd, Montreal, Canada}

\author[20]{Pengxin \snm{Yu}}
\address[20]{Infervision Medical Technology Co., Ltd. Beijing, China}

\author[21]{Hongbin \snm{Guo}}
\author[21]{Zhanyao \snm{Huang}}
\address[21]{Department of Biomedical Engineering, Shantou University, China}

\author[22]{Gengwan \snm{Li}}
\author[22]{Xueru \snm{Zhang}}
\address[22]{Independent researchers}

\author[23]{Yubo \snm{Fan}}
\author[23]{Han \snm{Liu}}
\address[23]{Department of Computer Science, Vanderbilt University, Nashville, USA}

\author[24]{Bowen \snm{Xin}}
\author[24]{Aaron \snm{Nicolson}}
\address[24]{Australian e-Health Research Centre, CSIRO, Herston, Queensland, Australia}

\author[25]{Lujia \snm{Zhong}}
\author[25]{Zhiwei \snm{Deng}}
\address[25]{Stevens Neuroimaging and Informatics Institute, Keck School of Medicine, University of Southern California (USC), Los Angeles, California, USA}

\author[26]{Gustav \snm{M{\"u}ller-Franzes}}
\author[26]{Firas \snm{Khader}}
\address[26]{University Hospital Aachen, Aachen, Germany}

\author[27]{Xia \snm{Li}}
\author[27]{Ye \snm{Zhang}}
\address[27]{Center for Proton Therapy, Paul Scherrer Institut, Villigen, Switzerland; Department of Computer Science, ETH Zurich, Zurich, Switzerland}

\author[28]{C\'edric \snm{H\'emon}}
\author[28]{Valentin \snm{Boussot}}
\address[28]{University Rennes 1, CLCC Eug\`ene Marquis, INSERM, LTSI, Rennes, France}

\author[29]{Zhihao \snm{Zhang}}
\author[29]{Long \snm{Wang}}
\address[29]{Subtle Medical, Shanghai, China}

\author[30]{Lu \snm{Bai}}
\author[30]{Shaobin \snm{Wang}}
\address[30]{MedMind Technology Co. Ltd., Beijing, China}

\author[31]{Derk \snm{Mus}}
\author[31]{Bram \snm{Kooiman}}
\address[31]{MRI Guidance BV, Utrecht, The Netherlands}

\author[32]{Chelsea A. H. \snm{Sargeant}}
\author[32]{Edward G. A. \snm{Henderson}}
\address[32]{Division of Cancer Sciences, The University of Manchester, United Kingdom}

\author[33]{Satoshi \snm{Kondo}}
\author[34]{Satoshi \snm{Kasai}}
\address[33]{Muroran Institute of Technology, Hokkaido, Japan}
\address[34]{Niigata University of Health and Welfare, Niigata, Japan}

\author[35]{Reza \snm{Karimzadeh}}
\author[35]{Bulat \snm{Ibragimov}}
\address[35]{Image Analysis, Computational Modelling and Geometry, University of Copenhagen, Denmark}

\author[36]{Thomas \snm{Helfer}}
\author[37,38]{Jessica \snm{Dafflon}}
\address[36]{IACS, Stony Brook University, NY, USA}
\address[37]{Data Science and Sharing Team, Functional Magnetic Resonance Imaging Facility, National Institute of Mental Health, Bethesda, USA}
\address[38]{Machine Learning Team, Functional Magnetic Resonance Imaging Facility National Institute of Mental Health, Bethesda, USA}

\author[39]{Zijie \snm{Chen}}
\author[39]{Enpei \snm{Wang}}
\address[39]{Shenying Medical Technology (Shenzhen) Co., Ltd., Shenzhen, Guangdong, China}

\author[11]{Zoltan \snm{Perko}\fnref{fn2}}
\address[11]{Delft University of Technology, Faculty of Applied Sciences, Department of Radiation Science and Technology, Delft, The Netherlands}

\author[2,3]{Matteo \snm{Maspero}\corref{cor1}\fnref{fn2}}
\ead{m.maspero@umcutrecht.nl}
\cortext[cor1]{Corresponding author: Heidelberglaan 100, 3508 GA, UMC Utrecht, P.O. Box 85500 Utrecht, The Netherlands, Tel.: +31-88 75 67492;}

\received{4 June, 2024}
\begin{abstract}

Radiation therapy plays a crucial role in cancer treatment, necessitating precise delivery of radiation to tumors while sparing healthy tissues over multiple days. Computed tomography (CT) is integral for treatment planning, offering electron density data crucial for accurate dose calculations. However, accurately representing patient anatomy is challenging, especially in adaptive radiotherapy, where CT is not acquired daily. Magnetic resonance imaging (MRI) provides superior soft-tissue contrast. Still, it lacks electron density information, while cone beam CT (CBCT) lacks direct electron density calibration and is mainly used for patient positioning.

Adopting MRI-only or CBCT-based adaptive radiotherapy eliminates the need for CT planning but presents challenges. Synthetic CT (sCT) generation techniques aim to address these challenges by using image synthesis to bridge the gap between MRI, CBCT, and CT. The SynthRAD2023 challenge was organized to compare synthetic CT generation methods using multi-center ground truth data from 1080 patients, divided into two tasks: 1) MRI-to-CT and 2) CBCT-to-CT. The evaluation included image similarity and dose-based metrics from proton and photon plans.

The challenge attracted significant participation, with 617 registrations and 22/17 valid submissions for tasks 1/2. Top-performing teams achieved high structural similarity indices (\(\geq0.87\)/\(0.90\)) and gamma pass rates for photon (\(\geq98.1\%/99.0\%\)) and proton (\(\geq97.3\%/97.0\%\)) plans. However, no significant correlation was found between image similarity metrics and dose accuracy, emphasizing the need for dose evaluation when assessing the clinical applicability of sCT.

SynthRAD2023 facilitated the investigation and benchmarking of sCT generation techniques, providing insights for developing MRI-only and CBCT-based adaptive radiotherapy. It showcased the growing capacity of deep learning to produce high-quality sCT, reducing reliance on conventional CT for treatment planning.

\end{abstract}

\end{frontmatter}


\section{Introduction}
\label{sec:introduction}
More than half of cancer patients receive radiotherapy as the standard care, providing effective local treatment \citep{Chandra2021}. Radiotherapy is typically delivered daily over several weeks \citep{Mitchell2013}, aiming to provide a high radiation dose to the target while minimizing the dose to the surrounding healthy tissue. To achieve conformal radiation treatment, obtaining an electron density map of the patient's anatomy is crucial to determine beam attenuation and local dose deposition \citep{Grgoire2011}. This electron density information is currently obtained through computed tomography (CT) \citep{Seco2006}. However, tumors are not always clearly visible on CT, and magnetic resonance imaging (MRI) has been proposed as its superior soft-tissue contrast offers improved visibility of tumor-boundaries and organs-at-risk (OARs) \citep{Schmidt2015}.
Moreover, throughout the treatment course, patient anatomy may vary. In adaptive radiotherapy, new treatment plans are generated weekly or daily while the patient is on the treatment couch to maintain dose conformality. During adaptive radiotherapy, typically cone-beam CT (CBCT) \citep{Nijkamp2008} or MRI \citep{Lagendijk2014} are the sole imaging modalities at hand. 
However, neither MRI nor CBCT allows for direct treatment plan optimization as accurate electron density information is lacking.
Techniques have been developed to generate synthetic CT (sCT) (also called pseudo-CT, virtual CT, surrogate CT) from MRI and CBCT to aid in determining local beam attenuation and dose deposition for treatment planning \citep{Edmund2017}.
The sCT generation has paved the way for MRI-based treatment planning (MRI-only radiotherapy) and CBCT-based adaptive radiotherapy, which avoid additional radiation exposure due to imaging and reduce the treatment centers' workload by omitting unnecessary scans. 

Although several approaches for obtaining sCT exist, including bulk density override and atlas-based methods, deep neural networks have recently shown promise in generating sCT \cite{SpadeaMaspero2021}. Neural networks can be broadly categorized into convolutional neural networks (CNNs), e.g., U-net \citep{unet}, generative adversarial networks (GANs), e.g., cycleGAN, pix2pix, \citep{gan,CycleGAN2017,pix2pix2017}, and, more recently, (vision-)transformers \citep{transformer,vit} and diffusion models \citep{diffusion}. Paired (supervised) and unpaired (unsupervised) training approaches have been suggested depending on the network architecture. The models were trained using 2-dimensional (2D) slices or 3D CT and MRI/CBCT volumes. Moreover, 2.5D approaches considering neighboring slices or perpendicular planes have been introduced to deal with spatial information and coherence while maintaining performance and feasible memory use. Most of these papers claim that their sCT generation method outperforms others. However, networks are often trained on different datasets and anatomies and evaluated using different metrics, making consistent methodological comparison difficult. Moreover, most sCT methods are evaluated based on image similarity metrics, whereas what matters is, ultimately, the effect of sCT on the treatment plan dose distribution, and image metrics do not necessarily reflect the dose accuracy \citep{Kieselmann2018}. This lack of a fair comparison hinders the identification of the best network design choices that should be implemented in clinical sCT tools.

To address these issues and provide a fair comparison, we organized the SynthRAD2023 Grand Challenge, held in conjunction with MICCAI 2023. In the challenge, we provided ground truth data and developed methods to facilitate fair model comparisons and increase the understanding of how different network designs influence performance. This challenge encourages the development and evaluation of state-of-the-art algorithms for generating accurate and clinically relevant sCT images from MRI and CBCT data. Two tasks were defined based on a new publicly available dataset~\cite{Thummerer2023synth}: 1) MRI-to-CT generation for MRI-only radiotherapy and MRI-guided radiotherapy and 2) CBCT-to-CT generation for image-guided adaptive radiotherapy (IGART) and online adaptive radiotherapy. 

This paper reviews the challenge participation, evaluation, and ranking of the submitted algorithms based on image similarity and dose assessment for sCTs compared to ground truth CTs. The analysis explores trends in submitted algorithms and their correlation with overall performance, focusing on the impact of variation within the dataset, the metrics chosen for evaluation, and examining ranking stability.

\section{Material and methods}
\subsection{Challenge setup}
\begin{figure*}[t]
\centering
\includegraphics[width=0.8\linewidth]{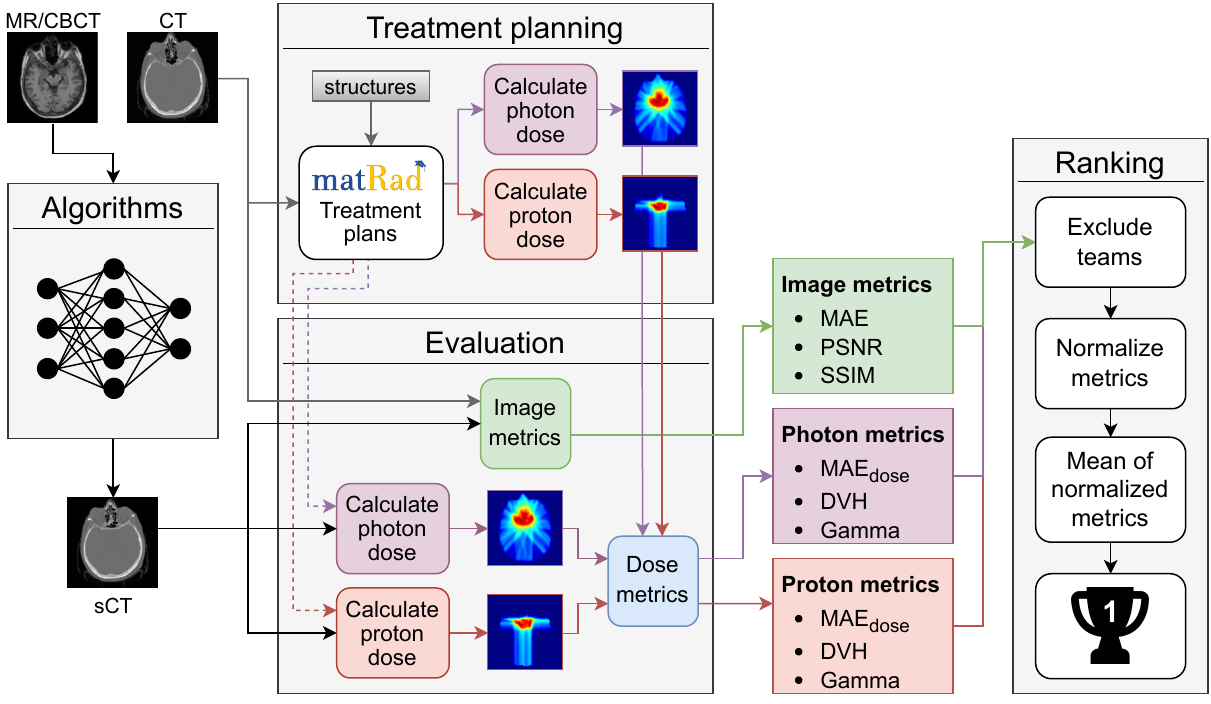}
\caption{\textbf{The SynthRAD2023 pipeline}. Left: the participants' algorithms generate sCT from input MRI or CBCT images. Middle block: the obtained sCT is evaluated with image similarity metrics (comparing sCT images to ground truth CT images) and dose metrics (comparing dose distributions recalculated on sCT and ground truth CT for pre-planned photon and proton treatment plans). Right: after calculating the metrics, the winner is determined by applying a ranking approach.}
\label{Fig:pipeline}
\end{figure*}

The SynthRAD (Synthesizing Computed Tomography for Radiotherapy) challenge allowed teams to test and compare their sCT algorithms. The challenge was hosted on the Grand Challenge website \url{https://synthrad2023.grand-challenge.org}. It consisted of two tasks: task 1 involved generating sCT from MRI data, while task 2 focused on developing sCT from CBCT data. Each task comprises two subtasks involving the brain and pelvis regions.

The organizing team arose from the ``Image synthesis \& reconstruction'' expertise subgroup of the Dutch deep learning in radiotherapy initiative \url{www.DLinRT.org}. The organizing group encompasses early-stage researchers, PhDs, postdocs, four assistant professors, and one associate professor from five Dutch University Medical Centers and three Dutch Technical Universities.

\autoref{Fig:pipeline} presents an overview of the SynthRAD2023 Grand Challenge design, including the algorithms developed by the participants and the evaluation and ranking procedures performed by the organizers. Participants in the challenge were tasked with developing and training models capable of generating accurate sCT images using only input MRI or CBCT. Participants could participate in either task 1, task 2, or both. Only fully automated methods trained from scratch on the provided data could be used; in other words, pre-trained models were not allowed. The submissions were automatically evaluated on the Grand Challenge environment.
Further details regarding participation rules and policies can be found in the appendix. As the sCTs are intended for radiotherapy, we analyzed photon and proton dose metrics alongside image similarity metrics, as described in section \ref{subsec: evaluation method}. To determine the winner of the challenge, we ranked the teams based on these metrics, for which we provide a further explanation in section \ref{subsec: ranking method}. To ensure transparency and enable further exploration of the methods employed during the challenge, the data preprocessing and evaluation code can be accessed at \url{https://github.com/SynthRAD2023}.

\subsection{Challenge phases}
The challenge was divided into four phases: training, validation, preliminary test, and test. Teams had two months to familiarize themselves with the challenge and begin training their algorithms, as the training data was released on April 1, 2023. The validation phase began on June 1, 2023, and was a Type-1 challenge in which participants were required to execute the inference locally and submit the corresponding sCTs. This phase allowed for up to two submissions every four days, and the submitted sCTs were automatically assessed using image similarity metrics. The results were then updated on an open leaderboard, allowing real-time comparison between participating teams. The ground truth CT images used for validation were not shared with the participants to prevent biased results. The final test phase was a Type-2 challenge in which teams had to upload a Docker image containing their method, which is inferred and evaluated on the Grand Challenge platform. The test data and ground truth CTs were kept hidden. To familiarize participants with a Type-2 challenge, we introduced the preliminary test phase, which started on May 1, 2023. The preliminary test phase used six cases; only image similarity metrics were evaluated. The final test phase started on July 16, 2023, and lasted five weeks. The preliminary test phase and test phase ended on August 22, 2023. Teams were required to upload a Docker image of their algorithm and a description of their methods. To minimize algorithm tweaking to the test data, each team could submit only twice during the testing phase, and only the last submission was counted. The second submission allowed participants to correct potential errors arising during the first submission. During this phase, the generated sCT images underwent an image similarity evaluation and a photon and proton dose evaluation to verify the most relevant metrics for radiotherapy. The image similarity metrics were calculated online on the platform provided by Grand Challenge, and the dose evaluation was performed offline due to the computational resources required. At the end of the testing phase, the final ranking was published to show the performance of the participating teams.
After the challenge, a post-challenge test phase was opened, and the preliminary and validation phases were reopened to enable continuous evaluation of algorithms until September 20, 2028.

\subsection{Dataset}
Data from 1080 patients undergoing radiotherapy treatment were included in the SynthRAD2023 dataset. The dataset consisted of imaging data from three Dutch University Medical Centers. Both task 1 (MRI-to-CT) and task 2 (CBCT-to-CT) included data from 270 patients for both the brain and pelvis anatomy (leading to $2 \times 2 \times 270$ image pairs). The 270 cases were divided into a training, validation, and test set of 180, 30, and 60 patients. The dataset consisted primarily of adult patients, with no gender restrictions applied. Only patients for whom the MRI or CBCT was acquired within two months of the CT were included to limit anatomical changes. It should be emphasized that the datasets for task 1 and task 2 did not contain the same patients. A detailed dataset description can be found in the publication by~\cite{Thummerer2023synth}. Ethical approval was obtained from the data-providing institutes' internal review boards/Medical Ethical committees. The data were released under the CC BY-NC (Attribution-Non-Commercial) license and made available via Zenodo at \url{https://zenodo.org/doi/10.5281/zenodo.7835406} (train), \url{https://zenodo.org/doi/10.5281/zenodo.7868168} (validation), and \finaltesturl (test, available from 01-01-2028).

The imaging protocols used to acquire the MRI and CBCT adhered to the clinical routines of the individual centers. As a result, variations in the MRI, CBCT, and CT imaging protocols were present between centers and between datasets, which are representative of real-world application scenarios. A comprehensive table detailing the imaging parameters was provided alongside the dataset~\cite{Thummerer2023synth}. For task 1, MRIs were acquired with scanners from two different vendors using different settings per site. Additionally, centers A and C used MRI scanners with field strengths of 1.5T and 3T, while center B exclusively utilized a 1.5T scanner. T1-weighted gradient echo was selected for all brain data. The datasets from centers B and C included T1-weighted MRI acquired after Gadolinium contrast agent injection, whereas those from center A were acquired without contrast agent injection.
The pelvis data comprised two-thirds of a T1-weighted gradient echo sequence and a T2-weighted spin echo sequence. For task 2, CBCTs were acquired with Linacs from two different vendors. The two sites that scanned CBCTs with Linacs from the same vendor had different acquisition protocols.

As described by \cite{Thummerer2023synth}, the data were preprocessed by resampling the voxel size to $1\times1\times1$mm\textsuperscript{3} for the brain and $1\times1\times2.5$mm\textsuperscript{3} for the pelvis patients, respectively. The face was intentionally removed for brain cases to protect patient privacy or proprietary information. The patient outline was automatically segmented on the MRI/CBCT using thresholding and morphological operations. This was followed by a dilation of 20 voxels in the axial plane and 2 in the superior-inferior directions. To ensure alignment between the MRI/CBCT and the CT, the field of view of the MRI/CBCT and CT was adjusted based on the patient outline, and rigid registration was performed. The resulting mask, including surrounding air, was provided and could be used by the participants for preprocessing.

\subsection{Baseline algorithms}
Two bulk-assignment baseline sCT models were used to provide insight into the evaluation metrics: “water” and “stratified”. The water approach assigned 0 HU to voxels within the dilated body contour mask and -1000 HU outside the mask (air). 
As suggested by \cite{maspero2017feasibility}, a stratified approach was employed to obtain images resembling bulk-assigned sCT without geometrical deformations by starting from ground truth CT. Stratified sCTs were obtained by classifying the ground truth CT data into five categories and assigning bulk density values for voxels within their specific HU ranges. Voxels were categorized into five classes based on HU intensity levels, mapping a range of density values to a population-derived HU value for this tissue \citep{maspero2017feasibility}, indicated as \(\langle\text{lower bound}, \text{upper bound}\rangle \text{ HU} \rightarrow \text{xx HU}\): 'air' ($\langle-\infty, -210\rangle$ HU $\rightarrow -968$ HU), 'adipose tissue' ($[-210, -20\rangle$ HU $\rightarrow -86$ HU), 'soft tissue' ($[-20, 120\rangle$ HU $\rightarrow$ 42 HU), 'bone marrow' ($[120, 555\rangle$ HU $\rightarrow$ 198 HU), and 'cortical bone' ($[555, \infty\rangle$ HU $\rightarrow$ 949 HU). The accuracy of the bone segmentation was further refined using a binary hole-filling algorithm to avoid soft tissue and air voxels within bone structures.

\subsection{Evaluation}
\label{subsec: evaluation method}
The sCTs generated by the participants were compared to the ground truth CTs based on metrics comparing image similarity and dose accuracy.
\subsubsection{Image similarity}
During the validation and test phases of the SynthRAD2023 Grand Challenge, the accuracy of the generated sCT images was evaluated using image similarity metrics within the dilated body contour masks \(\mathcal{B} = \{i \,|\, \mathcal{M}_i = 1\}\) provided with the dataset. This evaluation aimed to assess how closely the sCTs resembled the reference CTs. The mean absolute error (MAE), peak signal-to-noise ratio (PSNR), and structural similarity index measure (SSIM) were considered as image similarity metrics, as they are commonly used in medical image synthesis~\citep{SpadeaMaspero2021}.

\textbf{Masked MAE} was calculated to measure the average absolute difference between corresponding voxels in the sCT and CT, defined as
\begin{equation}
\operatorname{MAE}(\text{CT}, \text{sCT}) = \frac{1}{|\mathcal{B}|}\sum_{i\in \mathcal{B}} \left\vert\text{CT}_i - \text{sCT}_i\right\vert
\end{equation}
in which we sum over the voxels inside the body contour \(\mathcal{B}\) and normalized by the total number of masked voxels \(\,| \mathcal{B} \,|\).

\textbf{Masked PSNR} was calculated to quantify the ratio of maximum signal intensity over the noise level in the sCT compared to the CT, defined as

\begin{equation}
\text{PSNR}(\text{CT}, \text{sCT}) = 10 \log_{10}\left(\frac{Q^2}{\frac{1}{|\mathcal{B}|}\sum_{i\in \mathcal{B}}\left(\text{CT}_i -\text{sCT}_i\right)^2}\right),
\end{equation}
where \(Q\) is the dynamic range of the voxel intensities ([\(-1024\), \(3000\)] HU). The CT and sCT were clipped to the dynamic range to calculate the masked PSNR.

\textbf{Masked SSIM} was calculated to assess structural similarity between CT and sCT. The SSIM for a voxel \(i\) between two images \(x\) and \(y\) is computed by 
\begin{equation}
    \operatorname{SSIM}_{i}(x,y) = \frac{\left(2\mu_x^i\mu_y^i+c_1\right)\left(2\sigma_{xy}^i+c_2\right)}{\left(\left(\mu_x^i\right)^2+\left(\mu_y^i\right)^2 + c_1\right)\left(\left(\sigma_x^i\right)^2+\left(\sigma_y^i\right)^2+c_2\right)},
\end{equation}
where \(\mu_x^i\) and \(\sigma_x^i\) are the mean and variance, respectively, of \(x\) within an \(N\times N\times N\) window centered on voxel \(i\) and \(\sigma_{xy}^i\) is the covariance of \(x\) and \(y\) within an \(N\times N\times N\) window centered on voxel \(i\). \(N=7\) is the window size, and \(c_1 = (0.01 \cdot L)^2\) and \(c_2 = (0.03 \cdot L)^2\) are normalization constants, where \(L = (3000 - (-1024))\) HU is the dynamic range of the volumes. The final masked SSIM value is then obtained by computing 
\begin{equation}
    \operatorname{SSIM}(\text{CT}, \text{sCT}) = \frac{1}{|\mathcal{B}|}\sum_{i\in \mathcal{B}} \operatorname{SSIM}_{i}(\text{CT}, \text{sCT}),
\end{equation}
where the intensities of both the CT and sCT were clipped to [\(-1024\), \(3000\)] HU and then adjusted to be non-negative by adding 1024 HU.

\subsubsection{Dose distribution similarity}
Photon and proton intensity-modulated treatment plans were optimized based on the reference CT using the matRad treatment planning system~\citep{matrad}. The dose was prescribed to the planning target volume (PTV) for simplicity in both modalities, i.e., no robust optimization was performed for proton plans, with specific doses and isodose levels for the brain and pelvis. Only co-planar plans were considered, with photon plans utilizing 9-13 equi-angled 6~MV beams from a generic Linac model and proton plans utilizing 3–4 beams (from bilateral and opaque angles) from a generic proton system available in matRad. To reduce the dose to the healthy tissues and to ensure plan uniformity between patients, we used the same objective functions and constraints available in matRad per treatment site. OAR dose limits were treated as hard constraints whenever possible and were revised on a patient-specific basis when hard constraints were not achievable. For a few patients, the number of beams and some optimization parameters (e.g., optimizer, maximum number of iterations, and objective weights) were also fine-tuned to meet dose prescriptions and organ-at-risk (OAR) limits. All planning goals and OAR dose limits were based on international guidelines for the brain~\citep{lambrecht2018radiation} and pelvis~\citep{hall2021nrg} are summarized in~\autoref{tab:dose constraints}.

\begin{table}[]
\footnotesize
\caption{\label{tab:dose constraints}Dose constraints and planning objectives used in matRad for the brain and pelvis cases, respectively. Values are taken from \cite{lambrecht2018radiation} and \cite{hall2021nrg}.}
\centering
\begin{tabular}{ll|ll}
\toprule
\multicolumn{2}{c|}{\textbf{Brain}} & \multicolumn{2}{c}{\textbf{Pelvis}} \\
\multicolumn{2}{c|}{$30 \times 2.0$ Gy to 95\% of the PTV} & \multicolumn{2}{c}{$20\times3.0$ Gy to 95\% of the PTV} \\
\midrule
\textbf{Structure} & \textbf{Constraint} &  \textbf{Structure} & \textbf{Constraint}  \\
\midrule
\multirow{5}{*}{Brainstem} & & \multirow{5}{*}{Rectum} &  $V_{60 \, \text{Gy}} < 1\%$ \\

& \multirow{2}{*}{$D_{0.03 \, \text{cc}} < 60  \, \text{Gy}$ } & & $ V_{50 \, \text{Gy}} < 22\%$\\

& \multirow{2}{*}{$D_{0.03 \, \text{cc}} < 54 \, \text{Gy}$} & & $V_{40 \, \text{Gy}} < 38\%$ \\
& & & $V_{30 \, \text{Gy}} < 57\%$ \\
& & & $V_{20 \, \text{Gy}} < 85\%$ \\
\hline

\multirow{4}{*}{Chiasm} & \multirow{4}{*}{$D_{0.03 \, \text{cc}} < 55 \, \text{Gy}$} & \multirow{4}{*}{Bladder} & $V_{60 \, \text{Gy}} < 3\%$ \\
& & &  $V_{56.8 \, \text{Gy}} < 5\%$ \\
& & & $V_{48 \, \text{Gy}} < 25\%$ \\
& & & $V_{40 \, \text{Gy}} < 50\%$ \\

\hline

Optical Nerve & $D_{0.03 \, \text{cc}} < 55 \, \text{Gy}$ & Femur heads & $D_{\text{max}} < 37 \, \text{Gy}$ \\

\hline
Cochlea & $D_{\text{mean}} < 45 \, \text{Gy}$ & Colon & $D_{\text{max}} < 50 \, \text{Gy}$ \\
\hline

Brain & $V_{60 \, \text{Gy}} < 3 \, \text{cc}$ & Small bowel & $D_{\text{max}} < 40 \, \text{Gy}$ \\
\bottomrule

\end{tabular}
\end{table}

Throughout the dose evaluation process, the dose was recalculated on each sCT for both proton and photon treatment plans. This recalculation was carried out without propagating organ delineations or replanning, a deliberate measure taken to avoid potential differences arising from plan optimization. Subsequently, the differences between the planning dose distributions, originally calculated on CT, and the recalculated dose distributions on the sCT for both photon and proton plans were quantified using three specific metrics.
To ensure high reproducibility and facilitate fair comparisons for the SynthRAD2023 test set, the offline dose evaluation will be available at \finaltesturl at the time of the release of the test set.

\textbf{Relative mean absolute dose difference} within high dose regions \(\mathcal{H} = \{i \,|\, D_{CT,i} \geq 0.9\cdot D_{prescribed}\}\) were calculated to assess the difference in received dose in and around the target, defined as

\begin{equation}
\text{MAE}_{\text{dose}} = \frac{1}{|\mathcal{H}|}\sum_{i \in \mathcal{H}}\frac{D_{CT,i} - D_{sCT,i}}{D_{\text{prescribed}}},
\end{equation}
with $D_{(s)CT}$ being the dose distribution in the $(s)CT$ and $D_{prescribed}$ the prescribed dose.

\textbf{Dose-volume histogram (DVH) parameters} were calculated to assess the differences in the doses received by the PTV and OARs: the near-minimum dose in the PTV $D98_{\text{PTV}}$, the PTV volume receiving at least 95\% of the prescribed dose $V95_{\text{PTV}}$, the near-maximum dose of a given OAR $D2_{\text{OAR}}$, and the mean dose received by a given OAR $D\text{mean}_{\text{OAR}}$. Specifically, the use of the near-minimum and near-maximum was suggested by ICRU83 (\url{https://www.fnkv.cz/soubory/216/icru-83.pdf}). We included the relative absolute differences for all parameters as defined by

\begin{equation}
D98_{\text{PTV,CT}} = \frac{|D98_{\text{PTV,CT}} - D98_{\text{PTV,sCT}} + \epsilon|}{D98_{\text{PTV,CT}} + \epsilon},
\end{equation}

\begin{equation}
V95_{\text{PTV,CT}} = \frac{|V95_{\text{PTV,CT}} - V95_{\text{PTV,sCT}} + \epsilon|}{V95_{\text{PTV,CT}} + \epsilon},
\end{equation}

\begin{equation}
D2_{\text{OARs}} = \frac{1}{n_{\text{OARs}}}\sum_{\text{OAR}} \frac{|D2_{\text{OAR,CT}} - D2_{\text{OAR,sCT}} + \epsilon|}{D2_{\text{OAR,CT}} + \epsilon},
\end{equation}

\begin{equation}
D\text{mean}_{\text{OARs}} = \frac{1}{n_{\text{OARs}}}\sum_{\text{OAR}} \frac{|D\text{mean}_{\text{OAR,CT}} - D\text{mean}_{\text{OAR,sCT}} + \epsilon|}{D\text{mean}_{\text{OAR,CT}} + \epsilon},
\end{equation}
where $\epsilon=1\mathrm{e}{-12}$ to avoid division by zero and $n_{\text{OARs}}$ is the number of OARs. For each patient, we used the three OARs (if available) that had the highest average of $D5_{\text{OAR}}$ and $D\text{mean}_{\text{OAR}}$ to analyze dose differences in organs close to the target. We summed the four terms to obtain one final value for the DVH metric.

\textbf{Gamma pass rates} were calculated to compare the 3D spatial dose distributions from the sCTs with the dose obtained from the CT. This calculation followed the 3D gamma pass rate approach described by~\cite{low1998technique} with a dose-difference criterion ($\Delta D$) of 2\% and a distance-to-agreement criterion ($\Delta d$) of 2 mm. The gamma pass rate at each position vector in the sCT was determined by comparing it with the CT dose. Gamma pass rates were evaluated within regions receiving doses $\geq 10\%$ of the prescribed dose~\citep{ezzell2009imrt}.

\subsection{Eligibility and ranking}
\label{subsec: ranking method}
The nine metrics defined above were calculated for each test case and aggregated across all test cases for each participating team (\(\mu\pm\sigma\)). Teams were not considered in the ranking if their method did not outperform the water baseline for all three individual image similarity metrics. Moreover, the participants' method must complete the generation of a single sCT within 15 minutes on the Grand-Challenge platform, as described in the appendix.

Several methods exist for creating a ranking for a challenge with multiple metrics, including 1) calculating the mean over all metrics and ranking the aggregated scores (MeanThenRank), 2) calculating the median over all metrics and ranking the aggregated scores (MedianThenRank), 3) calculating the ranking for each metric and computing the mean of the aggregated ranks (RankThenMean), and 4) Calculating the ranking for each metric and computing the median of the aggregated ranks (RankThenMedian). Directly applying MeanThenRank and MedianThenRank to the nine metrics is inappropriate due to their lack of normalization and the differing orderings (ascending or descending). To fairly rank the submissions, each metric was normalized and scaled between zero (indicating the worst average team performance) and one (indicating the best average team performance). Subsequently, the normalized metrics are used to calculate the mean or median and rank the aggregated score. 

In the context of the SynthRAD2023 challenge, the MeanThenRank approach was employed to determine the winners. This method should account for variations in team performance, enabling a fair evaluation considering the diverse clinically relevant aspects of image similarity, photon dose, and proton dose metrics. To analyze biases introduced by the ranking method, we also studied how the other ranking approaches would affect the outcome to assess ranking stability.

\subsection{Analysis}
\subsubsection{Overall sCT performance}
Besides computing the aggregated metrics per submission (\(\mu\pm\sigma\)), we analyzed the significance of one team outperforming another in terms of individual metrics. To do so, we used the Wilcoxon signed-rank test \citep{wilcoxon1945} with Holm's adjustment for multiple testing \citep{holm1979simple} for each metric separately, offering insights into the pairwise performance differences between teams. The significance level for this test is set at $\alpha=0.05$. Additionally, we recorded the inference time of the participant's methods (\(\mu\pm\sigma\)) to synthesize the CT from the CBCT or MRI data on the Grand Challenge infrastructure.

\subsubsection{Model design predictors}
\label{subsec:model_design_pred}

We evaluated the model design choices adopted by participating teams thoroughly, aiming to identify the impact of these choices on overall ranking and performance. Statistical significance of the differences in SSIM performance within each subtask is determined using the Mann-Whitney U-test \citep{mann1947test} ($\alpha=0.01$), chosen for its suitability in comparing two independent samples that may not adhere to normal distribution. This test is particularly robust in the context of our analysis, providing reliable insights into the performance disparities associated with distinct design choices. To define predictors for sCT performance, we analyzed different design choices. We categorized them into five key aspects: 1) model and anatomy, 2) backbone architecture, 3) spatial configuration, 4) preprocessing, 5) data augmentation, and 6) postprocessing.

\paragraph{Model and anatomy}
Teams used different strategies to handle brain and pelvis data. Some teams used one collective model trained on both brain and pelvis patients (``One model'') or conditioned the collective model on the anatomical region (``One model, anatomy conditional"). In contrast, others trained the same model separately for the brain and pelvis subsets (``Two identical models"). Additionally, some teams used the same or a similar backbone architecture for both regions with distinctions in training parameters or network layers (``Two identical backbones, different training param." and ``Two similar models", respectively). Others employed entirely different models for the two regions (``Two different models").
\paragraph{Backbone architecture}
The base of a synthesis model involved using a CNN encoder-decoder model, which is a standard choice for image reconstruction and translation within deep learning. Teams also explored alternative architectures, such as GAN-based models that introduced a discriminator network and adversarial loss, transformer-based architectures that emphasized attention in the synthesis process, and diffusion model-based approaches that relied on an iterative diffusion process during inference. Moreover, some teams used an ensemble of multiple models to produce the final output.
\paragraph{Supervision}
Each team reported the supervision approach adopted. Supervised (paired) training was guided by directly comparing predictions (sCT) to ground truth (CT) from the same cases. Unsupervised (unpaired) training was guided by introducing cycle-consistency as introduced by \citep{CycleGAN2017}.
\paragraph{Spatial configuration}
The implementation of sCT generation models varied in different spatial configurations. Opting for fully 3D models was possible, considering the entire image volume as input. However, fully 3D models were often restricted in use by available computing resources; therefore, many studies employed 3D patch-based approaches, 2.5D models considering multiple consecutive 2D slices, or a combination of orthogonal slices, full slice 2D models, or 2D patch-based models.
\paragraph{Preprocessing}
A range of preprocessing techniques were used in the submitted algorithms, focusing on resizing and intensity normalization, necessary for stable and optimal model training. Resizing was used to achieve the desired voxel size, such as in the case of iso-resampling or the desired model input size. Intensity normalization was implemented linearly at the population level, at the patient level, or as standardization by ensuring well-distributed data based on a specific mean and standard deviation. Furthermore, some teams used intensity clipping to remove outliers, applied histogram matching or provided a specialized pipeline for the specific modality or anatomical region being processed.
\paragraph{Data augmentation}
Various data augmentation techniques ensured a diverse training set, potentially making the models more robust to unseen cases in the test set. Teams introduced randomness through random crop or patch selection, flipping, rotation, blurring, noise addition, and intensity transformations like bias field or contrast adjustments. Random deformations, whether affine or elastic, were also applied to enhance the diversity of the training set.
\paragraph{Postprocessing}
Some teams that implemented patch-based models averaged overlapping patches at test time. The multiple outputs of ensembled methods could be combined into a single sCT. Additionally, specific postprocessing steps were implemented considering prior knowledge of the modality, such as noise and artifact removal. The inversion of original preprocessing steps, such as normalization and padding, was crucial in obtaining the final sCT with accurate dimensions and voxel representation in Hounsfield units (HU).

\subsubsection{Data influence}
By examining the teams' performances, we analyzed the test dataset to identify the characteristics and features of the samples correlating with synthesized image quality. The analysis compared image similarity and dose metrics (\(\mu\pm\sigma\)) averaged for each task, center, and anatomy within the test set. In addition, for task 1, the influence of MR acquisition protocol and magnetic field strength on performance was investigated. Statistical significance between the groups was established using the Mann-Whitney U-Test \citep{mann1947test} ($\alpha=0.01$). Lastly, we extended our analysis to a patient level, allowing for detailed evaluation of low-performing patients.

\subsubsection{Metric correlations}
For clarification throughout the paper, we defined the term `metric group' to refer to one of the three categories of evaluation metrics: image similarity metrics (MAE, PSNR, and SSIM), photon dose metrics (MAE\textsubscript{dose}, DVH\textsubscript{metric}, and $\gamma$), and proton dose metrics (MAE\textsubscript{dose}, DVH\textsubscript{metric}, and $\gamma$).

Our objective was to analyze the correlation within and between metric groups. To achieve this, we employed visual assessments to illustrate correlations within a metric group. We used the Spearman rank correlation coefficient \(\rho\) \citep{spearman} to quantify correlations between all metrics. This coefficient considers the ordinal relationship between the ranks of single test case performances, providing robustness against variations in the scale and direction of the values. 

\subsubsection{Ranking stability and correlations}
We used Kendall's $\tau$ correlation coefficient \citep{kendall1938new} between the approaches to analyze the effect of ranking approach choice. This coefficient quantifies the correlations between the ranking approaches outlined in section \ref{subsec: ranking method}, assessing the similarity in the relative ordering of elements across different rankings. 

In addition, we investigated the stability of the final rankings at a patient level, as recommended by \cite{wiesenfarth2021methods}. This involved implementing bootstrapping to examine variations in the ranking positions of all teams. The ranking process was iteratively applied to 1000 bootstrap sets. Each bootstrapping set consisted of 120 randomly selected patients from the test set, with patients potentially being selected more than once. The MeanThenRank approach was employed to rank the teams by first normalizing the metric values based on the best and worst average performance of that metric per bootstrap sample.

\section{Participation}
\begin{table}
\small
\centering
\caption{Details on the challenge participation. Participants without a team are displayed as a one-person team.}
\label{tab:partipants}
\begin{tabular}{llll}
\toprule
& \textbf{Validation} & \textbf{Preliminary test} & \textbf{Test} \\
\midrule
\multirow{3}{*}{\textbf{Task 1}} & 275 valid & 77 valid & 18 included,\\
& submissions & submissions & 4 excluded, and\\
& (from 38 teams) &  (from 27 teams) & 2 failed teams \\
\midrule
\multirow{3}{*}{\textbf{Task 2}} & 207 valid & 36 valid & \multirow{2}{*}{14 included and}\\
 & submissions & submissions & \multirow{2}{*}{3 excluded teams}\\
 & (from 25 teams) &  (from 15 teams)& \\
 
\bottomrule
\end{tabular}
\end{table}
The SynthRAD2023 Grand Challenge witnessed substantial participation from research teams worldwide, showcasing various techniques and methodologies for sCT generation. By the end of the test phase, the training dataset had been downloaded 1797 times, and 617 researchers had registered for the challenge, forming 94 teams and 429 individual participants. Participation in the challenge phases decreased over time, resulting in 22 and 17 successful submissions in the test phase for tasks 1 and 2, respectively. Based on the criteria described in section \ref{subsec: ranking method}, 18 and 14 teams were included in the analysis for tasks 1 and 2, respectively (\autoref{tab:partipants}). Note that due to an unexpectedly large matrix size for one patient in the test set, the inference time limit was raised to accommodate the sCT generation.
Nine of the included teams participated in both tasks, primarily utilizing the same or similar models for both tasks. \Cref{tab:summary task1,tab:summary task2} show an overview of the proposed methods of all teams for tasks 1 and 2, respectively. More detailed descriptions of the methods implemented by the top five teams for both tasks are presented in sections \ref{method SMU} to \ref{method Pengxin Yu}. Detailed method descriptions of all other teams can be found in supplementary document A.

\subsection{SMU-MedVision (task 1 \& 2)}
\label{method SMU}
SMU-MedVision employed a hybrid 3D patch-based CNN and transformer Unet network with multi-scale structure extraction and preservation (MSEP) for task 1 \citep{10230486, 10268458}. In the encoder, they employed channel and spatial-wise attention to extract spatial information, allowing for varying input sizes. Additionally, a residual dilated Swin transformer (RDSformer) was integrated into each skip connection of the UNet to enhance the preservation of structural information in cross-modal features~\citep{liu2021swin}. Two identical models were created for both anatomical regions, including the masked MAE and VGG19 perceptual loss \citep{johnson2016perceptual}. Preprocessing involved Z-score normalization tailored to individual patient statistics and random horizontal and vertical flipping for data augmentation. At test time, overlapping patches were created by selecting every 80,000th voxel within the body mask as the central point of each patch. These overlapping patches were averaged to result in the full sCT. The model underwent training for 200 epochs using the Adam optimizer with a learning rate of $2\mathrm{e}{-4}$ and a poly decay scheduler. The final epoch used at test time was determined based on the best MAE in the sub-validation set created from the training set. 

For task 2, SMU-MedVision implemented a 2.5D Unet++ \citep{zhou2018unet++} with a ResNeXt101 backbone, with the loss function combining masked MAE loss, VGG19 perceptual loss \citep{johnson2016perceptual}, and L2 regularization. The model was trained using brain and pelvis data and then fine-tuned per region. Preprocessing involved resizing, clipping, and linear normalization of the CT. Training data augmentation included shift scale rotations with horizontal and vertical flipping, while test data underwent augmentation via horizontal and vertical flipping. Slices of $5 \times384\times384$ voxels were used for collective pretraining, and the model input sizes for fine-tuning were $5\times288\times288$ voxels for the brain and $5\times416\times416$ voxels for the pelvis. Postprocessing included the inversion of test-time augmentations. The model was collectively trained for 40 epochs and then fine-tuned using 5-fold cross-validation for 50 for the brain and 40 epochs for the pelvis, respectively, and optimized using an AdamW optimizer with a stepped decay learning rate schedule. The final result was based on an ensemble of all five folds (with the best validation MAE) and a model trained on the completely provided training set (for the number of epochs mentioned above).

\subsection{Jetta\_Pang (task 1)}
\label{method Jetta Pang}
Jetta\_Pang implemented two 3D patch-based nnU-Net \citep{isensee2021nnu} models with an MSE loss for task 1:
\afterpage{%
\begin{landscape}
\makeatother
\rowcolors{1}{}{lightgray}
\setlength{\tabcolsep}{4.5pt}

\begin{table}
\centering
\caption{\label{tab:summary task1}Ranking and model details task 1 (MRI-to-CT synthesis). When a check is used, this step is applied to both MRI and CT and brain (br) and pelvis (pel); otherwise, it is specified by the subgroup. All distinctions listed in the first two rows are described in section \ref{subsec:model_design_pred}. }
\resizebox{\linewidth}{!}{%
\begin{tabular}[]{c|c|cccccc|ccccc|cc|cccc|ccccccc|ccccccc|ccc}
\toprule
\textbf{Rank}  & \textbf{Team} & \multicolumn{6}{c|}{\textbf{Model \& anatomy}}  & \multicolumn{5}{c|}{\textbf{Backbone arch.}} & \multicolumn{2}{c|}{\textbf{Sup.}} & \multicolumn{4}{c|}{\textbf{Spatial config.}}  & \multicolumn{7}{c|}{\textbf{Preprocessing}} & \multicolumn{7}{c|}{\textbf{Data augmentation}} & \multicolumn{3}{c}{\textbf{Postprocess.}}\\
\rowcolor{white}
\midrule
& & 
  \rotatebox{270}{One model} &
  \rotatebox{270}{One model, anatomy conditional} &
  \rotatebox{270}{Two identical models} &
  \rotatebox{270}{Two identical backbones, different training param.} &
  \rotatebox{270}{Two similar models} &
  \rotatebox{270}{Two different models} &
  \rotatebox{270}{CNN encoder-decoder} &
  \rotatebox{270}{GAN} &
  \rotatebox{270}{Transformer} &
  \rotatebox{270}{Diffusion model} &
  \rotatebox{270}{Ensemble of multiple models} &
  \rotatebox{270}{Supervised} &
  \rotatebox{270}{Unsupervised} &
  \rotatebox{270}{3D} &
  \rotatebox{270}{3D patch-based} &
  \rotatebox{270}{2.5D} &
  \rotatebox{270}{2D} &
  \rotatebox{270}{Iso-resampling / resizing} &
  \rotatebox{270}{Clipping} &
  \rotatebox{270}{Patient level linear normalization} &
  \rotatebox{270}{Population level normalization} &
  \rotatebox{270}{Standardization} &
  \rotatebox{270}{Histogram matching} &
  \rotatebox{270}{Other (e.g. N4 correction, smoothing, limb removal)} &
  \rotatebox{270}{Random crop / patch} &
  \rotatebox{270}{Flipping} &
  \rotatebox{270}{Rotation} &
  \rotatebox{270}{Blurring} &
  \rotatebox{270}{Noise addition} &
  \rotatebox{270}{Intensity transform (bias field, contrast / histogram adj.)} &
  \rotatebox{270}{Deformation (affine or elastic)} &
  \rotatebox{270}{Average overlapping patches} &
  \rotatebox{270}{Noise / artifact removal} &
  \rotatebox{270}{Revert normalization / padding} \\
 \midrule

1 & SMU-MedVision &&&\ding{51}&&&  &&&\ding{51}&&   &\ding{51}&   &&\ding{51}&&& &&&&\ding{51}&&&  &\ding{51}&&&&&&  \ding{51}&&\ding{51} \\

2 & Jetta\_Pang &&&&\ding{51}&&   &\ding{51}&&&&   &\ding{51}&   &&\ding{51}&&& &&&&{\small MR}&&& &&&&&&& \ding{51}&& \\

3 & FAYIU &&&\ding{51}&&&   &&&\ding{51}&&   &\ding{51}&   &&\ding{51}&&& &&&\ding{51}&&&& \ding{51}&&&&&&& \ding{51}&&\ding{51} \\

4 & Elekta &&&\ding{51}&&&   &&\ding{51}&&&\ding{51}   &\ding{51}&   &&&\ding{51}&& \ding{51}&\ding{51}&&\ding{51}&&&& \ding{51}&&&{\small MR}&&{\small MR}&\ding{51}    &\ding{51}&&\ding{51} \\ 

5 & iu\_mia &&\ding{51}&&&&   &\ding{51}&&&&   &\ding{51}&   &&&&\ding{51}& {\small pel}&&&{\small CT}&{\small MR}&&& \ding{51}&&&&&&& \ding{51}&&\ding{51} \\ 

6 & ShantouBME &&\ding{51}&&&&    &\ding{51}&&&&    &\ding{51}&   &&&&\ding{51}&   &&{\small MR}&{\small CT}&&&&   \ding{51}&&&&&&&   &&\ding{51} \\ 

7 & FGH\_365 &&\ding{51}&\ding{51}&&& &&\ding{51}&&&\ding{51}   &\ding{51}& &&\ding{51}&& &&\ding{51}&{\small MR}&{\small CT}&&&    &\ding{51}&&&\ding{51}&\ding{51}&\ding{51}  &\ding{51}&\ding{51}&&\ding{51} \\ 

8 & USC-LONI &&&&\ding{51}&& &&&&\ding{51}&   &\ding{51}& &&&\ding{51}& &\ding{51}&&{\small MR}&{\small CT}&&& &&&&&&& &&\ding{51}&\ding{51} \\ 

9 & UKA &&&\ding{51}&&& &\ding{51}&&&&\ding{51}   &\ding{51}& &&&\ding{51}& &&\ding{51}&&{\small CT}&&& &&\ding{51}&&&&& &\ding{51}&&\ding{51} \\ 

10 & PSICPT\_AI4PT &&&\ding{51}&&& &\ding{51}&&&&   &\ding{51}& &&&\ding{51}& &&&&{\small CT}&{\small MR}&& &\ding{51}&\ding{51}&\ding{51}&&&& &&&\ding{51} \\ 

11 & SubtleCT &&&\ding{51}&&& &\ding{51}&&&&   &\ding{51}& &&&\ding{51}& &&\ding{51}&{\small MR}&{\small CT}&&& &&&&&&& &&&\ding{51} \\ 

12 & mriG &&&\ding{51}&&& &\ding{51}&&&&   &\ding{51}& &&\ding{51}&& &&&\ding{51}&&&& &\ding{51}&\ding{51}&\ding{51}&&&\ding{51}& &\ding{51}&&\ding{51} \\ 

13 & KoalAI &&&&&\ding{51}& &&\ding{51}&&&\ding{51}   &\ding{51}& &&\ding{51}&& &&&&{\small CT}&&{\small MR}&\ding{51} &\ding{51}&&\ding{51}&&&\ding{51}&\ding{51} &\ding{51}&&\ding{51} \\ 

14 & Breizh-CT &&&\ding{51}&&& &&\ding{51}&&&   &\ding{51}& &&\ding{51}&& &&&&\ding{51}&&& &&&&&&& &\ding{51}&&\ding{51} \\ 

15 & SKJP &&&&\ding{51}&& &\ding{51}&&&&   &\ding{51}& &&&\ding{51}& &&&{\small MR}&&&{\small MR}& &&&&&&& &&& \\ 

16 & reza.karimzadeh &&&\ding{51}&&& &&\ding{51}&\ding{51}&&   &\ding{51}& &&\ding{51}&& &&&&\ding{51}&&& &\ding{51}&&\ding{51}&&&\ding{51}& &\ding{51}&&\ding{51} \\ 

17 & thomashelfer &&&&&&\ding{51} &&{\small pel}&&{\small br}&   &\ding{51}&   &{\small br}&{\small pel}&& &&&&\ding{51}&&& &&&&&&& &&&\ding{51} \\ 

18 & X-MAN &\ding{51}&&&&& &&\ding{51}&&&   &\ding{51}& &&\ding{51}&& &&&{\small MR}&&&& &\ding{51}&&&&&\ding{51}& &\ding{51}&&\ding{51} \\ 

\bottomrule
\end{tabular}%
}
\end{table}
\end{landscape}
}
\afterpage{%
\begin{landscape}
\makeatother
\rowcolors{1}{}{lightgray}
\setlength{\tabcolsep}{4.5pt}

\begin{table}
\centering
\caption{\label{tab:summary task2}Ranking and model details task 2 (CBCT-to-CT synthesis). When a check is used, this step is applied to both CBCT (CB) and CT and the brain and pelvis (pl.). Otherwise, it is specified by the subgroup. All distinctions listed in the first two rows are described in section \ref{subsec:model_design_pred}.}
\resizebox{\linewidth}{!}{%
\begin{tabular}[]{c|c|ccccc|ccccc|cc|cccc|cccccc|cccccc|ccc}
\toprule
\textbf{Rank}  & \textbf{Team} & \multicolumn{5}{c|}{\textbf{Model \& anatomy}}  & \multicolumn{5}{c|}{\textbf{Backbone arch.}} & \multicolumn{2}{c|}{\textbf{Sup.}}  & \multicolumn{4}{c|}{\textbf{Spatial config.}}  & \multicolumn{6}{c|}{\textbf{Preprocessing}} & \multicolumn{6}{c|}{\textbf{Data augmentation}} & \multicolumn{3}{c}{\textbf{Postproc.}}\\  
\rowcolor{white}
\midrule
& & 
\rotatebox{270}{One model} &
  \rotatebox{270}{One model, anatomy conditional} &
  \rotatebox{270}{Two identical models} &
  \rotatebox{270}{Two identical backbones, different training param.} &
  \rotatebox{270}{Two similar models} &
  \rotatebox{270}{CNN encoder-decoder} &
  \rotatebox{270}{GAN} &
  \rotatebox{270}{Transformer} &
  \rotatebox{270}{Diffusion model} &
  \rotatebox{270}{Ensemble of multiple models} &
  \rotatebox{270}{Supervised} &
  \rotatebox{270}{Unsupervised} &
  \rotatebox{270}{3D patch-based} &
  \rotatebox{270}{2.5D} &
  \rotatebox{270}{2D} &
  \rotatebox{270}{2D patch-based} &
  \rotatebox{270}{Iso-resampling / resizing} &
  \rotatebox{270}{Clipping} &
  \rotatebox{270}{Patient level linear normalization} &
  \rotatebox{270}{Population level linear normalization} &
  \rotatebox{270}{Histogram matching} &
  \rotatebox{270}{Other (e.g. overflow correction, multi-windowing) } &
  \rotatebox{270}{Random crop / patch / translation} &
  \rotatebox{270}{Flipping} &
  \rotatebox{270}{Rotation} &
  \rotatebox{270}{Noise addition} &
  \rotatebox{270}{Intensity transform (contrast / histogram adj.)} &
  \rotatebox{270}{Deformation (affine or elastic)} &
  \rotatebox{270}{Average overlapping patches} &
  \rotatebox{270}{Noise / artifact / background removal} &
  \rotatebox{270}{Revert normalization / padding} \\
 \midrule

1 & SMU-MedVision &&\ding{51}&&\ding{51}&   &\ding{51}&&&&\ding{51}  &\ding{51}&   &&\ding{51}&&   &\ding{51}&\ding{51}&&\ding{51}&&   &&\ding{51}&\ding{51}&&&   &&&\ding{51} \\ 

2 & GEneRaTion    &&&\ding{51}&&   &&&\ding{51}&&\ding{51}  &\ding{51}&   &&&\ding{51}&   &&&&\ding{51}&&   &\ding{51}&\ding{51}&\ding{51}&&&   &&&\ding{51} \\ 

3 & iu\_mia       &&\ding{51}&&&   &\ding{51}&&&&   &\ding{51}&   &\ding{51}&&&   &{\small pel}&&&\ding{51}&&   &\ding{51}&&&&&   &\ding{51}&&\ding{51} \\ 

4 & FAYIU         &&&\ding{51}&&   &&&\ding{51}&&   &\ding{51}&   &\ding{51}&&&   &&&&\ding{51}&&   &\ding{51}&&&&&   &\ding{51}&&\ding{51} \\ 

5 & Pengxin Yu    &&&\ding{51}&&   &\ding{51}&&&&   &\ding{51}&   &\ding{51}&&&   &&&&\ding{51}&&   &\ding{51}&&&&&   &\ding{51}&&\ding{51} \\ 

6 & FGZ Medical Research &\ding{51}&&&&   &&&&\ding{51}&   &\ding{51}&   &&&\ding{51}&   &\ding{51}&\ding{51}&&\ding{51}&&   &&&&&&   &&&\ding{51} \\ 

7 & KoalAI        &&&&&\ding{51}   &&\ding{51}&&&\ding{51}   &\ding{51}&   &\ding{51}&&&   &&&&\ding{51}&&{\small pel}   &\ding{51}&&\ding{51}&&\ding{51}&\ding{51}   &\ding{51}&&\ding{51} \\ 

8 & FGH\_365      &&\ding{51}&\ding{51}&&   &&\ding{51}&&&\ding{51}   &\ding{51}&    &\ding{51}&&&   &&\ding{51}&{\small CB}&{\small CT}&&   &\ding{51}&\ding{51}&&\ding{51}&\ding{51}&\ding{51}   &\ding{51}&&\ding{51} \\ 

9 & UKA           &&&\ding{51}&&   &\ding{51}&&&&\ding{51}  &\ding{51}&   &&\ding{51}&&   &&\ding{51}&&\ding{51}&&   &&\ding{51}&&&&   &\ding{51}&&\ding{51} \\ 

10 & Breizh-CT    &&&\ding{51}&&   &&\ding{51}&&&   &\ding{51}&   &&&&\ding{51}   &&&&\ding{51}&{\small CB}&   &&&&&&   &\ding{51}&&\ding{51} \\ 

11 & MedicalMind  &&&\ding{51}&&   &&\ding{51}&&&   &\ding{51}&   &&&\ding{51}&   &\ding{51}&&&&&   &\ding{51}&\ding{51}&\ding{51}&&&   &&\ding{51}&\ding{51} \\ 

12 & RRRocket\_Lollies &&&\ding{51}&&   &&\ding{51}&&&   &&\ding{51}   &&&\ding{51}&   &\ding{51}&\ding{51}&&\ding{51}&&\ding{51}   &&&&&&   &&&\ding{51} \\ 

13 & SKJP         &&&&\ding{51}&   &\ding{51}&&&&   &\ding{51}&   &&\ding{51}&&   &&&&{\small CB}&&   &&&&&&   &&& \\ 

14 & X-MAN        &\ding{51}&&&&   &&\ding{51}&&&   &\ding{51}&   &\ding{51}&&&   &&&{\small CB}&&&   &\ding{51}&&&&\ding{51}&   &\ding{51}&&\ding{51} \\ 

\bottomrule
\end{tabular}%
}
\end{table}
\end{landscape}
}
Model-Brain and Model-Pelvis. Preprocessing involved Z-score normalization for MRI and no normalization for CT images. No resizing, rescaling, or data augmentation was applied. Model input sizes were $64\times128\times224$ voxels for brain patches and $112\times160\times128$ voxels for pelvis patches. Inference utilized the nnU-Net's default sliding window with half-patch size overlap, and no postprocessing steps were applied since the sCT was presented in HU. The models were trained for 1000 epochs using an SGD optimizer with a Nesterov momentum of 0.99, an initial learning rate of $1\mathrm{e}{-2}$, and a polyLR scheduler.

\subsection{GEneRaTion (task 2)}
\label{method GEneRaTion}
GEneRaTion employed a 2D restoration approach using a Swin transformer \citep{liang2021swinir} combined with a pre-trained masked autoencoder \citep{he2022masked} for task 2. The SwinV2 architecture \cite{liu2022swinV2} was enhanced by incorporating group propagation blocks \citep{yang2022gpvit}. Depending on the training stage, the model included either an L1, MSE, or perceptual loss \citep{johnson2016perceptual}. Two identical models were created for the brain and pelvis. (CB)CT was linearly normalized between [-1000, 3000] HU for the brain and [-1000, 2000] HU for the pelvis. In a self-supervised pretraining phase, an L1 loss and a learning rate of $1\mathrm{e}{-4}$ were applied to $8\times8$ random patches with at least 75\% of the patch within the provided body mask. Random $90^\circ$ rotations or horizontal or vertical flipping were part of this pretraining step. Subsequently, the models were fine-tuned for 100 epochs on axial slices randomly cropped to $160\times160$ voxels, utilizing three stages. These stages involved training with 1) L1 loss and a learning rate of $1\mathrm{e}{-4}$, 2) MSE loss with a learning rate of $2\mathrm{e}{-5}$, and 3) a perceptual loss with a learning rate of $1\mathrm{e}{-5}$. During test-time ensembling, the three sCTs were combined through a weighted average, with weights calculated by $\frac{sCT - \text{mean}(sCT)}{\text{max}(sCT)}$, and preprocessing steps were restored.

\subsection{FAYIU (task 1 \& 2)}
\label{method FAYIU}
Team FAYIU implemented a patch-based 3D Swin UNETR \citep{hatamizadeh2021swin} in MONAI \cite{cardoso2022monai} for both tasks and regions separately. The Swin UNETR architecture, incorporating a vision transformer-based encoder and CNN-based decoder, enabled the processing of 3D patches. The models used a masked L1 loss. MRI inputs were normalized by dividing by 1000, while (CB)CT inputs were first made non-negative and subsequently divided by 2000. For training, 20 random patches of $32\times96\times96$ voxels were selected per patient, and no other data augmentation techniques were applied. At inference time, patches overlapping by $28 \times 72 \times 72$ voxels were selected, and overlapping regions were averaged in a weighted manner, with the weights for adjacent patches decreasing linearly as the overlap distance increased. Furthermore, the CT normalization procedure was reverted to result in an sCT in HU. The models were trained for 4000 epochs using the Adam optimizer and step-wise learning rate decay from $5\mathrm{e}{-4}$ to $5\mathrm{e}{-5}$.

\subsection{iu\_mia (task 1 \& 2)}
\label{method iu mia}
Team iu\_mia employed a 3D patch-based ShuffleUNet \citep{chatterjee2021shuffleunet} model conditioned on the anatomical region for both tasks, with the L1 loss for both tasks. This model incorporates specialized 3D pixel unshuffling and shuffling modules to effectively handle the unique 3D aspects of medical imaging data. Z-score normalization was applied to the 3D MRI volumes, while (CB)CT volumes underwent linear scaling by $(\text{(CB)CT}-1024)/4024$. They selected random patches measuring $96\times96\times96$ voxels for training, and no other data augmentation techniques were applied. At test time, sCTs were generated from patches with a 62.5\% overlap and averaging using Gaussian weighting ($\sigma=0.125$), and the normalization process was inverted. The models were trained for 3000 epochs using the Adam optimizer with a linear learning rate scheduler initialized at $1\mathrm{e}{-3}$. 

\subsection{Elekta (task 1)}
\label{method Elekta}
Team Elekta only participated in task 1, where they employed a 2.5D pix2pix \citep{pix2pix2017} model using a ResUnet \citep{ResUnet2018} generator and a discriminator implemented similarly to the encoding part of the ResUnet. Spectral normalization \citep{miyato2018spectral} was applied after each convolutional layer, and instance normalization replaced group normalization. Two identical models were created per anatomical region, using the least squares GAN loss~\citep{mao2017least} with L1 regularization (weight of 50). Linear scaling of MRI and CT intensities was conducted to fit within the range of [-1, +1], with source ranges determined by percentiles for MRI and fixed as [-1000, +2200] HU or [-1000, +3000] HU for CT. Two networks were trained for both regions, one covering the full CT intensity range and the other focusing on a narrower range. The training involved randomly selecting axial patches of $5\times192\times192$ and augmented using affine transformations, synthetic multiplicative bias fields, blurring, sharpening, gamma contrast adjustments, and linear intensity transformations for MRI. For CT, only affine transformations were applied. During inference, patches with $4 \times 96 \times 96$ voxels overlap were combined through weighted averaging, with higher weights assigned to pixels near the center of the patch and lower weights to those near the edge. 
The model was trained using the Adam optimizer and learning rates of $1\mathrm{e}{-4}$ and $5\mathrm{e}{-5}$ for the generator and discriminator, respectively. Additionally, a slow-moving exponential moving average (EMA) of the generator parameters was tracked during training and used as the final model for inference. Each model was trained six times, resulting in a final sCT ensembled by averaging the results of the six models.

\subsection{Pengxin Yu (task 2)}
\label{method Pengxin Yu}
Pengxin Yu employed a 3D patch-based model inspired by \cite{ge2019stereo} for task 2, implemented separately for the brain and pelvis. The model architecture featured consecutive multiscale residual blocks, effectively extracting fine-grained spatial structures and integrating stereo-correlation and image-expression constraints alongside the L1 reconstruction loss to guide structural detail and scene content. CBCT was linearly normalized between [-1000, 2000] HU, and for CT, center and region-specific windows were set: brain center A: [0, 3000] HU, brain center B and C: [-1000, 2000] HU, pelvis center A: [0, 2000] HU, and pelvis center B and C: [-1000, 1000] HU, after 
\afterpage{%
\clearpage
\begin{landscape}
\makeatother
\rowcolors{1}{}{lightgray}
\setlength{\tabcolsep}{4.5pt}

\begin{table}
\centering
\caption{\label{tab:metric_results}All quantitative metrics \(\left(\mu\pm\sigma\right)\) produced by every participant in task 1 (MRI-to-CT) and task 2 (CBCT-to-CT). There were three image-based and six dose-based metrics: three for photon treatment and three for proton treatment. The best results per task per metric are marked in boldface. }
\subcaption*{Metric table 1: The quantitative metrics for task 1 (MRI-to-CT). Participants who scored worse than the water baseline on one image metric were excluded from the final ranking. }

\begin{adjustbox}{max width=1.25\textwidth}
\begin{tabular}[]{c|c|ccc|ccc|ccc}
\toprule
\textbf{Rank}  & \textbf{Team} & \multicolumn{3}{c|}{\textbf{Image metrics}}  & \multicolumn{6}{c}{\textbf{Dose metrics}}\\  
\rowcolor{white}
\midrule
 & &  \multicolumn{3}{c|}{} & \multicolumn{3}{c|}{\textbf{Photon}} & \multicolumn{3}{c}{\textbf{Proton}} \\
 \rowcolor{white}\midrule
 & & MAE (HU, \(\downarrow\))& PSNR (dB, \(\uparrow\)) & SSIM (\(\uparrow\))   & MAE (Gy, \(\downarrow\)) & DVH (\(\downarrow\)) & \(\gamma_\text{2\%/2 mm}\) (\(\uparrow\)) & MAE (Gy, \(\downarrow\)) & DVH (\(\downarrow\)) & \(\gamma_\text{2\%/2 mm}\) (\(\uparrow\)) \\ \midrule 
    1 &SMU-MedVision&   \(\mathbf{58.83\pm13.41}\)& \(\mathbf{29.61\pm1.79}\) &  \(\mathbf{0.885\pm0.029}\)   &  \(\mathbf{0.0040\pm0.0032}\) & \(\mathbf{0.0265\pm0.0382}\) &  \(98.23\pm4.45\) &  \(0.0326\pm0.0220\) &       \(\mathbf{0.2087\pm0.2604}\) & \(97.28\pm2.58\) \\ 
    2 &Jetta\_Pang  &   \(65.73\pm13.75\)  & \(28.38\pm1.68\) &  \(0.869\pm0.032\)  &  \(0.0040\pm0.0037\) & \(0.0268\pm0.0429\) &   \(\mathbf{99.00\pm1.98}\) & \(\mathbf{0.0316\pm0.0194}\) &        \(0.2208\pm0.2680\) & \(\mathbf{97.54\pm2.37}\) \\ 
    3 &FAYIU  &   \(61.72\pm13.32\)  & \(28.83\pm1.61\) &   \(0.876\pm0.030\)  & \(0.0041\pm0.0036\) & \(0.0273\pm0.0437\) &  \(98.18\pm4.24\) &  \(0.0320\pm0.0202\) &        \(0.2150\pm0.2732\) & \(97.25\pm2.56\) \\ 
    4 & Elekta &   \(62.76\pm13.06\)  &   \(28.80\pm1.60\) &   \(0.875\pm0.030\)  &  \(0.0040\pm0.0036\) & \(0.0286\pm0.0525\) &  \(98.15\pm4.21\) &  \(0.0332\pm0.0220\) &       \(0.2271\pm0.2756\) &  \(97.27\pm2.50\) \\ 
    5 &iu\_mia  &   \(62.83\pm13.77\)  &  \(28.70\pm1.59\) &  \(0.873\pm0.029\)  &  \(0.0040\pm0.0034\) & \(0.0278\pm0.0533\) &  \(98.07\pm4.65\) & \(0.0322\pm0.0187\) &       \(0.2173\pm0.2608\) &  \(97.31\pm2.50\) \\ 
    6 &ShantouBME  &   \(67.54\pm14.17\)  & \(28.05\pm1.55\) &  \(0.863\pm0.031\)  & \(0.0042\pm0.0034\) & \(0.0275\pm0.0408\) &  \(98.11\pm4.18\) & \(0.0334\pm0.0197\) &       \(0.2185\pm0.2575\) & \(97.23\pm2.34\) \\ 
    7 &FGH\_365  &   \(66.75\pm13.18\)  & \(28.59\pm1.51\) &  \(0.866\pm0.032\)  & \(0.0043\pm0.0039\) & \(0.0324\pm0.0598\) &  \(97.94\pm4.75\) & \(0.0349\pm0.0254\) &       \(0.2249\pm0.2674\) & \(96.91\pm2.82\) \\ 
    8 &USC-LONI  &     \(70.70\pm14.40\)  & \(27.77\pm1.74\) &  \(0.854\pm0.034\)  & \(0.0045\pm0.0037\) &  \(0.0331\pm0.0590\) &  \(97.97\pm4.32\) &  \(0.0365\pm0.0210\) &       \(0.2715\pm0.3116\) &  \(97.00\pm2.77\) \\ 
    9 &UKA  &   \(77.39\pm22.04\)  & \(27.83\pm2.16\) &   \(0.849\pm0.050\)  & \(0.0045\pm0.0036\) & \(0.0327\pm0.0504\) &   \(98.40\pm4.04\) & \(0.0394\pm0.0337\) &       \(0.3641\pm0.3683\) & \(96.79\pm3.05\) \\ 
   10 &PSICPT\_AI4PT  &    \(78.00\pm27.44\)  & \(27.55\pm2.16\) &  \(0.839\pm0.049\)  & \(0.0044\pm0.0031\) & \(0.0316\pm0.0487\) &  \(98.17\pm4.13\) &  \(0.0370\pm0.0263\) &       \(0.2326\pm0.2695\) &  \(97.18\pm2.10\) \\ 
   11 & SubtleCT &   \(66.53\pm14.63\)  & \(28.51\pm1.68\) &  \(0.869\pm0.029\)  & \(0.0054\pm0.0042\) &  \(0.0370\pm0.0624\) &   \(98.20\pm4.12\) & \(0.0443\pm0.0308\) &       \(0.2353\pm0.2807\) & \(96.17\pm2.99\) \\ 
   12 & mriG &   \(82.01\pm17.77\)  & \(26.38\pm1.53\) &  \(0.842\pm0.035\)  &  \(0.0046\pm0.0040\) & \(0.0305\pm0.0453\) &  \(97.77\pm4.24\) & \(0.0318\pm0.0182\) &       \(0.2363\pm0.2772\) &  \(97.08\pm2.70\) \\ 
   13 & KoalAI &   \(68.94\pm11.82\)  & \(28.14\pm1.37\) &  \(0.862\pm0.029\)  & \(0.0054\pm0.0045\) & \(0.0338\pm0.0432\) &  \(97.94\pm4.49\) &  \(0.0460\pm0.0314\) &        \(0.2450\pm0.2952\) &  \(95.86\pm3.60\) \\ 
   14 &Breizh-CT  &   \(93.57\pm19.01\)  & \(25.86\pm1.43\) &  \(0.806\pm0.032\)  &  \(0.0050\pm0.0041\) & \(0.0334\pm0.0491\) &  \(98.56\pm2.44\) & \(0.0454\pm0.0222\) &       \(0.3052\pm0.3153\) &  \(95.97\pm3.00\) \\ 
   15 &SKJP  &   \(88.42\pm26.89\)  & \(26.44\pm2.03\) &  \(0.815\pm0.043\)  & \(0.0063\pm0.0047\) & \(0.0428\pm0.0553\) &  \(97.72\pm4.97\) & \(0.0562\pm0.0319\) &       \(0.3445\pm0.4111\) & \(94.83\pm4.26\) \\ 
   16 & reza.karimzadeh &  \(113.38\pm20.35\)  & \(24.71\pm1.43\) &  \(0.764\pm0.034\)  & \(0.0083\pm0.0067\) & \(0.0542\pm0.0644\) &   \(97.02\pm4.70\) & \(0.0565\pm0.0288\) &       \(0.4068\pm0.6937\) &  \(94.47\pm3.80\) \\ 
   17 & thomashelfer &  \(126.32\pm17.01\)  & \(23.69\pm0.94\) &  \(0.756\pm0.029\)  &  \(0.0098\pm0.0070\) & \(0.0599\pm0.0643\) &  \(97.38\pm4.86\) & \(0.0791\pm0.0585\) &       \(0.3823\pm0.4215\) & \(94.53\pm3.66\) \\ 
   18 & X-MAN &  \(117.88\pm45.08\)  &  \(25.64\pm2.20\) &  \(0.774\pm0.097\)  & \(0.0117\pm0.0148\) & \(0.0736\pm0.0957\) &   \(96.42\pm5.00\) & \(0.0817\pm0.0759\) & \(3514.188\pm38491.0637\) & \(93.72\pm5.77\) \\ 
   19 & Water baseline &  \(332.93\pm89.53\)  & \(17.95\pm1.73\) &  \(0.552\pm0.127\)  & \(0.0166\pm0.0104\) & \(0.0972\pm0.1057\) &   \(96.70\pm5.36\) & \(0.1334\pm0.0573\) &       \(1.8218\pm8.2652\) & \(88.87\pm7.83\) \\
   - & Stratified baseline &  \(69.45\pm16.33\)  & \(28.42\pm1.92\) & \(0.854\pm0.027\)  & \(0.0049\pm0.0039\) & \(0.0347\pm0.0504\) & \(98.21\pm5.17\) & \(0.0515\pm0.0309\)  & \(0.4180\pm0.3567\) & \(95.43\pm3.76\)  \\
   \hline
   \end{tabular}
   \end{adjustbox}

\vspace{0.5cm}
\subcaption*{Metric table 2: The quantitative metrics for task 2 (CBCT-to-CT). Participants who scored worse than the water baseline on one image metric were excluded from the final ranking. }
\begin{adjustbox}{max width=1.25\textwidth}
\begin{tabular}[]{c|c|ccc|ccc|ccc}
\toprule
\rowcolor{white}

\textbf{Rank}  & \textbf{Team} & \multicolumn{3}{c|}{\textbf{Image metrics}}  & \multicolumn{6}{c}{\textbf{Dose metrics}}\\  
\rowcolor{white}
\midrule
 & &  \multicolumn{3}{c|}{} & \multicolumn{3}{c|}{\textbf{Photon}} & \multicolumn{3}{c}{\textbf{Proton}} \\
 \rowcolor{white}\midrule
 & & MAE (HU, \(\downarrow\)) & PSNR (dB, \(\uparrow\)) & SSIM (\(\uparrow\))   & MAE (Gy, \(\downarrow\)) & DVH (\(\downarrow\)) & \(\gamma_\text{2\%/2 mm}\) (\(\uparrow\)) & MAE (Gy, \(\downarrow\)) & DVH (\(\downarrow\)) & \(\gamma_\text{2\%/2 mm}\) (\(\uparrow\)) \\ \midrule 
   1 & SMU-MedVision &   \(\mathbf{49.95\pm11.78}\)  &  \(\mathbf{30.79\pm2.00}\) & \(\mathbf{0.906\pm0.036}\) & \(\mathbf{0.0038\pm0.0042}\) &  \(\mathbf{0.0240\pm0.0703}\) &  \(99.49\pm1.65\) & \(\mathbf{0.0283\pm0.0251}\) &       \(\mathbf{0.1663\pm0.2235}\) & \(\mathbf{97.57\pm3.12}\) \\
   2 & GEneRaTion &     \(55.50\pm11.00\) & \(30.48\pm1.72\)& \(0.897\pm0.033\)  &  \(0.0040\pm0.0042\) &  \(\mathbf{0.0241\pm0.0530}\) &   \(99.55\pm1.20\) & \(0.0294\pm0.0251\) &       \(0.1689\pm0.2188\) & \(97.42\pm3.11\) \\
   3 & iu\_mia &  \(50.79\pm11.81\)& \(30.58\pm1.95\)  & \(\mathbf{0.906\pm0.034}\)  & \(0.0045\pm0.0083\) & \(0.0326\pm0.1543\) &  \(98.99\pm4.57\) & \(0.0336\pm0.0408\) &       \(0.1728\pm0.2269\) &  \(97.00\pm4.72\) \\
   4 & FAYIU &   \(51.18\pm11.34\) &  \(30.40\pm1.93\)& \(0.903\pm0.034\)  &  \(0.0044\pm0.0080\) & \(0.0317\pm0.1415\) &  \(99.06\pm4.29\) & \(0.0333\pm0.0393\) &       \(0.1711\pm0.2197\) & \(97.09\pm4.42\) \\
   5 & Pengxin Yu &    \(54.05\pm12.30\) & \(30.56\pm1.95\)&   \(0.900\pm0.037\)  &  \(0.0043\pm0.0070\) & \(0.0304\pm0.1297\) &  \(99.19\pm3.72\) &  \(0.0320\pm0.0342\) &       \(0.1762\pm0.2296\) & \(97.16\pm4.18\) \\
   6 & FGZ Medical Research &   \(60.65\pm12.56\) & \(29.67\pm1.71\)& \(0.879\pm0.039\)  &  \(0.0040\pm0.0032\) & \(0.0251\pm0.0442\) &  \(\mathbf{99.57\pm1.07}\) & \(0.0307\pm0.0208\) &       \(0.2239\pm0.2718\) & \(97.46\pm2.85\) \\
   7 & KoalAI &   \(56.13\pm12.06\) & \(30.11\pm1.89\)& \(0.897\pm0.034\)  &  \(0.0055\pm0.0080\) & \(0.0385\pm0.1411\) &  \(98.99\pm4.38\) & \(0.0408\pm0.0373\) &       \(0.2106\pm0.2497\) & \(96.05\pm4.77\) \\
   8 & FGH\_365 &   \(56.29\pm11.08\) & \(30.24\pm1.79\)& \(0.896\pm0.035\)  & \(0.0058\pm0.0084\) &  \(0.0410\pm0.1517\) &  \(98.97\pm4.51\) & \(0.0432\pm0.0439\) &       \(0.2029\pm0.2387\) & \(95.94\pm5.08\) \\
   9 &UKA &   \(65.46\pm19.25\) & \(29.13\pm2.64\)& \(0.881\pm0.041\)  & \(0.0049\pm0.0092\) &  \(0.0364\pm0.1540\) &  \(98.98\pm4.89\) &  \(0.0365\pm0.0390\) &        \(0.218\pm0.2479\) & \(96.83\pm4.79\) \\
   10 &Breizh-CT &    \(71.28\pm13.60\) & \(28.43\pm1.65\)& \(0.863\pm0.041\)  & \(0.0052\pm0.0052\) & \(0.0347\pm0.0739\) &  \(99.25\pm2.23\) &  \(0.0449\pm0.0380\) &       \(0.2309\pm0.2495\) & \(96.06\pm4.15\) \\
   11 &MedicalMind &    \(68.40\pm13.48\) & \(29.18\pm1.63\)&  \(0.875\pm0.030\)  &  \(0.0094\pm0.0110\) & \(0.0665\pm0.1671\) &  \(98.42\pm5.38\) & \(0.0678\pm0.0569\) &       \(0.2395\pm0.2534\) & \(94.11\pm6.23\) \\
   12 &RRRocket\_Lollies &   \(71.58\pm13.79\) &  \(28.34\pm1.50\)& \(0.862\pm0.036\)  & \(0.0099\pm0.0095\) & \(0.0651\pm0.1497\) &  \(98.42\pm4.95\) &  \(0.0740\pm0.0481\) &       \(0.2744\pm0.2678\) & \(92.32\pm5.87\) \\
   13& SKJP&   \(78.63\pm18.88\) & \(27.98\pm1.71\)& \(0.853\pm0.033\)  & \(0.0113\pm0.0092\) & \(0.0784\pm0.1478\) &  \(98.67\pm4.76\) & \(0.0844\pm0.0533\) &       \(0.4572\pm0.4907\) &  \(91.50\pm6.38\) \\
   14 & X-MAN&   \(99.15\pm59.43\) & \(27.51\pm3.41\)& \(0.831\pm0.087\)  & \(0.0227\pm0.0384\) &  \(0.1150\pm0.2003\) & \(92.44\pm14.5\) & \(0.1035\pm0.1056\) &       \(0.3747\pm0.3803\) & \(91.70\pm10.8\) \\
   15 & Water baseline& \(344.26\pm125.32\) & \(17.97\pm2.08\)& \(0.546\pm0.149\)  & \(0.0191\pm0.0118\) & \(0.1255\pm0.1663\) &   \(96.33\pm5.80\) & \(0.1453\pm0.0525\) &    \(28.7704\pm309.6595\) & \(85.08\pm9.78\) \\
   - & Stratified baseline & \(69.99\pm18.93\) & \(28.65\pm2.25\)& \(0.837\pm0.057\)   & \(0.0046\pm0.0027\) & \(0.0332\pm0.0432\) & \(\mathbf{99.86\pm0.46}\) & \(0.0432\pm0.0282\) & \(0.3936\pm0.3608\) & \(95.93\pm2.97\)   \\ 
   \hline
\end{tabular}
\end{adjustbox}
\end{table}
\end{landscape}

\begin{figure*}[t]
\subfloat[Example sCTs of participants for task 1 (MRI-to-CT). Despite significant synthesis errors at the tissue boundaries, high gamma-pass rates are achieved for all methods.]{%
  \includegraphics[clip,width=0.99\linewidth]{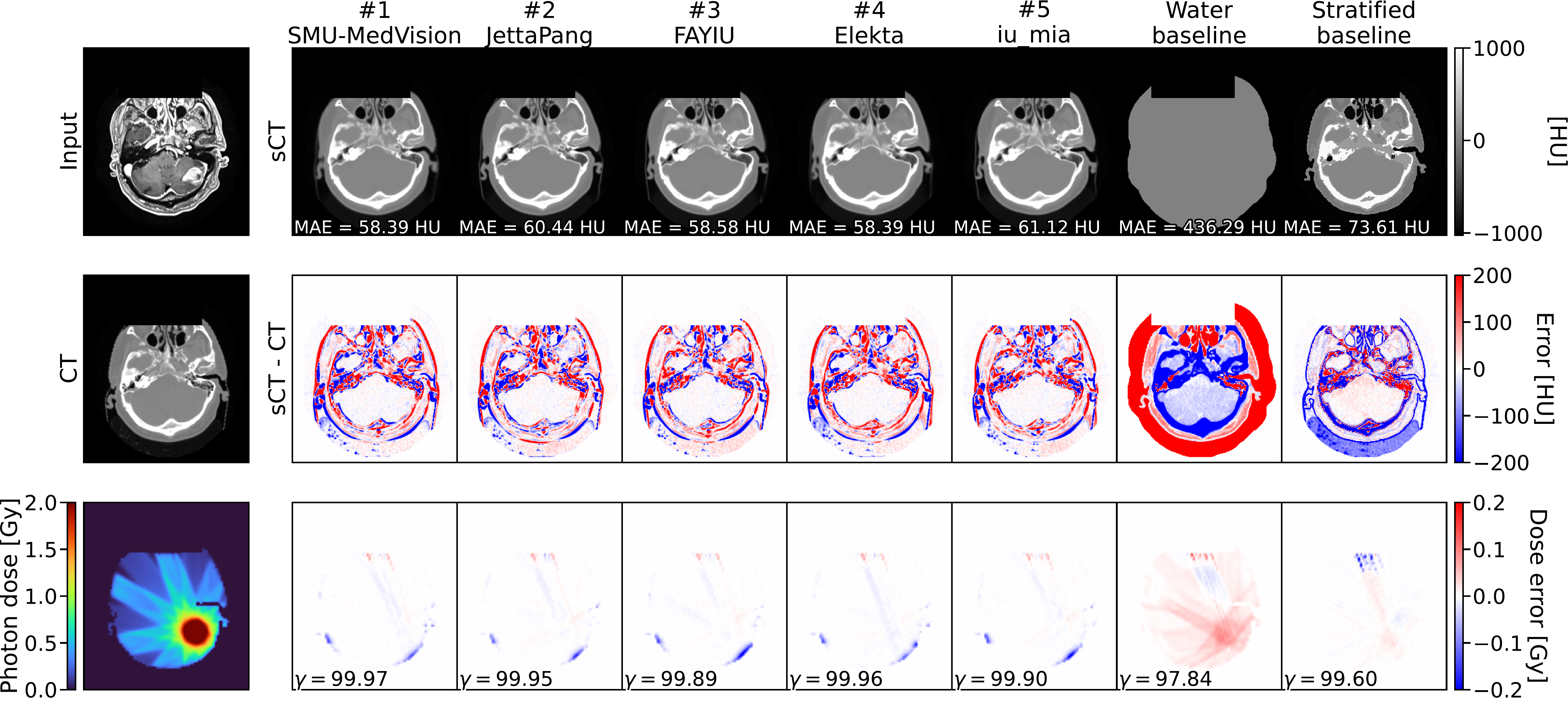}%
}
\vspace{0.5cm}
\subfloat[Example sCTs of participants for task 2 (CBCT-to-CT). The largest synthesis errors occur at the boundaries between tissue types (e.g., air/body contour boundary or soft-tissue/bone boundary). Moreover, the anatomy does not fit in the FOV of the CBCT, requiring synthesis of tissue outside the FOV]{%
  \includegraphics[clip,width=0.99\linewidth]{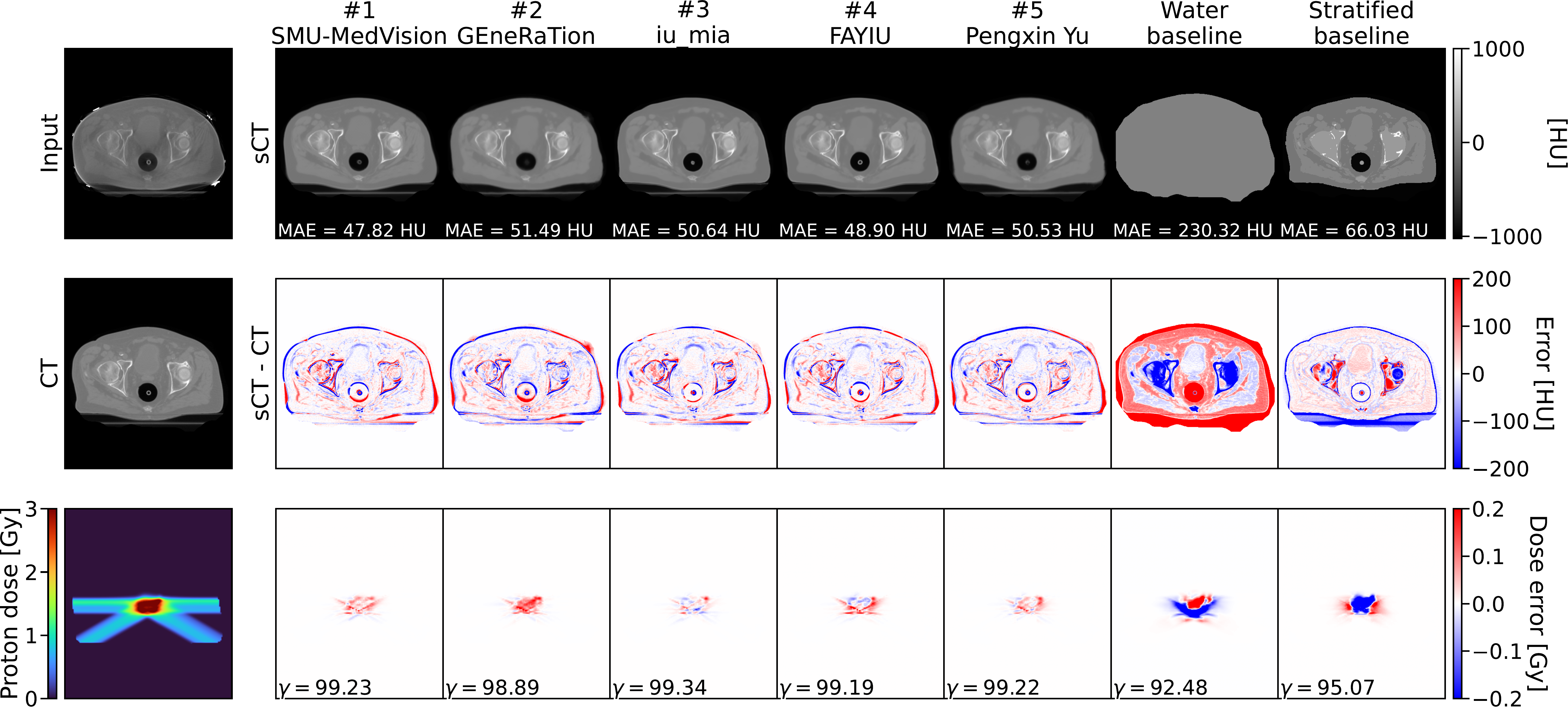}%
}
\caption{Examples of synthetic CTs for task 1 (MRI-to-CT; a) and task 2 (CBCT-to-CT: b). The model input is shown in the upper left, and the ground truth is in the center-left. The sCT of the top five participants for task 1 and task 2 are shown in the top row. The difference from ground truth CT is shown in the middle row. On the bottom left is the planned irradiation based on the CT for a photon (a) and proton (b) plan. The bottom row shows the dose difference when the treatment plan is applied to the sCT (CT dose - sCT dose). All values outside the body contour were masked.}
\label{fig:example_pats}
\end{figure*}
\clearpage
}
which the intensities were linearly normalized. During training, patches of $8\times180\times180$ voxels were created by randomly resizing, cropping, and horizontal flipping. At test time, overlapping patches were selected with an overlap of $2 \times 32 \times 48$ voxels. The models were trained for 1000 epochs with the AdamW optimizer with an initial learning rate of $3\mathrm{e}{-4}$ and reducing the learning rate by a factor 10 when the validation loss has not decreased for 10 epochs in a row. The final epoch used at test-time was determined based on the best PSNR on the sub-validation set, created from the training set.

\section{Results}
\label{sec:results}
\subsection{Overall sCT generation performance }
\label{subsec:results overall performance}

\autoref{tab:metric_results} presents the final ranking and quantitative results of the 18 eligible teams for task 1 and 14 eligible teams for task 2, along with the two baseline algorithms. All eligible teams outperformed the water baseline in both tasks based on the image similarity metrics. Almost all teams also outperform the water baseline based on the dose metrics. However, one team (X-MAN) did not outperform the water baseline when considering the $\gamma_{\text{photon}}$ and DVH\textsubscript{proton} metrics in task 1 and the MAE\textsubscript{photon} and $\gamma_{\text{photon}}$ metrics in task 2.
On the other hand, 11/18 and 14/18 teams outperform the stratified baseline based on image similarity for tasks 1 and 2, respectively. Regarding the dose metrics in task 1, 10/18 and 14/18 outperformed the stratified baseline for the photon and proton gamma pass rate, respectively. In task 2, the stratified baseline outperformed all teams based on the photon gamma pass rate, while 11/15 teams achieved a higher proton gamma pass rate than the stratified baseline. Interestingly, a higher image similarity did not automatically lead to an improved dose distribution. For example, comparing SMU-MedVision (rank 1) and FGZ Medical Research (rank 6) for task 2, we observe a large difference in MAE of $49.95 \pm 11.78$ and $60.65 \pm 12.56$ HU, while a subtle difference in photon gamma pass rates of $99.49 \pm 1.65$ and $99.57 \pm 1.07$ is seen. Such differences motivated us to perform an in-depth statistical analysis examining the significance of one team outperforming another based on individual metrics (Figures 1 and 2 in supplementary document B). Based on the image similarity metrics, high-ranking teams robustly outperform lower-ranked teams. Statistical significant improvements were observed when comparing all image metrics between a team and another team ranked at least seven places lower for task 1, or six placed lower for task 2.
However, for the dose metrics, this relation is weaker. In task 1, no statistical significant differences were observed between the top fourteen teams regarding the photon dose metrics and top eleven teams regarding the proton dose metrics.
In task 2, no statistically significant differences were observed between the top eight teams regarding the photon and proton dose metrics, except for the fifth team (Pengxin Yu), which significantly outperforms the seventh team (KoalAI) regarding the proton DVH metric.

Overall, the teams successfully generated high-quality sCTs, accurately synthesizing soft-tissue density. However, visual examples in \autoref{fig:example_pats} show more pronounced errors at transitions between tissue densities, such as the boundaries between air and soft tissue or soft tissue and bone. These errors at the boundaries of the input with the ground truth CT appear consistent across teams and lead to increased dose error when a beam passes through these regions. Moreover, in the pelvic cases, the anatomy does not always fit within the field-of-view of the CBCT, requiring participants to synthesize anatomy not present in the model input.

The average inference time per case was $5.2 \pm 2.8$ minutes, with teams utilizing an average of $4.0\pm4.8$ GB of GPU RAM. The maximum observed inference time for a single case was 21.8 minutes. There was a notable spread in resource usage between teams, and a detailed overview per team per subtask is available in Figure 3 in supplementary document B.

\subsection{Model design predictors}
\label{subsec: results model design}
Of all the teams that participated in both tasks, the challenge winner, SMU-MedVision, was the only team to implement two different model architectures for each task. Most teams used the same model architecture for the brain and the pelvis but trained it separately for both regions. Therefore, the limited number of teams that chose similar or different models/parameters for the brain and pelvis did not allow for visible trends in the rankings (\Cref{tab:summary task1,tab:summary task2}). Nevertheless, teams that used one model conditioned on the anatomy consistently secured relatively high ranks for both tasks. Still, the team that trained one collective model without conditioning on the anatomical region ranked last in both tasks.

\begin{figure*}[htbp]
    \centering
    \includegraphics[width=\linewidth]{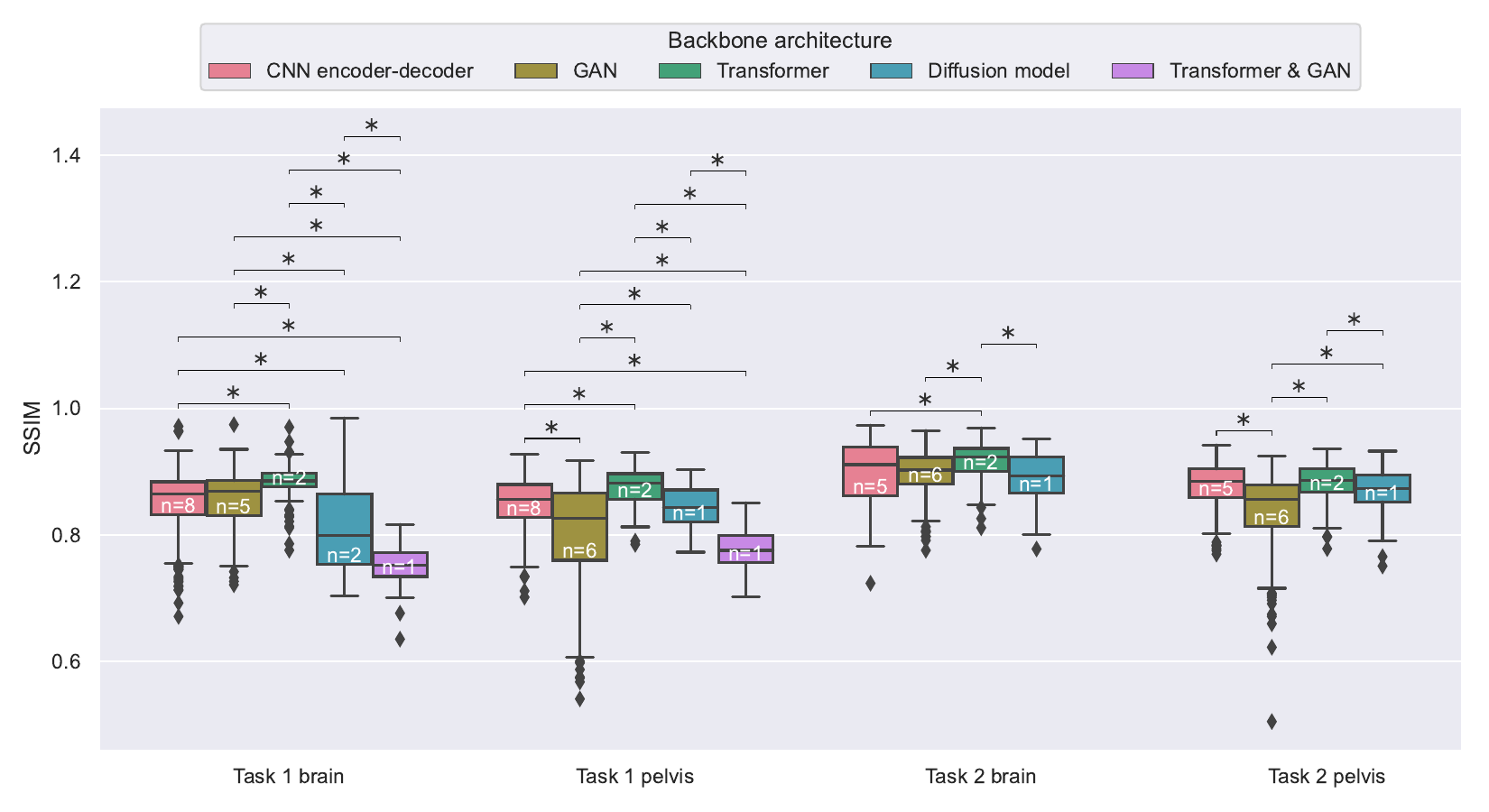}
    \caption{Boxplots of SSIM values for all patients in each subtask, i.e., task 1 (MRI-to-CT) or 2 (CBCT-to-CT) and brain or pelvis, grouped by the model backbone choice of each team. The $n$ in the boxes indicates the number of teams represented in that box. 
    An asterisk indicates significant differences within one subtask.}
    \label{fig:backbone_performance}
\end{figure*}

In addition, plain CNN decoder-encoder and GAN-based models were prevalent among the teams. However, the teams that placed first and third in task 1 and second and fourth in task 2 used transformer-based approaches. These transformers showed significantly better performance in both regions for task 1 and in the brain for task 2, achieving average SSIM values of $0.88 \pm 0.03$ for task 1 and $0.90 \pm 0.03$ for task 2 (\autoref{fig:backbone_performance}). Following the transformers, CNN encoder-decoder models were the next best-performing, yielding SSIM values of $0.85 \pm 0.04$ and $0.89 \pm 0.04$ for tasks 1 and 2, respectively. Conversely, teams using GANs tended to rank lower (\Cref{tab:summary task1,tab:summary task2}), with SSIM values of $0.83 \pm 0.07$ for task 1 and $0.87 \pm 0.05$ for task 2. Notably, GANs showed a significant performance drop, especially for the pelvis cases (\autoref{fig:backbone_performance}). Finally, the diffusion model, rarely adopted in this challenge, achieved SSIM values of $0.82 \pm 0.06$ and $0.88 \pm 0.04$ for tasks 1 and 2, respectively.

Only one team (RRRocket\_Lollies, task 2) implemented an unsupervised approach which placed them close to the bottom of the ranking (12\textsuperscript{th} out of 14). Due to the lack of unsupervised methods we could not extend the analysis on the supervision level.

\begin{figure*}
    \centering
    \includegraphics[width=0.98\linewidth]{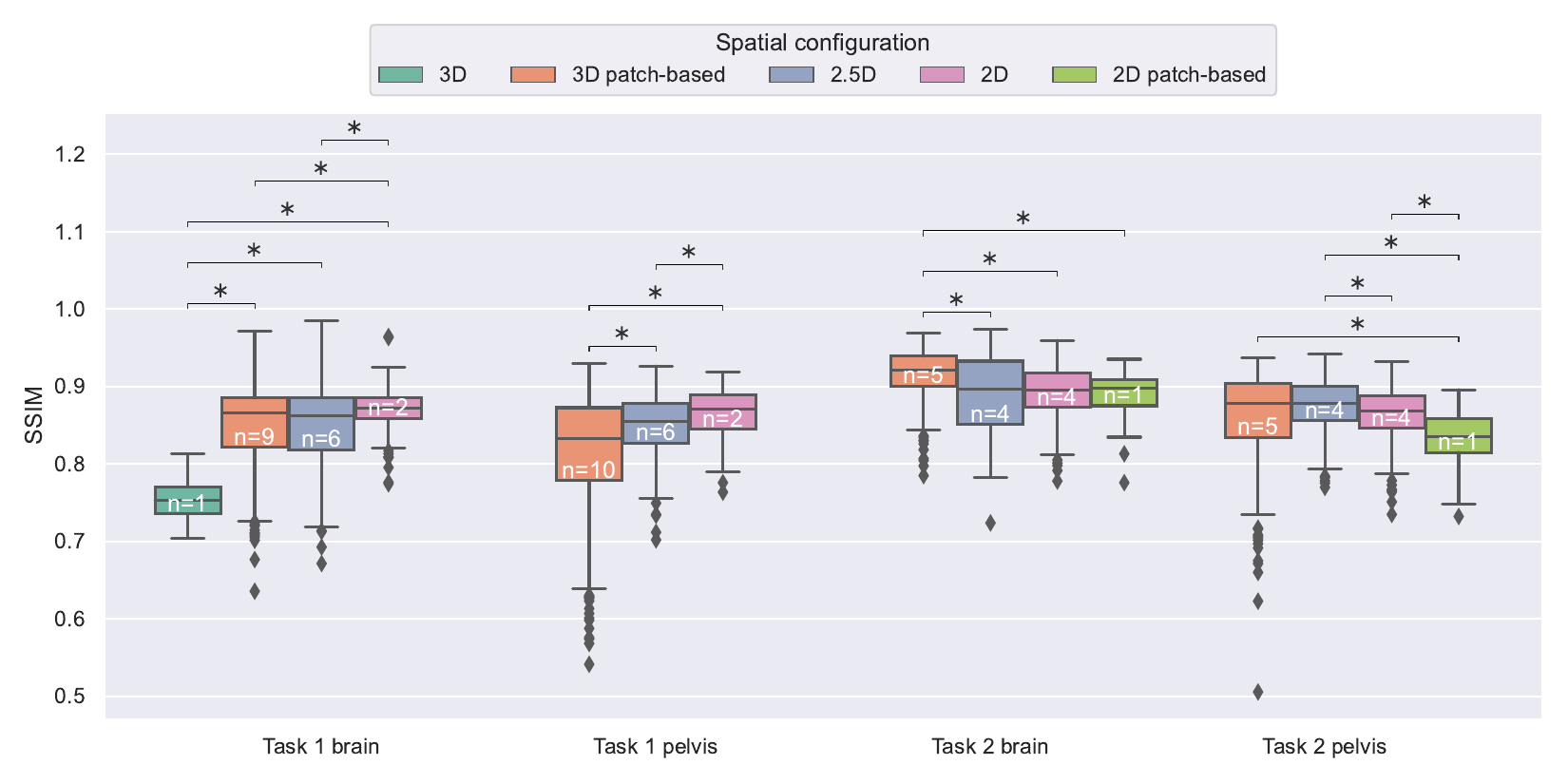}
    \caption{Boxplots of SSIM values for all patients in each subtask, i.e., task 1 (MRI-to-CT) or 2 (CBCT-to-CT) and brain or pelvis, grouped by the spatial configuration of the models designed by the team. The $n$ in the boxes indicates the number of teams represented in that box. An asterisk indicates significant differences within one subtask.}
    \label{fig:dimension_per_task}
\end{figure*}

Spatial configuration exhibited opposing trends in the two tasks (\autoref{fig:dimension_per_task}). For task 1 (MRI-to-CT), 3D patch-based, 2.5D, and 2D models achieved SSIM values of $0.83 \pm 0.06$, $0.85 \pm 0.04$, and $0.87 \pm 0.03$, respectively, with 2D models significantly outperforming the others. In contrast, for task 2 (CBCT-to-CT), 3D patch-based models (significantly) outperformed other models, with SSIM values of $0.89 \pm 0.06$, $0.88 \pm 0.04$, and $0.88 \pm 0.04$ for 3D patch-based, 2.5D, and 2D models, respectively.

No significant differences in choices for preprocessing, data augmentation, and postprocessing and, consequently, no trends in ranking were observed (\Cref{tab:summary task1,tab:summary task2}). The numerous combinations of processing steps and substantial differences in model design prevent definitive conclusions about the importance of specific processing steps.

\subsection{Data influence}
\label{subsec: results data influence}
\begin{figure*}
    \centering
    \includegraphics[width=\linewidth]{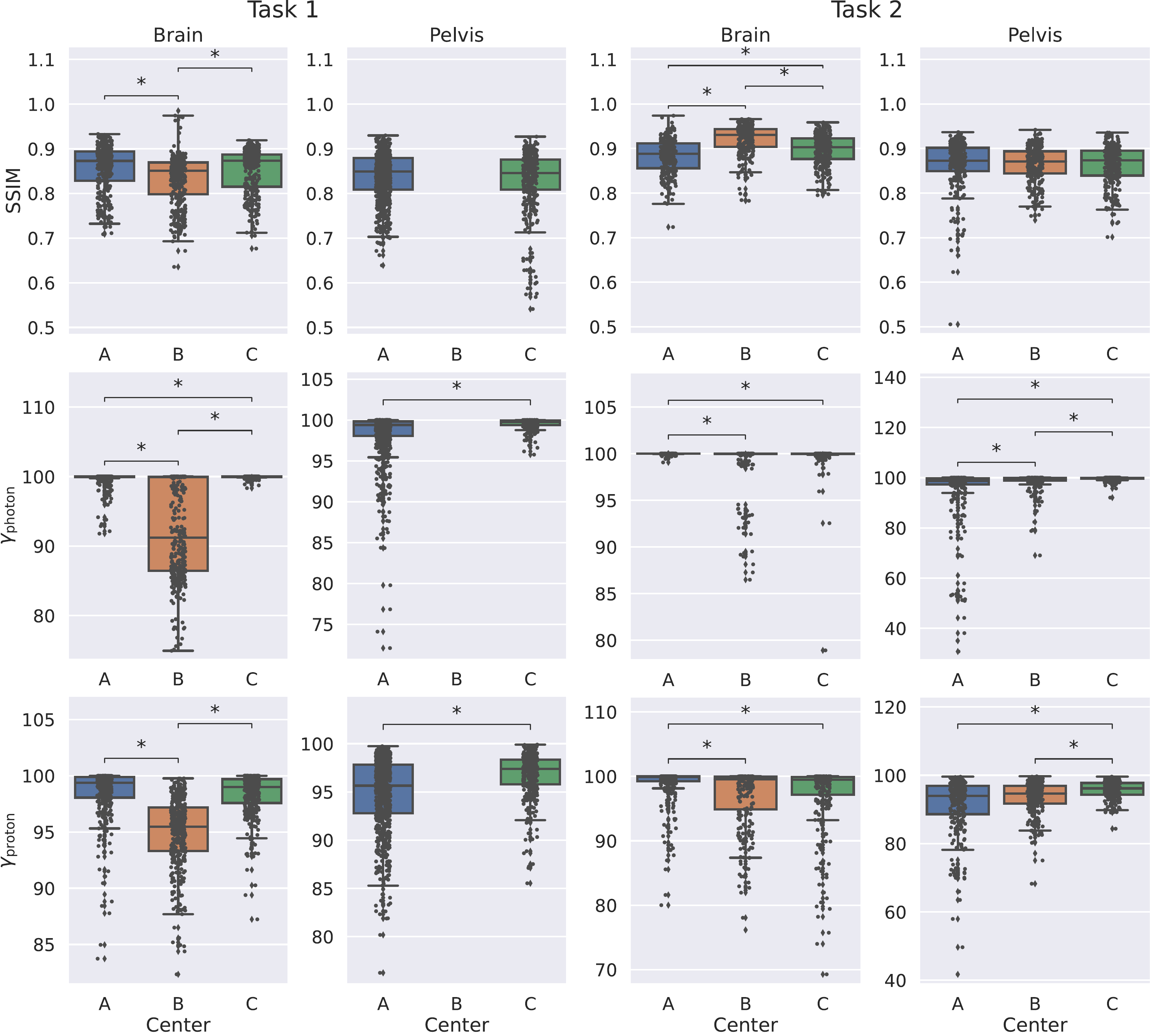}
    \caption{Boxplots of the teams' performance in terms of SSIM and gamma pass rates for photon and proton, grouped by different subsets in our dataset, analyzing the differences between task, anatomical region and center. Asterisks indicate significant differences.}
    \label{fig:data_invest}
\end{figure*}
The image quality of the brain patients was significantly different between the centers (\autoref{fig:data_invest}). In task 1, the participants generated sCTs for centers A, B, and C  with an SSIM of \(0.857\pm0.052\), \(0.831\pm0.056\), and \(0.852\pm0.050\), respectively. For task 2, the participants generated sCTs for centers A, B, and C with an SSIM of \(0.883\pm0.039\), \(0.921\pm0.034\), and \(0.897\pm0.035\), respectively. No statistically significant differences in image similarity were observed between centers for the pelvis data. The sCTs in task 2 showed a better image similarity than those in task 1, with an MAE of \(79.40\pm28.30\) HU for task 1 versus \(63.50\pm24.34\) HU for task 2. 
When considering the dose metrics for brain cases in task 1, center B (\(\gamma_\text{photon}=92.03\pm6.84\)) underperforms compared to centers A (\(\gamma_\text{photon}=99.65\pm1.09\)) and C (\(\gamma_\text{photon}=99.93\pm0.17\)). On the other hand, for pelvis cases in task 1, center A (\(\gamma_\text{photon}=98.29\pm3.00\)) underperforms relative to center C (\(\gamma_\text{photon}=99.55\pm0.58\)). For brain cases in task 2, minor dose differences were observed between the centers, with \(\gamma_\text{proton} = 98.80\pm2.83\), \(96.87\pm4.75\), and \(97.34\pm4.94\) for centers A, B, and C, respectively. 

For task 1, each center employed consistent MRI scanning protocols for each anatomical region. Consequently, a comparison at the level of the MRI scan sequence yields identical results, as illustrated in \autoref{fig:data_invest}. Moreover, the absence of variability in magnetic field strengths for centers B and C constrained this analysis to center A (Figure 4 in supplementary document B).
For the brain, the only significant difference was observed for \(\gamma_\text{photon}\), which decreased from \(98.99 \pm 1.43\) for 1.5T to \( 97.33 \pm 3.23\) for 3T. In contrast, for the pelvis, a significant increase in performance was observed for 3T compared to 1.5T. Specifically, the SSIM increased from $0.83 \pm 0.05$ to $0.84 \pm 0.05$, \(\gamma_\text{photon}\) increased from $97.51 \pm 3.45$ to $98.75 \pm 2.59$, and \(\gamma_\text{proton}\) increased from $93.29 \pm 4.05$ to $ 95.64 \pm 3.42$. 

A further investigation of the performance at the patient level, including a visual analysis of outlier patients, is presented in section 1.2 of supplementary document B.

\subsection{Metric correlations}
\label{subsec: results metric correlations}

    \begin{figure*}[t]
\subfloat[Image similarity metrics\label{figa:correlation_image}]{%
  \includegraphics[clip,height=7.1cm]{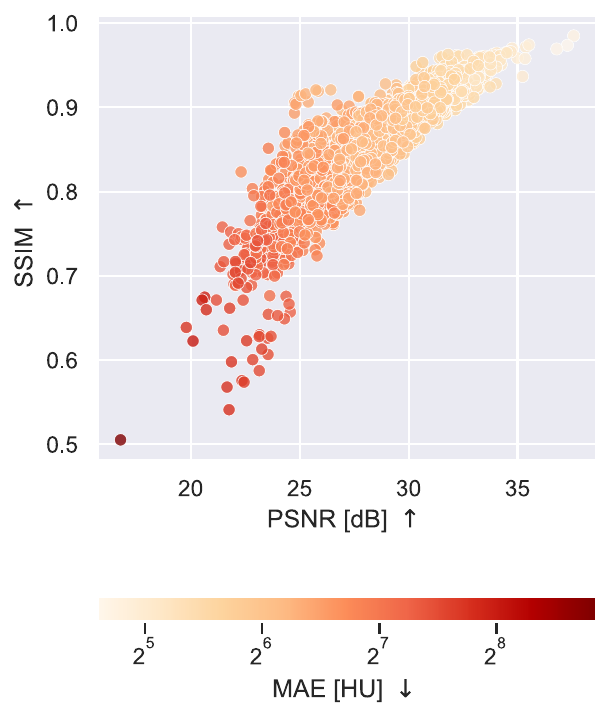}%
}
\hfill
\subfloat[Photon metrics\label{figb:correlation_photon}]{%
  \includegraphics[clip,height=7.1cm]{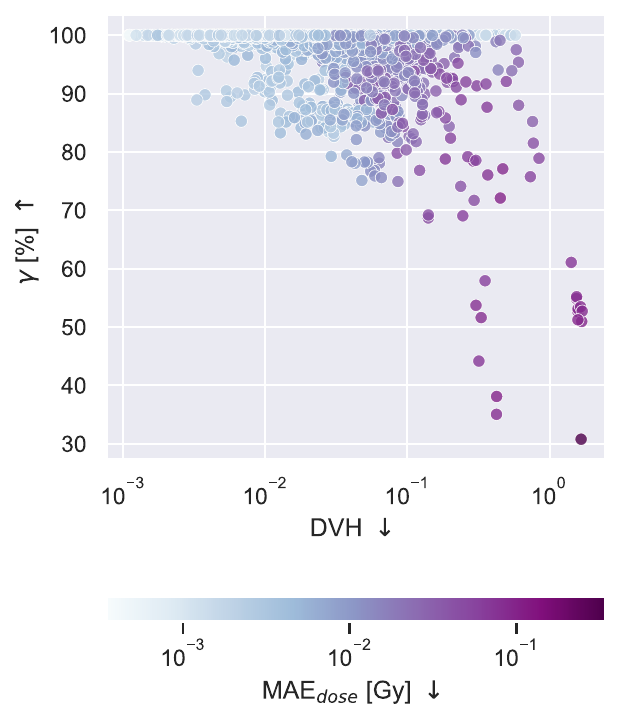}%
}
\hfill
\subfloat[Proton metrics\label{figc:correlation_proton}]{%
  \includegraphics[clip,height=7.1cm]{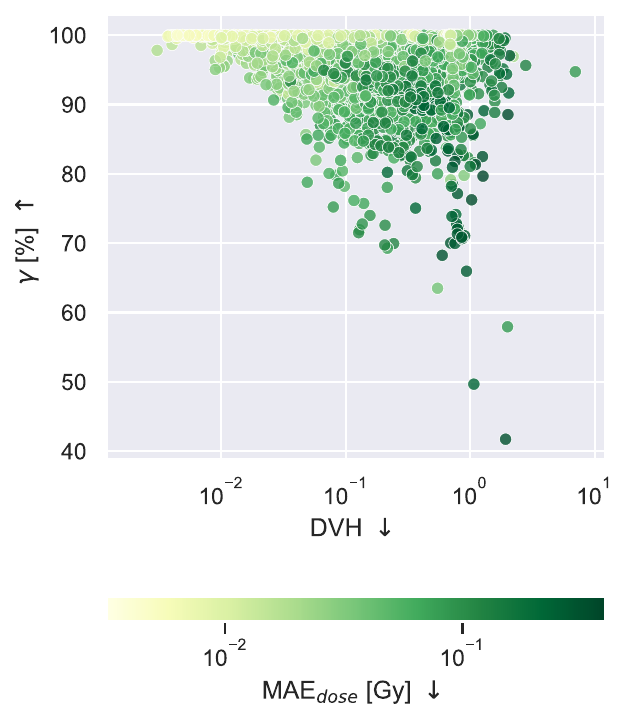}%
}
\caption{Correlation plots among metrics in the three categories: a) image metrics, b) photon metrics, and c) proton metrics. Each data point indicates a team's performance for one patient in either task 1 or 2. Note that some metrics are presented using a logarithmic scale, and one extreme outlier for the proton DVH metric (in the order of \(1 \times 10^{5})\) is excluded from the plot.}
\label{fig:correlation_metrics}
\end{figure*}
\begin{figure}
    \centering
    \includegraphics[width=\linewidth]{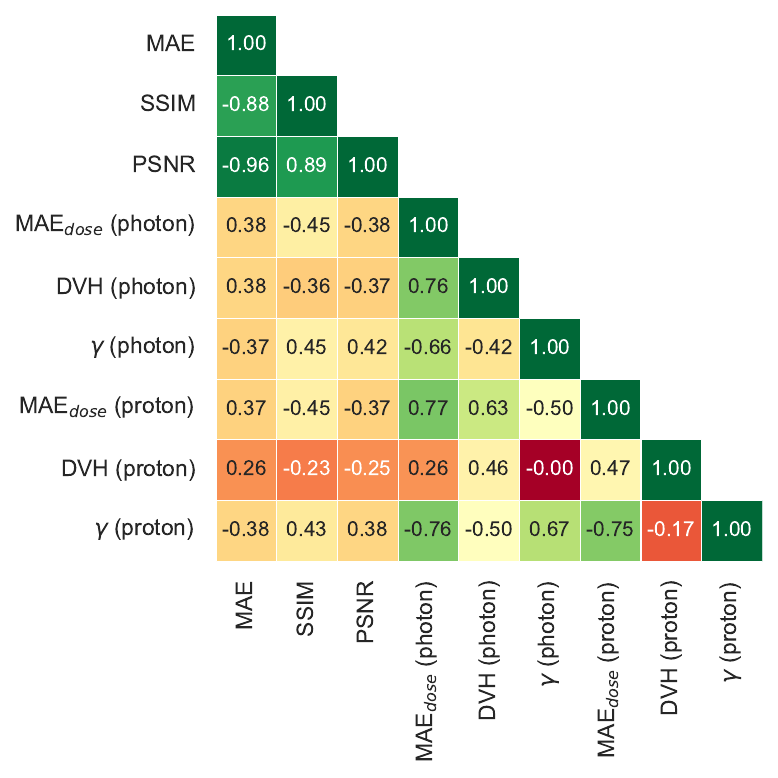}
    \caption{Spearman rank correlation coefficient \(\rho\) between the different metrics. Note that the interpretation of the correlation coefficient is contingent upon whether both compared metrics exhibit concordant trends.}
    \label{fig:spearman}
\end{figure}

\autoref{fig:correlation_metrics} highlights the correlations within the three metric groups. We observe strong correlations within the image similarity metric group, with the absolute inter-metric Spearman correlation coefficients \(\vert\rho\vert\) ranging from 0.88 to 0.96 (\autoref{fig:spearman}). These values consistently measure the underlying aspects of all three image metrics. In contrast, the photon and proton metrics show weaker correlations within their groups. Among the dose metrics, the MAE\textsubscript{dose} (photon) shows the highest correlation with the other dose metrics, such as $\gamma$ pass rate, with coefficients of -0.66 and -0.75 for photons and protons, respectively. While the correlation between the MAE\textsubscript{dose} (photon) and DVH (photon) is strong (0.76), the correlation between MAE\textsubscript{dose} (photon) and DVH (proton) is significantly lower (0.26). The proton DVH metric shows poor correlation with all metrics, highlighting the complex relationship between these metrics (\Cref{figb:correlation_photon,figc:correlation_proton,fig:spearman}).

Furthermore, the metric groups correlate moderately with each other. The average absolute coefficients between image similarity metrics and photon metrics were $0.40\pm 0.03$, while those with proton metrics were $0.47\pm 0.08$. Moreover, the average absolute correlation coefficient between photon and proton metrics was $0.50\pm 0.23$, with the large standard deviation introduced by a correlation coefficient of zero between DVH (proton) and $\gamma$ (photon). Overall, the results strongly suggest that an sCT similar to the ground truth CT does not directly translate into a dose distribution similar to the reference distribution, highlighting that the different metrics focus on different aspects in evaluating sCTs.

\subsection{Ranking stability and correlations}
\label{subsec: results ranking stability}
\begin{figure}
\subfloat[Task 1]{%
  \includegraphics[trim={0 0 0 1.5cm},clip,width=0.5\linewidth]{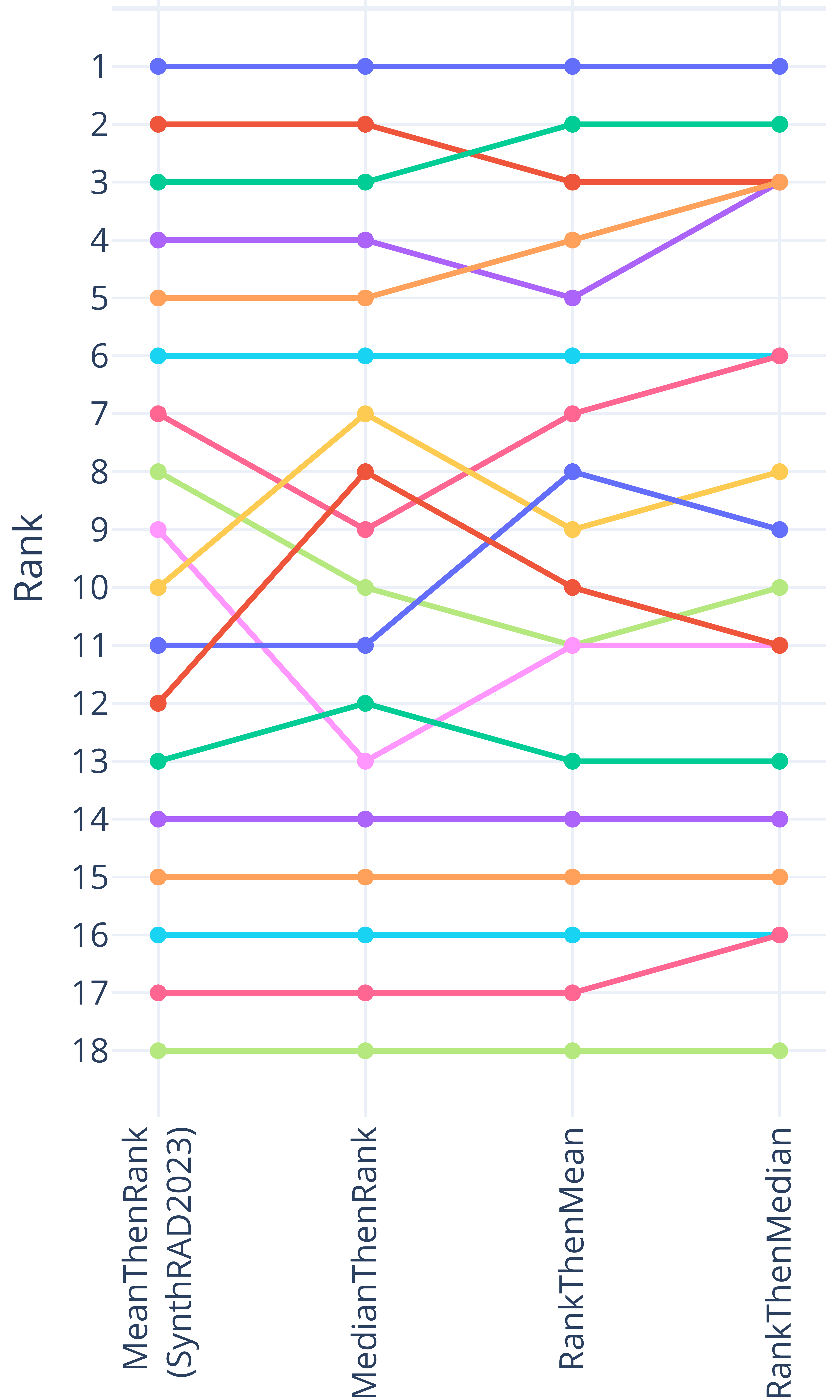}%
}
\subfloat[Task 2]{%
  \includegraphics[trim={0 0 0 1.5cm},clip,width=0.5\linewidth]{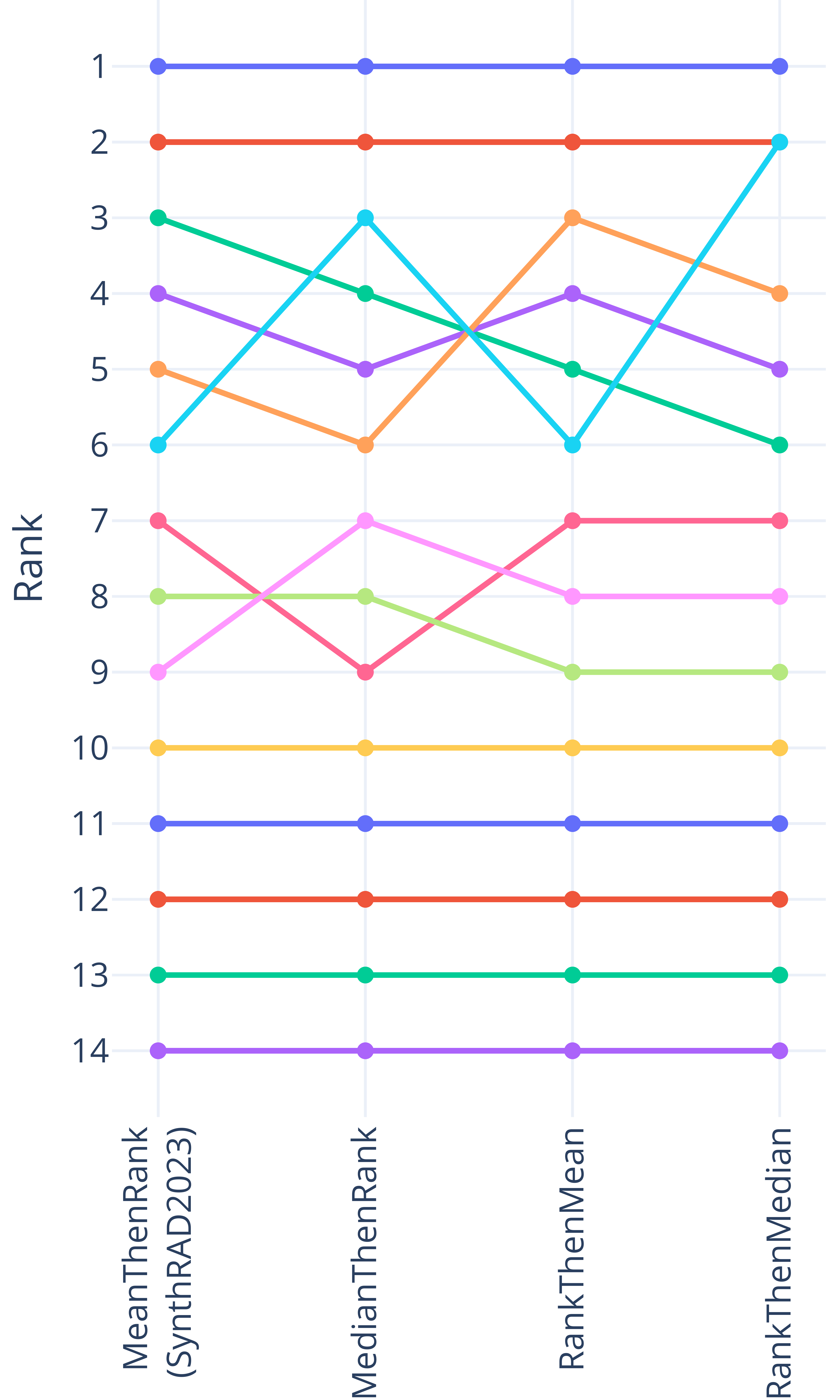}%
}
\caption{Stability of the chosen ranking approach (MeanThenRank) compared to the other three. For all approaches, we used the mean over all test patients to obtain one average value per metric per team.}
\label{fig:ranking stability different agg}
\end{figure}

\autoref{fig:ranking stability different agg} illustrates that the challenge winner also secured the top position for both tasks under the three other ranking approaches, and teams at the bottom of the rankings are also stable across the ranking approaches. However, the middle-ranked teams experience notable shifts. For task 1, transitioning from MeanThenRank to MedianThenRank caused substantial changes for UKA ($9 \rightarrow 13$) and mriG ($12 \rightarrow 8$). Conversely, task 2's largest shifts occurred when changing from MeanThenRank to RankThenMedian for FGZ Medical research ($6 \rightarrow 2$) and iu\_mia ($3 \rightarrow 6$). Despite these variations, all approaches strongly correlated with MeanThenRank approach, as indicated by Kendall's $\tau$ correlation coefficient (\autoref{tab:kendall ranking approaches}).

\begin{table}[h]
\small
\centering
\caption{Kendall's $\tau$ correlation coefficients for the ranking obtained from MeanThenRank compared to the other three ranking approaches.}
\label{tab:kendall ranking approaches}
\begin{tabular}{lcc}
\toprule
\textbf{Comparison} & \textbf{Task 1} & \textbf{Task 2} \\
\midrule
MeanThenRank vs MedianThenRank & 0.88 & 0.87 \\
MeanThenRank vs RankThenMean   & 0.88 & 0.91 \\
MeanThenRank vs RankThenMedian & 0.91 & 0.84 \\
\bottomrule
\end{tabular}
\end{table}

\begin{figure}
    \subfloat[Task 1]{%
    \includegraphics[width=\linewidth]{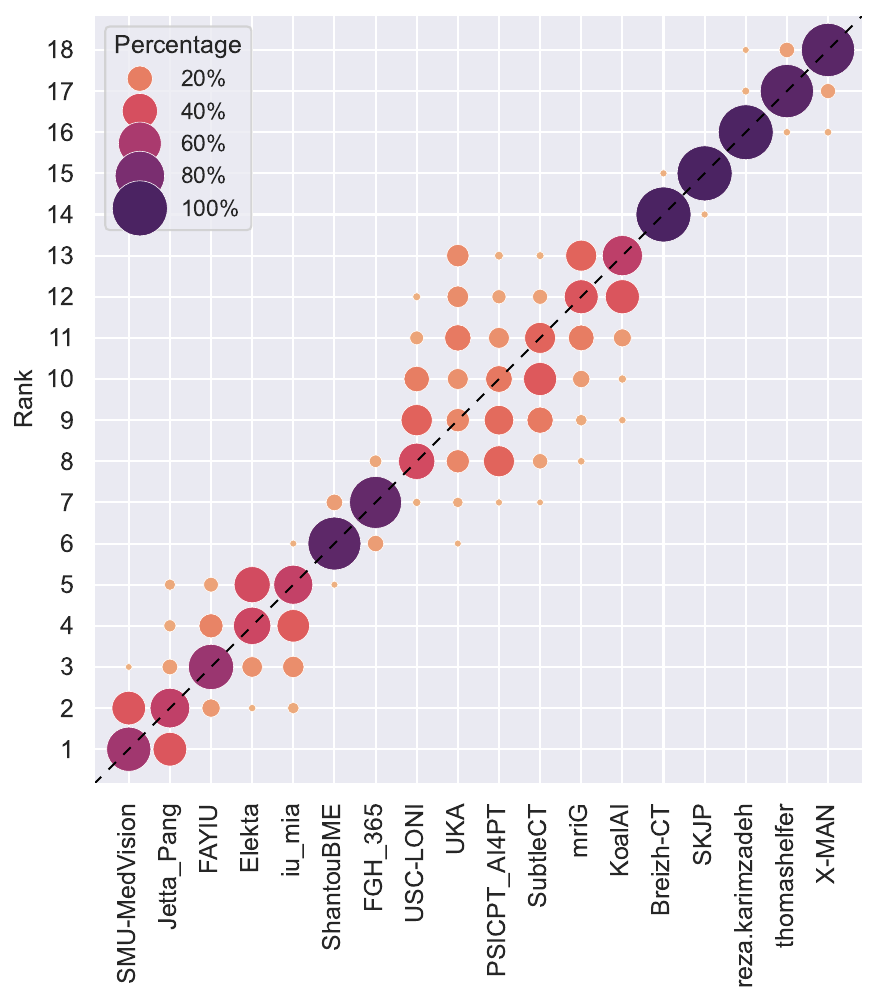}{}}
    \hfill
    \subfloat[Task 2]{%
    \includegraphics[width=\linewidth]{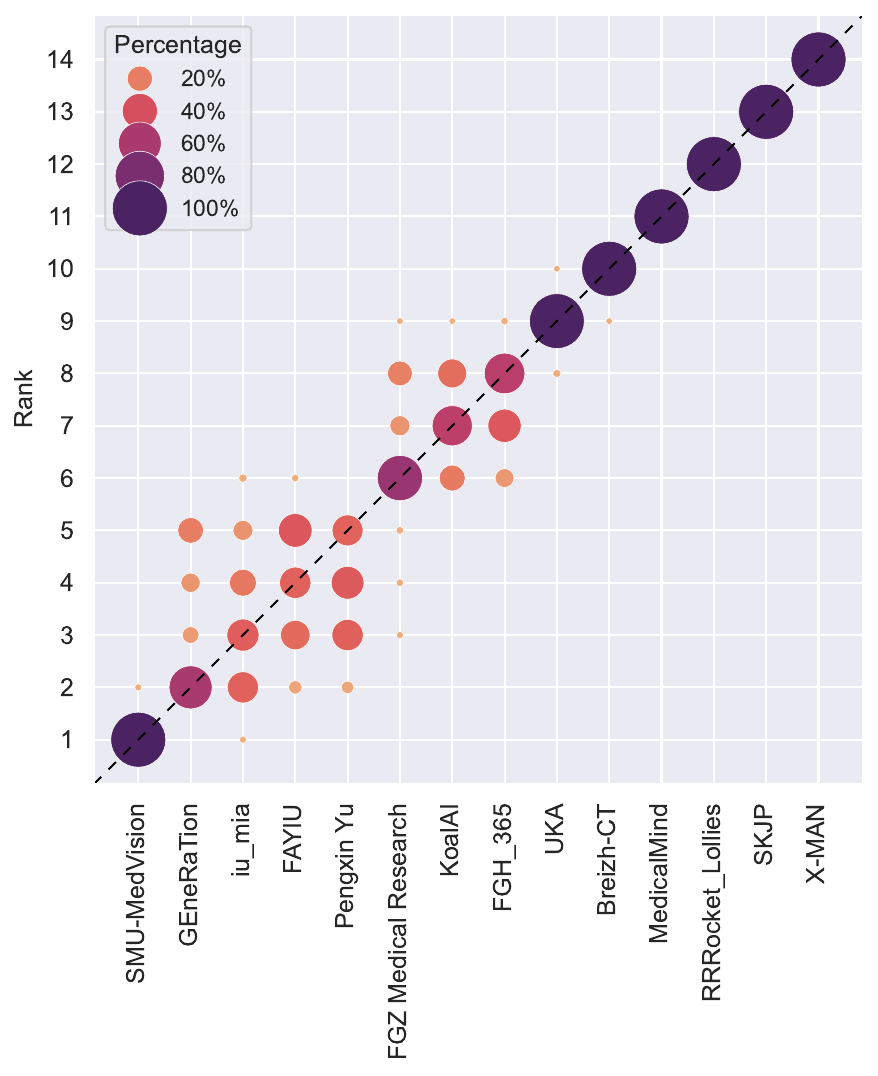}{}}
    \caption{Visualization of ranking stability. Blob size is proportional to the frequency of the rank achieved based on bootstrapping (N=1000).}
    \label{fig:ranking_stability_bootstrap}
\end{figure}

\autoref{fig:ranking_stability_bootstrap} demonstrates that the final rankings (determined by MeanThenRank) were relatively stable. There was high confidence in top-performing teams securing higher ranks and underperforming teams obtaining lower ranks, with the teams showing a maximum shift of 4 and 3 positions for tasks 1 and 2, respectively. SMU-MedVision had a 63.7\% certainty of being the winner for task 1 and 99.7\% for task 2, while certainties for the second to fifth places were lower, ranging from 45.0\% to 66.7\% for task 1 and from 30.5\% to 60.4\% for task 2. Teams in the middle of the rankings again showed some level of uncertainty, while it was inevitable that teams at the bottom of the ranking received the correct rank. This specifically holds for the last five teams for task 1 and the last six teams for task 2, with average certainties of $97.0 \pm 3.5\%$ and $99.8 \pm 0.4\%$, respectively.

\section{Discussion}
\label{sec:discussion}
SynthRAD2023 allowed the comparison of deep learning techniques for synthesizing CT from MRI or CBCT. It is the first large-scale, multi-center challenge for generating \textit{in-vivo} synthetic CT and garnered significant participation among the community, consisting of 617 participants, generating 39 valid submissions. The participants were generally able to synthesize high-quality sCT, outperforming the baseline algorithms in terms of image quality and dose accuracy.

The top five teams performed well, with SSIM values of at least 0.87 and 0.90 for tasks 1 and 2, respectively. Additionally, they exceeded gamma pass rates (2mm/2\%) of 98.07\% for photon and 97.25\% for proton treatment plans in task 1, and at least 98.99\% for photon and 97.00\% for proton treatment plans in task 2. These results indicate a high level of correspondence to the ground truth CTs. Nevertheless, despite the excellent performance, challenges remain for image synthesis. Difficulties for MRI-to-CT synthesis were encountered at air-tissue boundaries, potentially due to low MRI signal and magnetic susceptibility artifacts \citep{krupa2015artifacts}. Additionally, in our dataset, the limited field-of-view of CBCT compared to CT introduced challenges in accurately synthesizing the complete body contour in the sCT. 

Our analysis revealed that transformers \citep{transformer} outperform CNN encoder-decoder models (e.g., U-Net \citep{unet}), which in turn outperform GANs \citep{gan}. Notably, recent architectures like diffusion models \citep{diffusion} and transformer-GAN combinations performed worse than the architectures mentioned above. These findings contrast recent reviews considering sCT generation \citep{SpadeaMaspero2021,dayarathna2023deep}, which either found no correlation between model architecture and performance or suggested that diffusion models hold promise in this field. Despite the statistically significant performance differences observed in our challenge, the differences were marginal, and the sample size was limited. In addition to comparing model architectures, it is important to acknowledge the potential impact of variations in training methodologies, including the reliability of hyperparameter search, on the observed performance differences among different approaches. Therefore, whether the observed differences stem solely from architectural choices or are significantly influenced by other aspects of the complex end-to-end pipeline, including preprocessing, data augmentation, postprocessing, and training procedures, remains inconclusive. 
For instance, previous literature suggests that data augmentation generally benefits generative models, suggesting that this step may play an essential role in model performance \citep{Taylor2018,steiner2021train}.

The 2D models outperformed 2.5D and (patch-based) 3D models for MRI-to-CT synthesis, while the 3D models outperformed the 2(.5)D approaches for CBCT-to-CT synthesis. These results hold for both pelvis and brain cases in both tasks. However, it has been shown that for MRI-to-CT synthesis 2.5D (multi-view) models outperform 2D models \citep{Spadea2019,Maspero2020}, and that 3D models outperform 2D models \citep{Sun2022}. We did not identify the cause of these contrasting results between MRI-to-CT and CBCT-to-CT synthesis. Future work could investigate why the impact of spatial dimension differs between these imaging modalities for synthetic CT generation.

We found that the image similarity metrics are highly correlated among themselves (\(\vert\rho\vert \geq 0.88\)) and that the MAE of the photon and proton dose distribution are moderately correlated to their respective gamma pass rates (\(\vert\rho\vert \geq 0.66\)). Specifically, the photon and proton DVH metrics are weakly correlated with the respective gamma pass rates (\(\vert\rho\vert \leq 0.42\)) (\autoref{fig:spearman}). Furthermore, the average correlations between the image similarity metrics and dose metrics are low (\(\vert\rho\vert \leq 0.47\)) despite the similar goal of measuring correspondence between the sCT and ground truth CT. The difference in correlations observed within and between metric groups may be attributed to the distinct regions where each metric is measured: image similarity was assessed within the dilated body contour, while dose metrics were calculated within high-dose regions or specific organs. These findings suggest that image similarity metrics should not be solely relied upon to determine the clinical suitability of a model, as they are not a reliable surrogate for clinically relevant dose metrics. Previous literature aligns with the finding, corroborating the poor correlation between image similarity and dose accuracy~\citep{Kieselmann2018,Peng2020}. This highlights the need to perform thorough dose evaluations when clinically testing sCT generation approaches.

Two teams, i.e., UKA and PSCICP\_4AI4PT, scored unexpectedly low in the final test phase compared to the validation phase due to implementation errors or misinterpreting data details. After (re-)opening the post-challenge phases, the two teams submitted corrected versions of the algorithms. Based on image similarity metrics alone, UKA climbed seven positions (9 \(\rightarrow\) 2)  in the rankings for both tasks. Similarly, PSCICP\_4AI4PT climbed six positions (10 \(\rightarrow\) 4) for task 1. 
During the open test phase, the teams could not resubmit their algorithms, as they ran successfully on the platform. The low scores of the erroneous algorithms underscore the fairness of the adopted rules.

\subsection{Clinical impact}
Despite the similarity between the sCTs generated by the participants and the ground truth CTs, there remains a lack of consensus regarding the criteria determining the clinical acceptability of an sCT \citep{Vandewinckele2020}.
In radiotherapy, treatment planning is defined to meet specific dose prescriptions and constraints. In this sense, dose-related metrics may be considered clinically significant.
Some works have investigated clinical acceptance criteria for synthetic CTs. For example, \cite{https://doi.org/10.1002/mp.13716} considered photon gamma pass rates greater than 98\% acceptable using the 2mm/2\% criterion. On the other hand, \cite{Korsholm2014} proposes that treatments with a DVH difference of \(<2\%\) are clinically acceptable. However, these criteria were proposed for breast, head-and-neck, and thorax sCT generation; it is unclear whether these criteria translate to different anatomical regions. 
Before addressing the clinical impact of the challenge results, it is crucial to consider the quality of the treatment plans adopted for SynthRAD2023 evaluation. 
Treatment planning techniques may differ between institutes. The planning techniques chosen have been based on constraints adopted in clinical guidelines \citep{hall2021nrg,lambrecht2018radiation}, making the results of the challenge of clinical relevance. The Linac and proton systems used in the treatment planning were generic; however, studies have demonstrated their effectiveness by showing that gamma pass rates deviate by a maximum of 0.5\% when compared to dose engines adopted in clinical systems, independent of the irradiation type \citep{matrad}.

To indicate clinically acceptable sCT, we propose considering an average gamma pass rate (2 mm/2\%) above 99\% and 97.5\% for photon and proton irradiation in regions receiving at $\geq10\%$ of the prescribed dose, respectively. For the SynthRAD2023 challenge, only one team (Jetta\_Pang) met these criteria for the MRI-to-CT task. For the CBCT-to-CT task, one team (SMU-MedVision) met both criteria. In contrast, five other teams (GEneRaTion, FAYIU, Pengxin Yu, FGZ Medical Research, and Breizh-CT) met only the photon criterion. 
As previously mentioned, the evaluation was affected by differences in patient positioning between the imaging sessions.
Still, when considering the results on a population level, we do not expect to observe any systematic dose differences unless the sCT generation method introduced geometrically consistent distortion \citep{ADJEIWAAH2019}. The lack of systematic dose differences suggests that the solutions offered by the participants are promising and of high quality. Before implementing any proposed clinical solutions, evaluating them according to the clinical standards specific to each facility using the commissioned treatment planning system is advisable.

Currently, commercial solutions to generate sCT from MRI or CBCT are available \citep{mrcat, van2019bonemri, cronholm2020mriplanner, archambault2020ethos}. It would be interesting to compare the algorithms submitted to the SynthRAD2023 challenge with these commercial solutions. Some of these commercial solutions require a dedicated imaging protocol to generate accurate sCT data \citep{florkow2020deep,bratova2019validation, liu2023review}, making the comparison challenging. 
Exploring the necessity of specialized imaging protocols, or in simpler terms, assessing the ability of sCT algorithms to generalize across different input variations as in \cite{Nijskens2023}, could be worthwhile.

\subsection{Limitations of the SynthRAD2023 dataset and setup}
A substantial multi-center dataset was gathered for the SynthRAD2023 challenge. However, the dataset can be further improved despite its size and diversity. For example, the dataset consists solely of patients treated at Dutch hospitals, which may limit the dataset's heterogeneity, possibly resulting in low performance for case outliers in the data distributions. Additionally, it is important to acknowledge that the included MRIs represent only a subset of the magnetic field strengths commonly used in clinical practice (1.5T and 3T). This limits the generalizability of the findings to the broader spectrum of clinical MRI applications, where other field strengths are routinely employed.
While the inclusion of three centers represents a commendable starting point, extending the dataset with international data may improve the generalization capabilities of the submitted models and increase the clinical impact. 

A model ideally should generalize across different centers without conditional fine-tuning, as current commercial solutions are not center-specific. While some participants incorporated center-based prediction and optimization using information shared in the training and validation sets, effective models should extend beyond the provided centers to make a clinical impact. Future challenges may consider whether circumventing such information may lead to designing more general approaches.

Furthermore, the SynthRAD2023 dataset contained rigidly registered image pairs, resulting in residual anatomical mismatch after registration, as mentioned above. Reducing the registration error, e.g., recurring to deformable registration, may improve performances \citep{florkow2020deep}. However, it may also confound possible geometrical distortion to the input images the models may introduce, which is undesirable in a clinical scenario \citep{Pappas2017}. The impact of residual misregistration is corroborated by the paired nature of the dataset, and could be mitigated by performing unpaired synthetic CT generation. However, unpaired sCT generation prohibits a dosimetric evaluation of the generated sCT, and limits the use of established image similarity metric, such as the SSIM or PSNR.

An additional dataset limitation stems from the automated processing pipeline, where, for all brain patients from center B in task 1, the treatment table was included within the dilated body contour. At the same time, the table was successfully excluded from the dataset by the other centers, leading to inconsistent table representation in the dose evaluations. Moreover, for two out of sixty pelvis patients in task 2, the field-of-view of the CBCT was smaller than the body contour. Such patients can be considered outliers, and for future challenges, it would be beneficial to revise case selection and exclude them from the test set. The inclusion of the table in the mask and limited CBCT field-of-view had minor impact on the image similarity evaluation, which was computed within the provided mask, but could be more substantial for the dose evaluation due to beam attenuation. Note that the inconsistency was present for all teams, leaving the challenge ranking unbiased.

Another limitation arose from the absence of dose evaluation during the validation phase, hindering teams from optimizing their models for this radiotherapy-related metric. On the other hand, the lack of dose metrics during validation may have compelled participants to develop general methods that could function irrespective of the chosen planning strategy.

\subsection{Future direction}
SynthRAD2023 has set out to advance the state-of-the-art in MRI-to-CT and CBCT-to-CT generation. While the results are promising, these tasks have not yet been solved during this challenge. 
The dataset only included brain and pelvis patients. Other, maybe more challenging, anatomical regions could benefit from sCT generation, such as the thorax, head-and-neck, breast, or abdomen \citep{SpadeaMaspero2021}. In addition, it would be of interest to examine the generalizability of the models by including test data from centers that were not present in the training data \citep{texier2023computed}.

The positive reception to SynthRAD2023 has spurred the development of SynthRAD2025, which aims to expand the challenge beyond the Dutch national domain into more unexplored anatomical regions, such as the head-and-neck and abdomen.

Furthermore, addressing the limitations in data preparation and image registration discussed earlier will enhance the analysis of future challenges.

Lastly, we anticipate that the post-challenge phases will offer opportunities to validate and enhance the statistical robustness of the challenge's conclusions, enabling other researchers to compare their methods with the results of SynthRAD2023.

\section{Conclusion}
\label{sec:conclusion}

While synthetic CT generation has already become a clinical reality \citep{SpadeaMaspero2021}, the SynthRAD2023 Grand Challenge represents a pivotal advancement in image synthesis for radiotherapy planning. The challenge marks the first multi-center challenge with a substantial dataset, serving as a catalyst for further innovation in radiotherapy. Participants showcased their ability to generate high-quality sCTs, demonstrating high image similarity and accurate dose distributions. These achievements highlight the potential of deep learning for enhancing sCT generation. However, it's important to recognize that solely relying on image similarity metrics may not adequately capture the clinical applicability of sCTs. Nonetheless, these significant strides hold promise for reducing reliance on conventional CT and improving efficiency in radiotherapy.

\section*{Authors contribution}
\subsection*{Organizers}
\textbf{Evi Huijben:} Conceptualization, Formal Analysis, Investigation, Methodology, Software, Validation, Visualization, Writing – original draft, Writing – review \& editing.
\textbf{Maarten Terpstra:} Conceptualization, Formal Analysis, Investigation, Methodology, Software, Validation, Visualization, Writing – original draft,  Writing – review \& editing.
\textbf{Arthur Jr. Galapon:} Conceptualization, Data curation, Formal Analysis, Investigation, Resources, Visualization, Writing – review \& editing.
\textbf{Suraj Pai:} Conceptualization,  Formal Analysis, Investigation, Methodology,  Project administration, Software, Visualization, Writing – review \& editing.
\textbf{Adrian Thummerer:} Conceptualization, Data curation, Methodology, Software, Resources, Writing – review \& editing.
\textbf{Peter Koopmans:} Conceptualization, Methodology, Writing – review \& editing.
\textbf{Manya Afonso:} Conceptualization, Funding acquisition, Writing – review \& editing.
\textbf{Maureen van Eijnatten:} Conceptualization, Funding acquisition, Writing – review \& editing.
\textbf{Oliver Gurney-Champion:} Conceptualization, Writing – review \& editing.
\textbf{Zoltan Perko:} Conceptualization, Data curation, Software, Writing - review \& editing, Validation.
\textbf{Matteo Maspero:} Conceptualization, Investigation, Data curation, Software, Methodology, Validation, Funding acquisition, Project administration, Resources, Supervision, Writing – original draft, Writing – review \& editing.

\subsection*{Participants}
All the participants: Methodology, Software, Writing – review \& editing.\\
\textbf{SMU-MedVision} (Zeli Chen, Yiwen Zhang, Kaiyi Zheng, Chuanpu Li),
\textbf{Jetta\_Pang} (Haowen Pang, Chuyang Ye),
\textbf{GEneRaTion} (Runqi Wang, Tao Song),
\textbf{FAYIU} (Fuxin Fan, Jingna Qiu, Yixing Huang),
\textbf{iu\_mia} (Juhyung Ha, Jong Sung Park),
\textbf{Elekta} (Alexandra Alain-Beaudoin, Silvain B\'eriault),
\textbf{Pengxin Yu},
\textbf{ShantouBME} (Hongbin Guo, Zhanyao Huang),
\textbf{FGZ Medical Research} (Gengwan Li, Xueru Zhang),
\textbf{FGH\_365} (Yubo Fan, Han Liu),
\textbf{KoalAI} (Bowen Xin, Aaron Nicolson),
\textbf{USC-LONI} (Lujia Zhong, Zhiwei Deng),
\textbf{UKA} (Gustav M{\"u}ller-Franzes, Firas Khader),
\textbf{PSICPT\_AI4PT} (Xia Li, Ye Zhang),
\textbf{Breizh-CT} (C\'edric H\'emon, Valentin Boussot),
\textbf{SubtleCT} (Zhihao Zhang, Long Wang),
\textbf{MedicalMind} (Lu Bai, Shaobin Wang),
\textbf{mriG} (Derk Mus, Bram Kooiman),
\textbf{RRRocket\_Lollies} (Chelsea Sargeant, Edward Henderson),
\textbf{SKJP} (Satoshi Kondo, Satoshi Kasai),
\textbf{reza.karimzadeh} (Reza Karimzadeh, Bulat Ibragimov),
\textbf{thomashelfer} (Thomas Helfer, Jessica Dafflon),
\textbf{X-MAN} (Zijie Chen, Enpei Wang).

\section*{Acknowledgments}
The organizers would like to thank Eric van der Bijl for leading the data collection at Radboud UMC Nijmegen, Cornelis (Nico) AT van den Berg for supporting the challenge offering facilities and hosting the administrative account within UMC Utrecht, Joost JC Verhoeff for supporting the data retrieval at UMC Utrecht, Stefan Both supporting the data retrieval at UMC Groningen.
We thank the European Society of Radiation Oncology (ESTRO), \url{https://www.estro.org/} and the Nederlandse Vereniging voor Klinische Fysica (NVKF) \url{https://www.nvkf.nl/} for endorsing the event.

\section*{Funding}
SynthRAD2023 was funded by the TU/e, WUR, UU, UMC Utrecht (EWUU) alliance (\url{https://ewuu.nl}).

\section*{Declaration of generative AI}
While preparing this work, the authors used Writefull, DeepL Write, and ChatGPT to enhance the writing structure and refine grammar. After using these tools, the authors reviewed and edited the content as needed and took full responsibility for the publication's content.

\section*{Appendix: Participation rules and prize policies}
To ensure fairness and transparency in SynthRAD2023, organizers, data providers, and contributors were prohibited from participating in the challenge since data providers and organizers had access to the data, including the test set ground truth CTs. However, members affiliated with the organizers' institutes were allowed to participate, provided they had not co-authored any publications with the organizers in the year preceding the challenge.

Participants were required to develop fully automated methods that run in the Amazon Web Services (AWS) cloud environment using a single \texttt{g4dn.2xlarge} instance. This instance includes a GPU with 16 GB VRAM, an 8-core CPU, and 32 GB RAM. In this environment, the inference time for generating an sCT for a single case (one patient) is constrained to a maximum of 15 minutes.

Teams receiving a prize had to present their methodology at MICCAI 2023, sign all necessary prize acceptance documents, and submit a detailed paper in LNCS format outlining their methods. Additionally, participants committed to citing both the data challenge paper \citep{Thummerer2023synth} and this challenge overview paper in subsequent publications, whether scientific or non-scientific. Although sharing codes was strongly encouraged, it was not mandatory. The challenge results and rankings were publicly announced after the test phase concluded. The top five teams for both tasks were awarded a total of €10,000, with the following distribution: €2200, €1250, €850, €500, €200.

The complete challenge design can be found at \citep{SynhtRADChallengeDes}.

\bibliographystyle{model2-names.bst}\biboptions{authoryear}
\bibliography{refs}


\newpage
\onecolumn
\setcounter{section}{1}
\section*{Supplementary Document A: Participant methods}
Each subsection briefly describes the methods used by the participating teams. Top five methods were presented in the main paper. The team names correspond to the submission reported on the leaderboard at \url{https://synthrad2023.grand-challenge.org/evaluation/test/leaderboard}.
\subsection{ShantouBME (task 1)}
ShantouBME employed a 2D U-net \citep{unet} for task 1, incorporating an additional convolutional layer in the bottleneck. The model was trained using L1 loss, with separate models for the brain and pelvis. Magnetic resonance (MR) images were normalized at the patient level, while computed tomography (CT) images were normalized using the fixed range of $[-1024, 3000]$ Hounsfield units (HU). Random patches of $224\times224$ pixels were sampled to augment the training data. The model processed the full-size 2D slices in one go during testing, and the normalization procedure was reverted. The models were trained for 300 epochs using the Adam optimizer with a learning rate of $3\mathrm{e}-4$ and a step descent learning rate scheduler. The epoch with the lowest validation loss was selected for inference.

\subsection{FGZ Medical Research (task 2)}
FGZ Medical Research employed one collective 2D denoising diffusion probabilistic model (DDPM) \citep{diffusion} with a U-Net architecture \citep{unet} for brain and pelvis data in task 2. The DDPM was conditioned on the cone beam CT (CBCT) image and trained using the mean squared error (MSE) loss. Training used 500 diffusion steps, while inference used 15 diffusion steps, as implemented in denoising diffusion implicit models (DDIM) \citep{song2020denoising}. Preprocessing involved resizing slices to $256 \times 256$ pixels, clipping intensities to $[-1024,2000]$ HU, and normalizing to $[-1, 1]$. No data augmentation was applied. The normalization and resizing steps were reversed to produce a synthetic CT (sCT) in HU. The epoch with the lowest validation mean absolute error (MAE) was selected for inference. The model was trained for 200 epochs using AdamW optimizer and a learning rate of $1\mathrm{e}-4$ with a warm-up scheduler.

\subsection{FGH\_365 (task 1 \& 2)}
Participating in both tasks, team FGH\_365 employed a modality-, anatomy- , and site- (MAS)-specific strategy to synthesize sCT images across multiple modalities (tasks 1 and 2), anatomical regions (brain and pelvis), and sites (centers A, B, and C).
Their approach consisted of two MAS-specific solutions. For solution \#1, separate 3D pix2pix models \citep{pix2pix2017} were trained for each of the 11 MAS configurations present in the dataset. Solution \#2 consisted of one unified model conditioned on the MAS and was trained on the collective datasets of both tasks. The latter was based on the 3D pix2pix model but incorporated MAS-conditioned dynamic convolution layers \citep{zhang2021dodnet,liu2022moddrop++} at the first two and last two layers.
For all models, the generator loss included the L1 loss, the adversarial loss \citep{pix2pix2017}, and an edge-aware loss \citep{luo2021edge,fan2023temporal}, and the discriminator loss was the binary cross entropy (BCE) loss. 
MRI intensities were clipped to the 99.5\textsuperscript{th} percentile, and CT images were clipped to $[-1000, 3000]$ HU. All modalities were linearly normalized to $[-1,1]$ at the patient level. Data augmentation consisted of random patch selection, random affine transformations, flipping, noise addition, and random contrast adjustment. 
The models of solution \#1 were trained on 3D patches of $128\times128\times128$ voxels for brain and $256\times128\times64$ voxels for pelvis, while the model of solution \#2 was trained on 3D patches of $192\times192\times128$ voxels.
During inference, patches overlapping by 40\% (for volumes $\geq 500\times280\times140$ voxels) or 60\% (for volumes $< 500\times280\times140$ voxels)  were selected using a sliding window and overlapping regions were averaged to create the full-size sCT, which was then linearly rescaled from $[-1,1]$ to $[-1000,3000]$ HU. FGH\_365 proposed an uncertainty-based site prediction algorithm since the acquisition center was unavailable for the test data. This algorithm considered the MAE between the sCTs obtained from solutions \#1 and \#2 and assumed the correct center to have the lowest MAE.
Solutions \#1 and \#2 were trained for 5000 and 800 epochs, respectively, and both solutions used the AdamW optimizer with a constant learning rate of $2\mathrm{e}-4$. The final models were selected based on the epoch with the lowest validation MAE, and the final output was the average of the sCT predictions from the two MAS-specific solutions.

\subsection{KoalAI (task 1 \& 2)}
Team KoalAI used a locally enhanced (LE) generative adversarial network (GAN), training four models for each subtask. The models consisted of a 3D patched-based generator and a mixture of 3D and 2D patch discriminators \citep{pix2pix2017}. The generator loss consisted of the L1 loss with a weight of 100 and an adversarial loss (MSE) with a weight of 1. The team included different model architectures for the generator, including ResNet \citep{johnson2016perceptual}, UNet \citep{unet}, and DynUNet \citep{isensee2019automated}. Two architectures were considered for the discriminator, including a patch discriminator (PatchD) \citep{pix2pix2017} and a LE discriminator (LED) combining a 3D and a 2D PatchD. Specifically, task 1 pelvis used the ResNet generator and LED, and took inputs of $256\times 256\times 56$ voxels. The other three subtaks used an ensemble of three model architectures, presented as `generator \& discriminator' in the following. Task 1 brain considered 1) ResNet \& LED with an input size of $256\times256\times56$ voxels, 2) ResNet \& PatchD with an input size of $256\times56\times256$ voxels, and 3) ResNet \& LED with an input size of $56\times256\times256$ voxels.  Task 2 pelvis used 1) DynUNet \& PatchD with an input size of  $128\times128\times128$ voxels, 2) ResNet \& LED  with an input size of $256\times256\times56$ voxels, and 3) UNet \& PatchD with an input size of  $448\times448\times64$ voxels. Lastly, task 2 brain used ResNet \& LED with three different input sizes of $256\times256\times56$, 2) $256\times56\times256$, and 3) $56\times256\times256$ voxels. 
MRI data were preprocessed by histogram matching with a random MRI sample, N4 bias field correction, smoothing with a gradient anisotropic diffusion filter and applying the provided body mask. In addition, the arms on the pelvic MRI were removed using a 2D-connected component algorithm. CBCT data were preprocessed with lower-bound intensity scaling (0 to -1024), applying the provided body mask, and clipping intensities to $[-1000, 3000]$ HU. A thresholding algorithm was also applied, followed by denoising using a 2D connected component algorithm to remove bright spots surrounding the body in the pelvis CBCT data. MRI volumes were normalized to $[-1, 1]$ at the patient level, while (CB)CT inputs were normalized to $[-1, 1]$ using the fixed range $[-1024, 3000]$ HU. Data augmentation for both tasks included random patch selection, affine and elastic deformations, random intensity shifts, random contrast adjustments, and random histogram shifts. The models were trained using the Adam optimizer and learning rate $2\mathrm{e}-4$. The training was stopped when the validation MAE did not improve for 100 epochs, and the final model was selected based on the best validation MAE. At test time, output volumes were generated from patches with a 25\% overlap and averaged using equal weighting. The normalization process was inverted.

\subsection{USC-LONI (task 1)}
USC-LONI participated in task 1 and employed a 2.5D diffusion model \citep{diffusion} followed by two 2D U-Nets \citep{unet} acting as refinement networks. The diffusion model, which considered multiple axial slices, was trained using the L1 loss and a consistency loss assessing the difference between adjacent slices in the 2.5D data. One refinement network considered axial slices, while the other considered frontal slices, and both were trained using the L1 loss. CT data were normalized to $[-1,1]$ using a fixed range of $[-1024, 3000]$ HU, and MRI data were normalized at the patient level to $[0,1]$. The image slices were resized to $256\times256$ for the diffusion model, which considered 3 consecutive slices. The original size was used for the 2D refinement networks. No data augmentation was applied. During inference, overlapping 2.5D inputs were selected with a stride of 1, and DDIM sampling \citep{song2020denoising} with 20 steps was used for the diffusion model. Overlapping slices were averaged, resizing and normalization steps were reversed, and the two refinement networks were applied to the 2D axial and frontal slices. The diffusion model was trained for 140,000 iterations with a batch size of 16, using a learning rate of $2\mathrm{e}{-4}$ for the first 100,000 iterations and $5\mathrm{e}{-5}$ for the last 40,000 iterations.  The refinement U-nets were trained for 20,000 iterations with batch size 16 and learning rate $2\mathrm{e}{-4}$. All models were trained with the Adam optimizer.

\subsection{UKA (task 1 \& 2)}
UKA used a multiplanar approach consisting of three identical 2D U-Net models \citep{unet} without skip connections. Separate models were trained for each task and anatomical region using 2D axial, sagittal, or coronal slices. The loss function used a masked average of L1 loss and structural similarity index measure (SSIM) calculated at full and half resolution.
MRI intensities within the body mask were clipped to the 99\textsuperscript{th} percentile and normalized to $[-1,1]$ at the patient level. CBCT intensities were first adjusted to be non-negative and CBCT and CT images were clipped to $[0,3000]$ and $[-1024,3000]$ HU, respectively. The (CB)CTs were normalized to $[-1,1]$ using these fixed ranges.
To ensure that the input size was a multiple of 8, zero padding was applied, and data augmentation consisted of random flipping. At test time, horizontal and vertical flipping were applied to each input slice, and the predictions were flipped back and averaged with the unaugmented prediction. Finally, the multi-plane 2D predictions were averaged, and the intensities were rescaled to $[-1024, 3000]$ to produce the final 3D sCT. 
The models were trained using the AdamW optimizer with a $1\mathrm{e}-4$ learning rate. An early stopping criterion was employed to terminate the training process when the validation loss did not decrease for five epochs. The epoch with the lowest validation loss was selected for inference.

\subsection{PSICPT\_AI4PT (task 1)}
Team PSICPT\_AI4PT participated in task 1 for which they employed a 2.5D patch-based nnU-Net \citep{isensee2021nnu} for both regions separately. The models were trained using the L1 loss. MRI inputs were linearly normalized at the patient level, while CT inputs were normalized using the fixed range of $[-1000, 2000]$. For training, 64 sampling planes of $4\times128\times128$ voxels were randomly sampled for each patient by applying rotation, flipping, and random clipping. At inference time, patches overlapping by $2\times64\times64$ voxels were selected, and overlapping regions were averaged. Furthermore, the CT normalization procedure was reverted to result in an sCT in HU. The models were trained for 200 epochs using the Adam optimizer. A warm-up and cosine scheduler adjusted the learning rate to $2\mathrm{e}-4$. 

\subsection{Breizh-CT (task 1 \& 2)}
BreizhCT participated in both tasks using a pix2pix model \citep{pix2pix2017} with a six-block residual network \citep{he2016deep} generator and a patchGAN discriminator \citep{pix2pix2017}. Tasks 1 and 2 utilized a 2D and 3D patch-based model, respectively, and two separate models were trained for the brain and pelvic regions. The loss function combined the cGAN loss \citep{mirza2014conditional} and a custom perceptual loss \citep{johnson2016perceptual} using the ConvNext-tiny architecture \citep{Liu_2022_CVPR} pre-trained on ImageNet. The perceptual loss consisted of a style term and a content term, and for both tasks, the style term was computed by comparing sCT and CT; for task 2, the content term was also computed by comparing sCT and CBCT. The perceptual loss leveraged paired training data but without direct voxel-wise supervision between sCT and CT to avoid registration inaccuracies.
Preprocessing involved adjusting CBCT intensities using histogram matching with the ground CT during training and the CT from the training set with the highest mutual information during testing. (CB)CT intensities were clipped to $[-1024, 3000]$ HU and divided by 1000, and MRI intensities were clipped to [0, 2000] and divided by 1000. No data augmentation was applied. The model input sizes for task 1 were 2D patches of $224\times224$ pixels for the pelvis and $168\times168$ pixels for the brain, and the input sizes for task 2 were 3D patches of $32\times224\times224$ voxels for the pelvis and $66\times168\times168$ for the brain. For training and testing, patches were selected with a stride equal half the patch. Postprocessing involved taking the median of overlapping regions, reverting preprocessing steps, and clipping to $[-1024,3000]$. The models were trained for 200 epochs using the AdamW optimizer with a constant learning rate of $1\mathrm{e}-4$. The epoch with the best validation MAE was selected for inference.

\subsection{SubtleCT (task 1)}
SubtleCT participated in Task 1, using a custom 2.5D U-Net \citep{unet} with residual blocks \citep{he2016deep} replacing the convolutional blocks. They adopted a two-stage approach where the model was first trained with an enhanced CT (eCT) and then with the ground truth CT. The model used the L1 loss in combination with the SSIM loss, and two identical models were trained, one for the brain and one for the pelvis. 
The eCT was created by setting the window width/level of the CT to 1000/350 HU. MRIs were normalized using min-max normalization at the patient level, while CT and eCT were normalized using fixed ranges of $[-1024,3000]$ and $[-150,850]$ HU, respectively. No data augmentation was applied. Input for the model consisted of five adjacent slices, which were padded to a size of $288\times 288$ voxels for the brain and $512\times512$ voxels for the pelvis. During inference, the normalization and resizing procedures were reversed.
Both models were trained for 200 epochs, with the first 100 epochs devoted to the first stage, and the following 100 epochs to the second stage. The Adam optimizer was used with an initial learning rate of $1\mathrm{e}-4$ and an adaptive learning rate decay scheduler. The epoch with the best validation peak signal-to-noise ratio (PSNR) was selected for inference.

\subsection{MedicalMind (task 2)}
MedicalMind implemented two 2D models inspired by multi-scale gradients (MSG)-GAN \citep{karnewar2020msg} for task 2: Model-Brain and Model-Pelvis. The generator was a U-Net \citep{unet} with a ResNet-50 encoder \citep{he2016deep} and a decoder that included an AdaIn block \citep{karras2019style} before consecutive convolutions. The generator predicted an sCT at five different resolutions, and the loss considered MAE and MSE for each resolution. The discriminator was similar to a VGG network \citep{simonyan2014very} but considered all five resolutions as input, concatenated the low-resolution inputs with the down-scaled large-resolution features, and used the BCE loss. Preprocessing involved resizing 2D axial slices to $512\times 512$ voxels for both brain and pelvis, clipping CT intensities to [-1000, 2048] HU, and no normalization was applied. During training, random rotation, scale, shift, and flip operations were used to augment the data, and at test time, the sCT with the largest resolution was used and resized back to the original input size. Intensities outside the body mask were set to -1000 HU, and no other postprocessing steps were applied. Both models were trained for 106 epochs using the Adam optimizer with an initial learning rate of $6\mathrm{e}{-4}$ and a StepLR scheduler.

\subsection{mriG (task 1)}
Team mriG employed a patch-based 3D U-Net \citep{unet} for task 1 for both regions separately. A combined L1 and SSIM loss was used for training. For preprocessing, a combination of rigid (for bones) and deformable (for soft tissue) image registration techniques \citep{klein2009elastix} was used. Registration for the pelvis cases was guided by bone segmentations \citep{kuiper2021ct}, with the individual bones segmented using the method outlined by \cite{liu2021deep}. MRI inputs were normalized to $[-1, 1]$ on a case basis using the minimum value and the 99\textsuperscript{th} percentile, while CT inputs were first made non-negative and then divided by 3000. The data were also reordered to the canonical orientation. Data augmentation included spatial (random zoom and rotation) and intensity-based (random contrast adjustments) augmentations. Additionally, patches of $96\times96\times64$ voxels were sampled randomly during training. During inference, a sliding window approach was used with half-overlapping patches in each dimension, and Gaussian weighting was applied to the edges. Furthermore, the CT normalization procedure was reverted to result in an sCT in HU. The models were trained with the AdamW optimizer for 100,000 iterations with a batch size 12. The learning rate started at $1\mathrm{e}-4$ and ended at $1\mathrm{e}-5$.

\subsection{RRRocket\_Lollies (task 2)}
RRRocket\_Lollies employed a multi-channel 2D cycleGAN \citep{CycleGAN2017} with an auxiliary fusion network for task 2. The discriminator networks were as described by \cite{CycleGAN2017} with BCE loss; however, the generator architectures were modified to include U-Net-like long-range skip connections \citep{unet} between corresponding down- and up-sampling convolutional levels to preserve contextual information. Also, attention gates were added to the skip connections to emphasize salient features propagated forward from earlier in the network \cite{Schlemper2019}. An auxiliary fusion network was added onto the cycleGAN to assist in multi-channel recombination. This had an identical architecture to the generators but contained only a single residual block and short-range residual connections across convolutional layers. MSE loss was used for both the generators and fusion networks. Individual models with identical architectures were trained for each anatomical region.
Preprocessing involved resizing ($n_{\text{slices}}\times448\times 448 $ voxels for the pelvis and $n_{\text{slices}}\times304\times304$ for the brain), clipping, outlier correction to correct high-intensity voxels on the surface of the patient and multi-channel normalization. The CT and CBCT scans were normalized into three channels using windowing to enhance the contrast of anatomical structures. The full width of the image range $[-1024, 3000]$ HU was captured in channel one. In channel two, a contrast setting was used to view soft tissue structures; for the CT, this was $[-150, 150]$ HU and $[-100, 100]$ HU for the pelvis and brain, respectively. An automated peak finder was implemented to set the level to the CBCT soft tissue peak, with a fixed window width of  150 HU or $\pm$100 HU. In the final channel, the CT and CBCT images were clipped to $[600, 3000]$ HU to capture information about the high-density structures. Using min-max normalization, each channel was independently scaled to $[0,1]$. 
Postprocessing included reversal of preprocessing steps and multi-channel combination. To recombine the channels, the full-width image (channel one sCT for the pelvis, fusion network output sCT for the brain) underwent modifications based on specific conditions: values within the narrow range ( $-150$ to $150$ HU) were substituted with channel two values, and values $> 600$ HU were replaced with channel three values.
No data augmentation was applied during training. The models were optimized using the Adam optimizer and initial learning rates of $1\mathrm{e}-4$ and $2\mathrm{e}-4$ for the generators and discriminators, respectively. After 5 epochs, the learning rate was reduced to 80$\%$ of the learning rate every 2 epochs for both generator and discriminator. The models were trained for a maximum of 200 epochs; however, early stopping was applied when total generator loss did not improve for 20 epochs. The optimal model is chosen based on image similarity metrics (MAE, PSNR, and SSIM) calculated on train-time validation data.

\subsection{SKJP (task 1 \& 2)}
For both tasks, SKJP employed a 2.5D U-Net \citep{unet} as the basis of the synthesis network and replaced its encoder by EfficientNet-B7 \citep{tan2019efficientnet} with multi-slice inputs and single-slice outputs. The same architecture was used for both brain and pelvis data, but separate models were trained using the L1 loss. In task 1, data preprocessing involved histogram normalization and linear scaling of MRI intensities, while in task 2, linear scaling of CBCT intensities was applied. Furthermore, axial slices were cropped or padded to $320 \times 320$ voxels for the brain and $480 \times 640$ voxels for the pelvis, with the model considering three consecutive slices as input. No data augmentation was performed. During training, 32 2.5D input volumes were randomly sampled from each 3D volume. The initial learning rates were set to $1\mathrm{e}{-3}$, $5\mathrm{e}{-4}$, $1\mathrm{e}{-4}$, $5\mathrm{e}{-5}$ for task 1 brain, task 1 pelvis, task 2 brain, and task 2 pelvis, respectively. The model was trained for 100 epochs using AdamW optimizer, and the learning rate was decreased at every epoch with cosine annealing. The epoch with the lowest validation loss was used as the final model.

\subsection{Reza Karimzadeh (task 1)}
Reza Karimzadeh participated in task 1, employing a 3D patch-based pix2pix \citep{pix2pix2017} with a Swin UNETR \citep{hatamizadeh2021swin} generator. Two identical models were trained for the brain and pelvis. The training loss consisted of the L1 loss weighted by the ground truth CT, an SSIM loss, and an adversarial loss. Preprocessing involved linear normalization to $[-1, 1]$ at the patient level for MRI data and at the population level for CT data. During training, random patches of $64\times64\times64$ voxels were sampled and augmented using random rotations. During inference, patches with $50\%$ overlap were selected, and the final result was obtained by averaging predictions from overlapping regions. During postprocessing, the normalization procedure was reversed to obtain sCT outputs in HU. The model was trained for 1000 epochs using the AdamW optimizer and the fixed learning rate of $1\mathrm{e}-5$. The epoch with the best validation loss was used as the final model for inference.

\subsection{thomashelfer (task 1)}
Team thomashelfer only participated in task 1, where they used different models for brain and pelvis. For the brain, they employed a 3D latent diffusion model (LDM) \citep{esser2021taming, diffusion}, combining an autoencoder consisting of a U-Net \citep{unet} generator with a d diffusion model trained on the latent space of the autoencoder. In addition, ControlNet \citep{zhang2023adding} was used to ensure the generation of CT images conditioned on the MRI images. The LDM and ControlNet were implemented using MONAI generative models \citep{pinaya2023generative}. The autoencoder was trained with a combination of L1 loss, perceptual loss \citep{zhang2018unreasonable}, a patch-based adversarial objective \citep{rombach2022high}, and a KL regularization of the latent space. The diffusion model and ControlNet were trained using the losses suggested by \cite{pinaya2023generative}. 
The input data was center-cropped to $192\times192\times192$ voxels. MRI were normalized by dividing by 3000, and CT images by subtracting $-1024$ and then by 4024. No data augmentation was applied. Input images were encoded into a latent space of $48 \times 48 \times 48$ voxels. At test time, sCT intensities were rescaled to the original range, and no other postprocessing was applied.
The autoencoder was trained for 1000 epochs using Adam optimizer with learning rates of $5\mathrm{e}-5$ and $1\mathrm{e}-4$ for the generator and discriminator, respectively. The diffusion model was trained for 1000 epochs using AdamW optimizer with a learning rate of $2.5\mathrm{e}-5$, and the ControlNet was trained for 500 epochs using Adam optimizer with a learning rate of $2.5\mathrm{e}-5$. The epochs with the best validation loss were used at test time.

For the pelvis, the team employed a 3D (patch-based) pix2pix \citep{pix2pix2017} with a U-Net \citep{unet} generator. The generator loss consisted of the L1 loss with a weight of 100 and the adversarial loss (BCE) with a weight of 1. The pelvis data underwent normalization like the brain data, and no data augmentation was applied. The model processed half of the 3D input volume at a time, allowing for varying input sizes, and combined the two halves without overlap. Intensities were rescaled to the original intensity range to produce the final sCT output. The model was trained for 400 epochs using Adam optimizer with a learning rate of $1\mathrm{e}-3$ for both the generator and discriminator. The last epoch was used as the final model for inference.

\subsection{X-MAN (task 1 \& 2)}
X-MAN used a 3D patch-based cGAN \citep{liu2022synthetic} with a nine-block ResNet12 \citep{he2016deep} generator and a PatchGAN \citep{pix2pix2017} discriminator for both tasks. One collective model was trained for each task, combining brain and pelvis patients, using L1 loss and adversarial loss. MRI and CBCT intensities were normalized linearly at the patient level, while CT intensities were not normalized. The resulting sCT intensities were scaled from $[-1,1]$ to $[-2000, 2000]$ HU before calculating the loss. During training, random patches of $160\times160\times32$ voxels were selected. Data augmentation included random intensity shifts between $-10\%$ and $10\%$ and random gamma adjustments with gamma ranging from 0.5 to 1.5. At test time, patches overlapping by $32\times32\times8$ voxels were selected using a sliding window, and overlapping regions were averaged. The models were trained for 100 epochs using the Adam optimizer with initial learning rates set to $2\mathrm{e}-4$ and linearly decreasing to zero over all epochs.
\newpage
\setcounter{section}{1}
\setcounter{subsection}{0}
\renewcommand{\figurename}{Supplementary Fig.}
\setcounter{figure}{0}    
\section*{Supplementary Document B: Supplementary analyses and results}

This document provides additional analyses and results from the SynthRAD2023 challenge, which defined two tasks: 1) magnetic resonance imaging (MRI) to computed tomography (CT) synthesis and 2) cone beam CT (CBCT) to CT synthesis. These additional analyses include comparing the performance differences between all teams for each evaluation metric and investigating their statistical significance. In addition, we analyze the runtime of each team's algorithm and examine the average performance per patient in the test set. Finally, we visually analyze the results for two low-performing patients.

\subsection{Teams' performance and significance}
\label{app:sct_perf}

\begin{figure}[h]
    \centering
    \includegraphics[width=\linewidth]{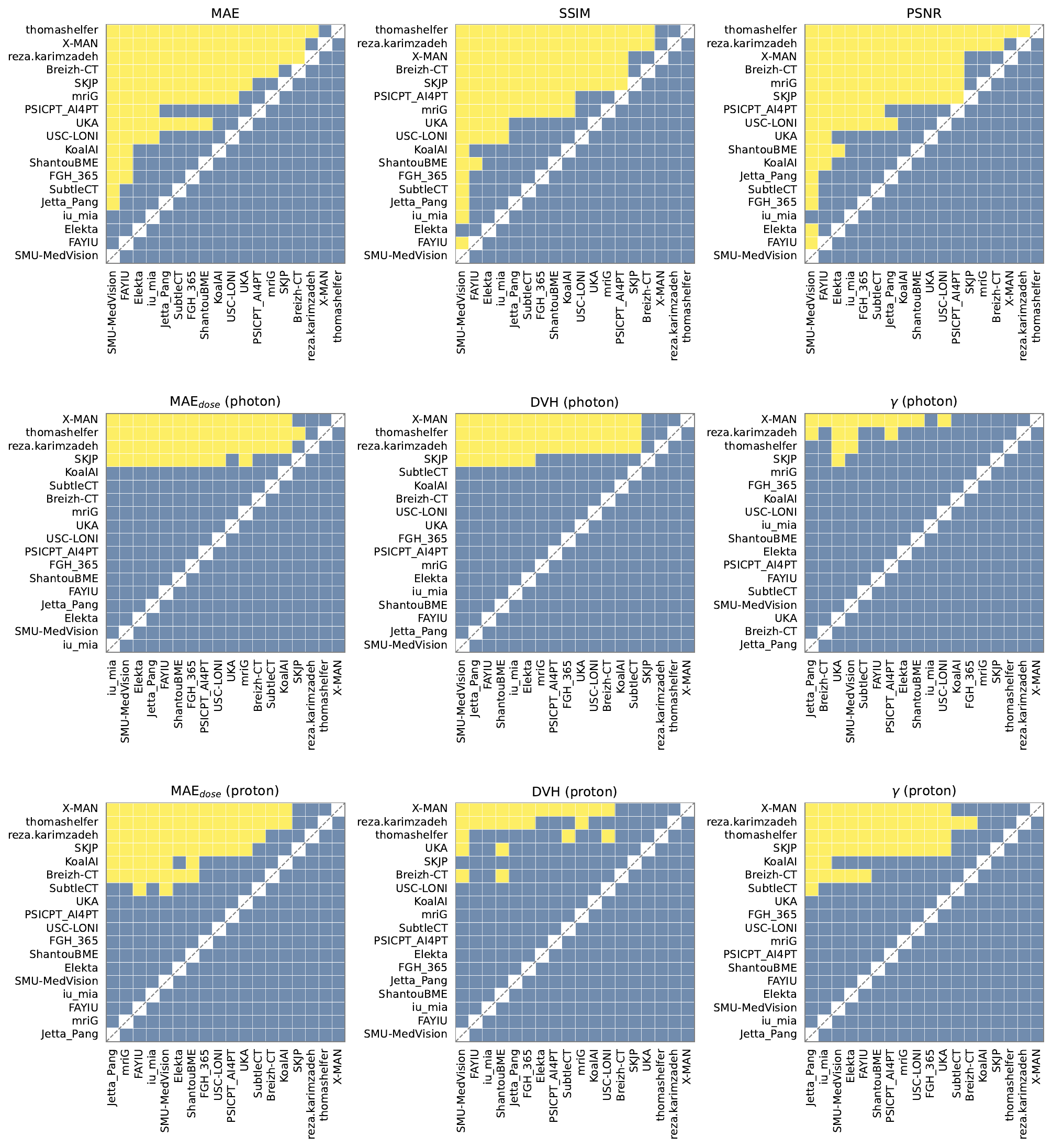}
    \caption{Significance map of task 1. Visualizing pairwise comparisons of team performances for the individual metrics, where teams are sorted based on the performance for the respective metric. Yellow shading in the upper triangle indicates that the team on the x-axis performs significantly better than the y-axis, while blue indicates no significantly better performance. Blue in the lower triangle indicates that the team on the x-axis performs significantly worse than the team on the y-axis.}
    \label{fig:significance_maps_appendix_task1}
\end{figure}

\begin{figure}[h]
    \centering
    \includegraphics[width=\linewidth]{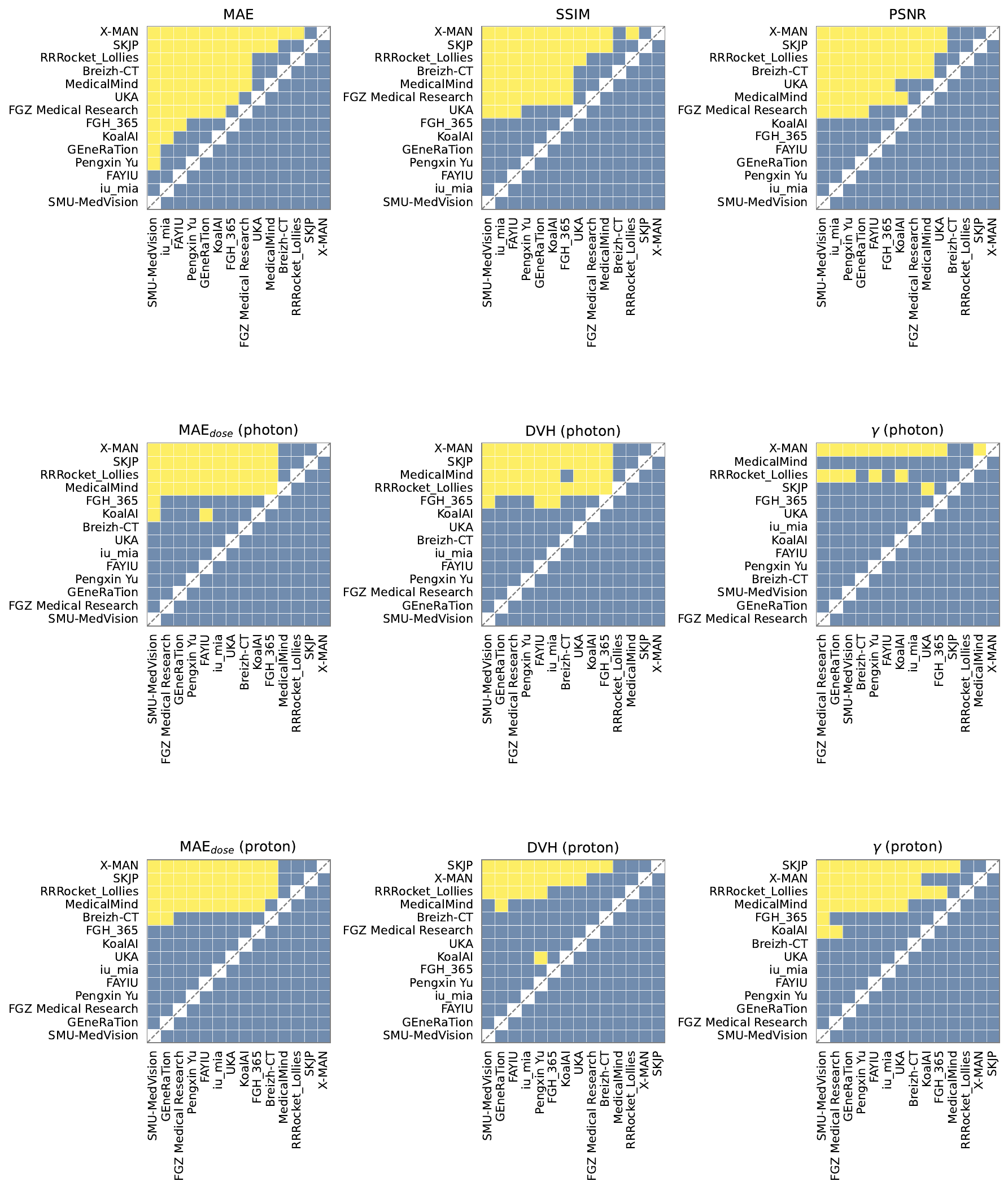}
    \caption{Significance map of task 2. Visualizing pairwise comparisons of team performances for the individual metrics, where teams are sorted based on the performance for the respective metric. Yellow shading in the upper triangle indicates that the team on the x-axis performs significantly better than the y-axis, while blue indicates no significantly better performance. Blue in the lower triangle indicates that the team on the x-axis performs significantly worse than the team on the y-axis.}
    \label{fig:significance_maps_appendix_task2}
\end{figure}

\begin{figure*}[h]
    \centering
    \includegraphics[width=\linewidth]{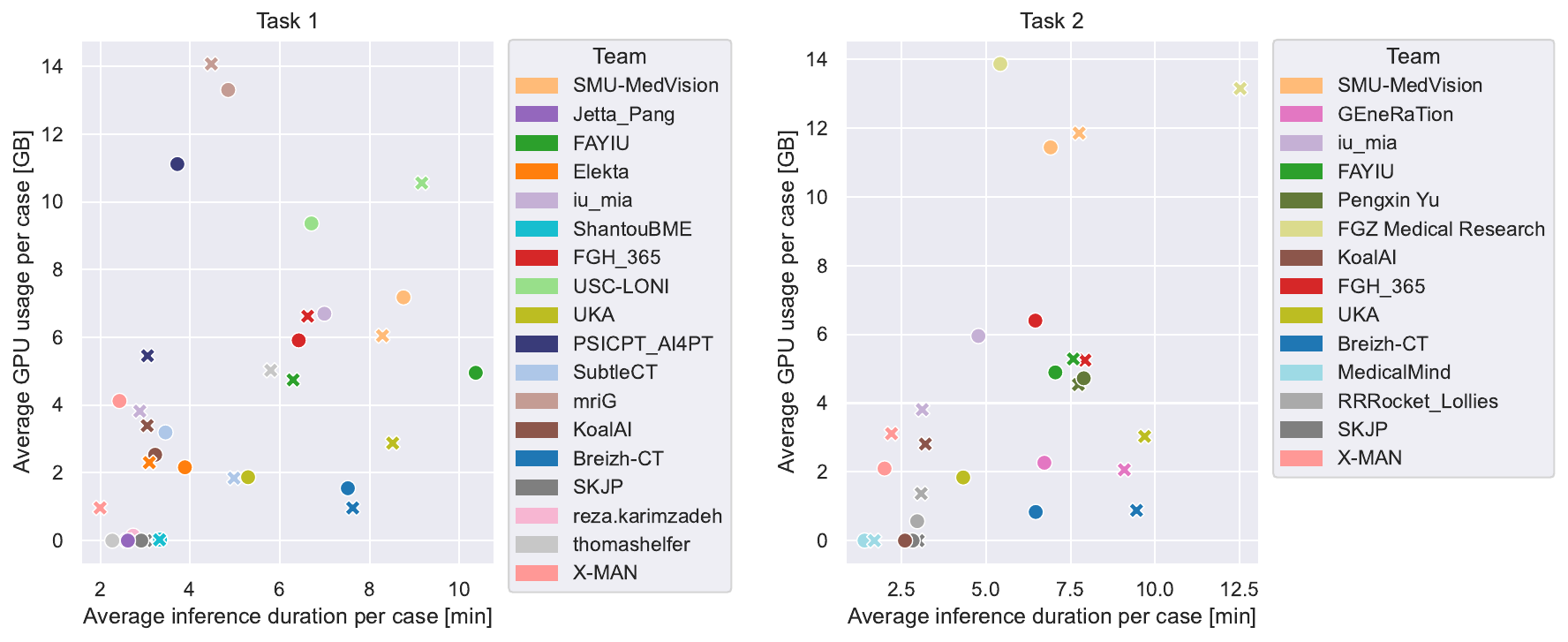}
    \caption{Resource utilization during inference for tasks 1 (left) and 2 (right), represented by inference time and GPU usage. Values are averaged over 60 patients per subtask for each team, with different teams distinguished by colors. A cross represents brain averages, while a circle represents pelvis averages.}
    \label{fig:GPU}
\end{figure*}

To state the significance of a team outperforming another in terms of individual metrics, we employed the Wilcoxon signed-rank test \citep{wilcoxon1945} with Holm's adjustment for multiple testing \citep{holm1979simple} for each metric separately (\Cref{fig:significance_maps_appendix_task1,fig:significance_maps_appendix_task2}). The significance level for this test is set at $\alpha=0.05$.  Based on the image similarity metrics, high-ranking teams robustly outperform lower-ranked teams. Statistical significant improvements were observed when comparing all image metrics between a team and another team ranked at least seven places lower for task 1, or six placed lower for task 2.
However, for the dose metrics, this relation is weaker. In task 1, no statistical significant differences were observed between the top fourteen teams regarding the photon dose metrics and top eleven teams regarding the proton dose metrics.
In task 2, no statistically significant differences were observed between the top eight teams regarding the photon and proton dose metrics, except for the fifth team (Pengxin Yu), which significantly outperforms the sixth team (KoalAI) regarding the proton DVH metric. Furthermore,  \autoref{fig:GPU} shows a detailed overview of the resource utilization per team per subtask.

\subsection{Data influence}
\label{app: data influence}
\begin{figure}
    \centering
    \includegraphics[width=0.9\linewidth]{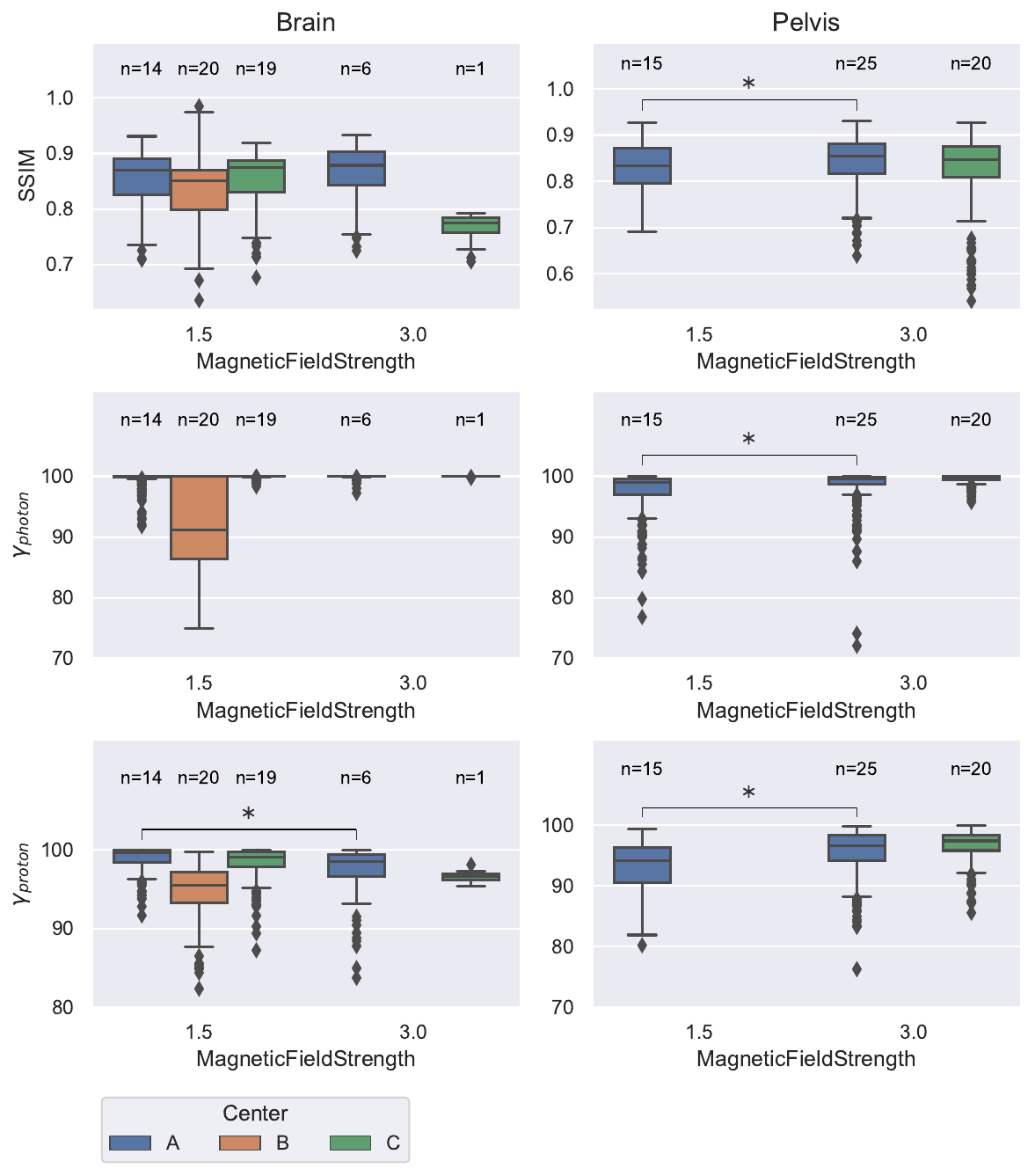}
    \caption{Boxplots of the teams’ performance for task 1 in terms of SSIM and gamma pass rates for photon and proton, grouped by different region, acquisition center and magnetic field strength. The number of test cases per subgroup is indicated by “n=x”. Statistical significance (Mann-Whitney U-test, $\alpha=0.01$) was calculated for center A between different field strengths of the same region and is indicated by an asterisk if significant.}
    \label{fig:fieldstrength}
\end{figure}

For task 1, we performed a more detailed analysis of the influence of magnetic field strength on sCT generation performance. However, the absence of variability in magnetic field strengths for centers B and C constrained this analysis to center A (\autoref{fig:fieldstrength}).
For the brain, the only significant difference was observed for \(\gamma_\text{photon}\), which decreased from \(98.99 \pm 1.43\) for 1.5T to \( 97.33 \pm 3.23\) for 3T. In contrast, for the pelvis, a significant increase in performance was observed for 3T compared to 1.5T. Specifically, the SSIM increased from $0.83 \pm 0.05$ to $0.84 \pm 0.05$, \(\gamma_\text{photon}\) increased from $97.51 \pm 3.45$ to $98.75 \pm 2.59$, and \(\gamma_\text{proton}\) increased from $93.29 \pm 4.05$ to $ 95.64 \pm 3.42$. 

\begin{figure}[h]
    \centering
    \includegraphics[width=0.5\linewidth]{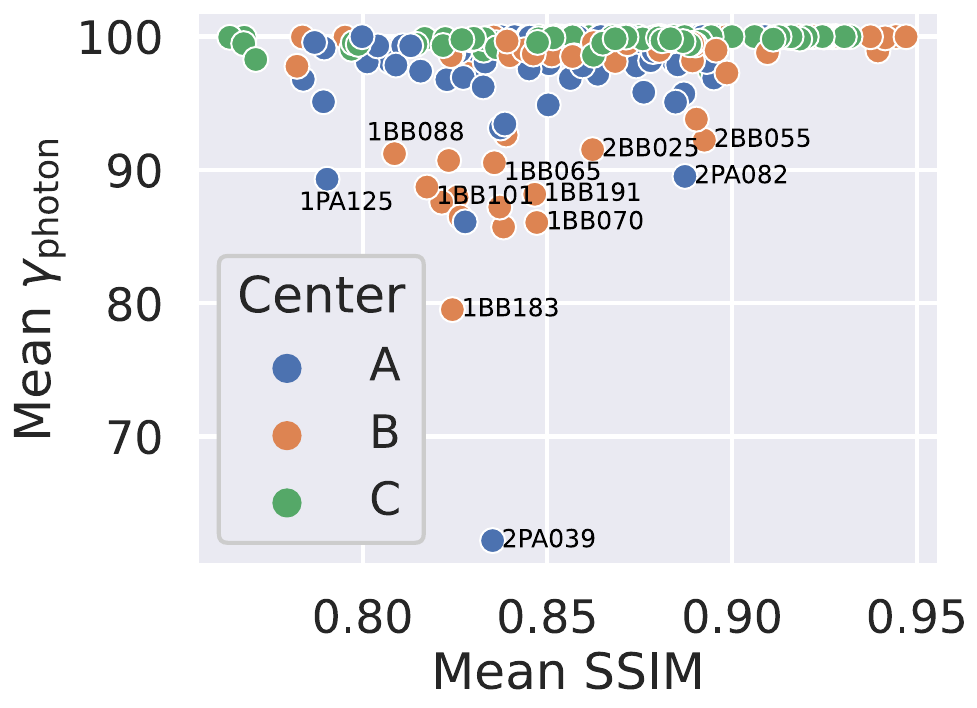}
    \caption{Mean SSIM versus mean photon gamma pass rate per patient, averaged over all participants. The hue encodes the center. Two patients are clear outliers, while multiple patients of center B are underperforming.}
    \label{fig:data_invest_outlier}
\end{figure}

\begin{figure}[h]

\subfloat[Example sCTs for outlier patient 1BB183. The image quality is high, but the CT still contains the table, which is difficult to synthesize if not present in the input. The image error map has been masked with the provided mask.]{%
  \includegraphics[clip,width=0.99\linewidth]{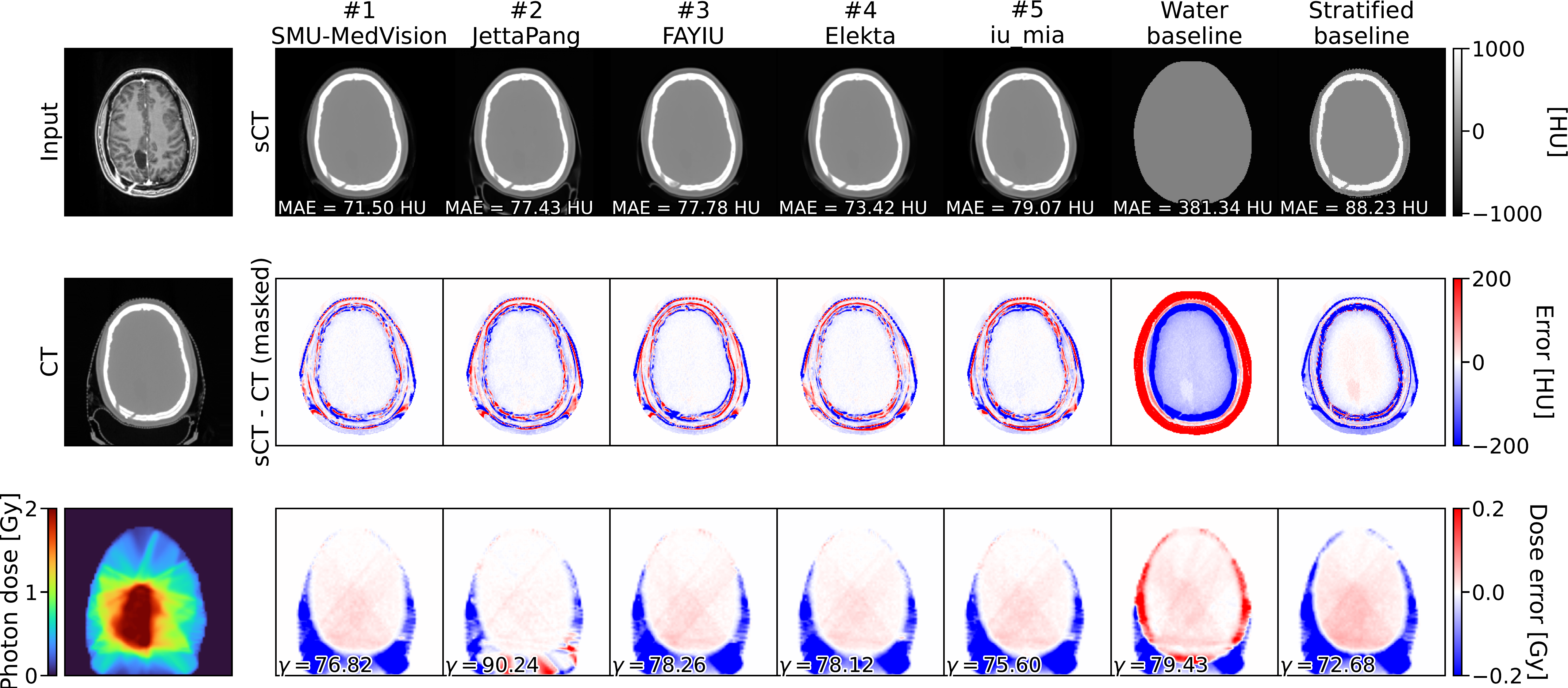}%
}

\vspace{1cm}
\subfloat[Example sCTs for outlier patient 2PA039. The challenging anatomy and artifacts make accurate sCT generation difficult. The image error map has been masked with the provided mask.]{%
  \includegraphics[clip,width=0.99\linewidth]{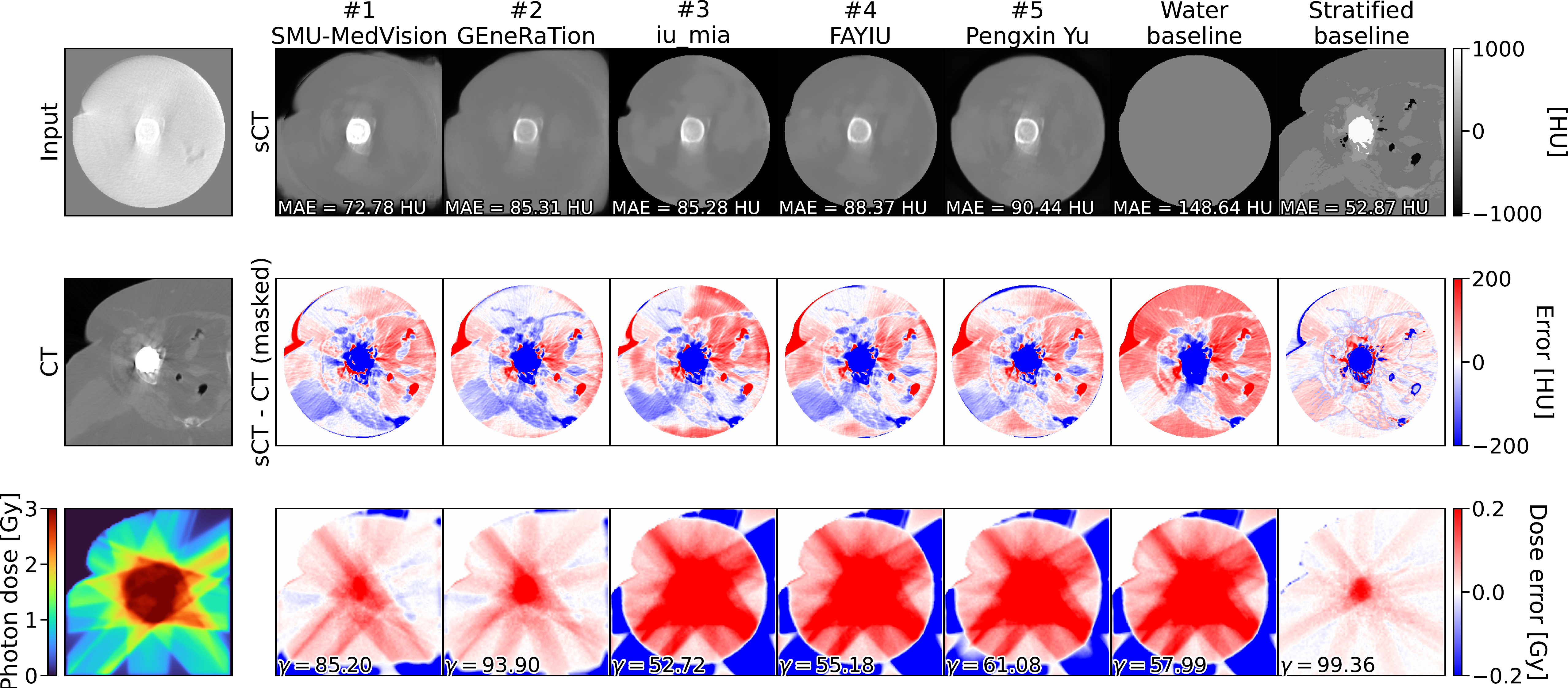}%
}

\caption{Examples of underperforming patients: patient 1BB183 for task 1 (MRI-to-CT; a) and patient 2PA039 for task 2 (CBCT-to-CT: b). The model input is shown in the upper left, and the ground truth is in the center-left. The sCT of the top five participants for task 1 and task 2 are shown in the top row. The difference from ground truth CT after masking with the provided mask is shown in the middle row. On the bottom left is the planned irradiation based on the CT for a photon (a) and proton (b) plan. The bottom row shows the dose difference when the treatment plan is applied to the sCT (CT dose - sCT dose).}
\label{fig:qualitative_evaluation_underperformance}
\end{figure}

In analyzing the synthetic CT (sCT) generation performance at the patient level for both tasks, we present the mean SSIM and mean photon gamma pass rate per patient in \autoref{fig:data_invest_outlier}. One pelvis patient in center A for task 2 is severely underperforming in terms of photon gamma pass rate, while multiple brain patients from center B are underperforming in task 1. A visual inspection of two of these patients (1BB183 and 2PA039) (\autoref{fig:qualitative_evaluation_underperformance}) reveals that for patient 1BB183 (\autoref{fig:qualitative_evaluation_underperformance}a), the image quality of sCTs is high. However, the table of the CT scanner is still visible in the ground truth CT but not in the MRI. Jetta\_Pang was able to synthesize a table, increasing the gamma pass rate (\(\gamma=90.24\%\)) while the other teams did not synthesize a table, which led to large dose errors (\(\gamma=75-79\%\)). This leads to higher dose deposition within the target while omitting the dose deposition in the table, which received more than 10\% of the target dose and is therefore included in the gamma pass rate analysis, causing low gamma pass rates for these patients. Further investigation shows this happened for multiple brain patients from center B in task 1. Furthermore, patient 2PA039 (\autoref{fig:qualitative_evaluation_underperformance}b) suffers from a low-quality CBCT with a central artifact combined with a small field-of-view, making image synthesis difficult.

\end{document}